\documentclass[12pt,a4paper]{book}

\selectfont
\usepackage{amssymb}
\usepackage[english]{babel}
\usepackage{fancyhdr}
\usepackage{booktabs}
\usepackage{cite}
\usepackage{amsmath}
\usepackage{dsfont}
\usepackage{amsfonts}
\usepackage{epsfig}
\usepackage{latexsym}
\usepackage{slashed}
\usepackage{xcolor}
\usepackage{cancel}
\usepackage{verbatim}
\usepackage{nicefrac}
\usepackage{amsbsy}
\usepackage{graphicx}
\usepackage[heightrounded]{geometry}
\usepackage{comment}
\usepackage{url}
\usepackage{stmaryrd}
\usepackage[T1]{fontenc}

\allowdisplaybreaks[1]

\usepackage{titlesec}
\titleformat{\part}[display]{\bf\LARGE\filcenter}{\titlerule[1.5pt] \vspace{1pc} \Huge\MakeUppercase{\partname} \thepart}{1pc}{\titlerule[1.5pt] \vspace{1pc} \Huge}
\titleformat{\chapter}[display]{\bf\Large\filcenter}{\titlerule[1pt] \vspace{1pc} \LARGE\MakeUppercase{\chaptertitlename} \thechapter}{1pc}{\titlerule[1pt] \vspace{1pc} \LARGE}

\usepackage[small,bf]{caption}
\usepackage{subfig}

\renewcommand{\chaptermark}[1]%
         {\markboth{\thechapter.\ #1}{}}
\renewcommand{\sectionmark}[1]%
         {\markright{\thesection\ #1}}
\makeatletter
\def\cleardoublepage{\clearpage\if@twoside \ifodd\c@page\else
    \hbox{}
    \thispagestyle{plain}
    \newpage
    \if@twocolumn\hbox{}\newpage\fi\fi\fi}
\makeatother \clearpage{\pagestyle{plain}\cleardoublepage}

\makeatletter
\renewcommand\tableofcontents{%
    \if@twocolumn
      \@restonecoltrue\onecolumn
    \else
      \@restonecolfalse
    \fi
    \chapter*{\contentsname
        \@mkboth{%
           \contentsname}{\contentsname}}%
    \@starttoc{toc}%
    \if@restonecol\twocolumn\fi
    }
\makeatother

\newcommand{\LMUTitle}[9]{
  \thispagestyle{empty}
  \vspace*{\stretch{1}}
  {\parindent0cm
   \rule{\linewidth}{.7ex}}
  \begin{flushright}

    \vspace*{\stretch{1}}
    \sffamily\bfseries\Huge
    #1\\
    \vspace*{\stretch{1}}
    \sffamily\bfseries\large
    #2
    \vspace*{\stretch{1}}
  \end{flushright}
  \rule{\linewidth}{.7ex}
  \vspace*{\stretch{5}}
  \begin{center}
    \includegraphics[width=2in]{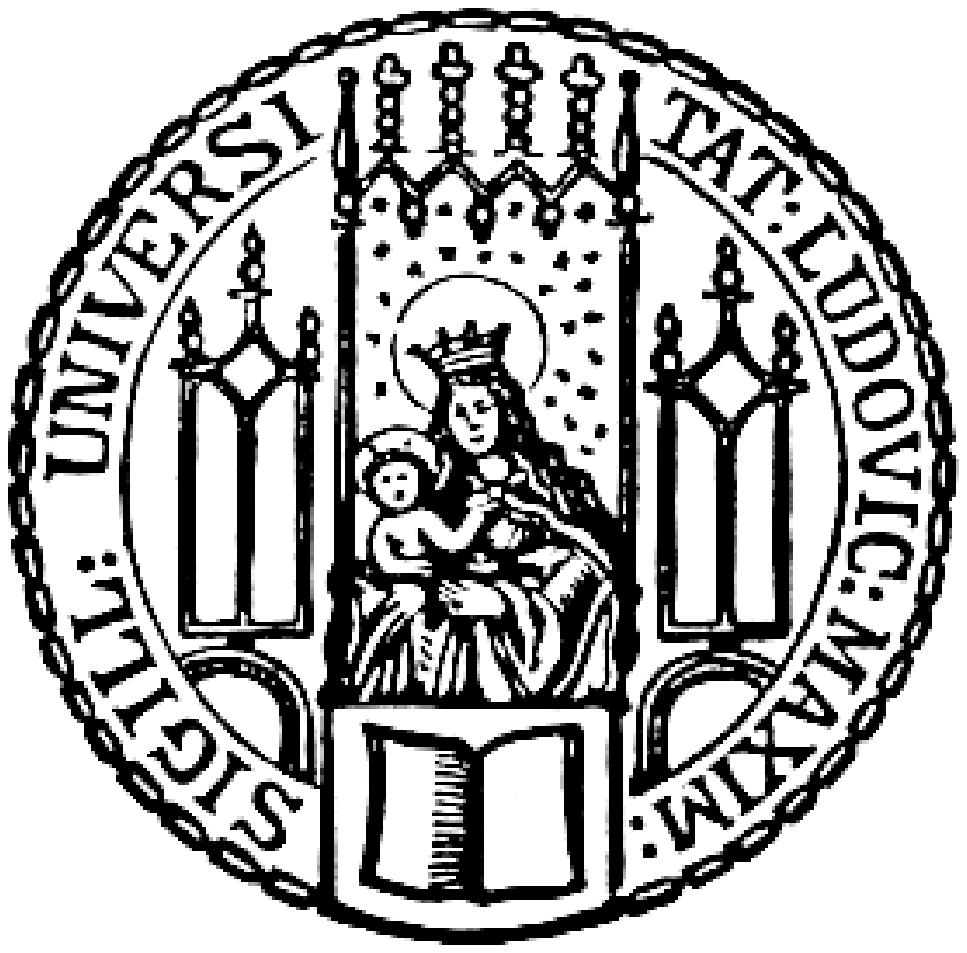} \hspace{2cm}
  \end{center}
  \vspace*{\stretch{1}}
  \begin{center}\sffamily\LARGE{#5}\end{center}
  \newpage
  \thispagestyle{empty}

  \cleardoublepage
  \thispagestyle{empty}

  \vspace*{\stretch{1}}
  {\parindent0cm
  \rule{\linewidth}{.7ex}}
  \begin{flushright}
    \vspace*{\stretch{1}}
    \sffamily\bfseries\Huge
    #1\\
    \vspace*{\stretch{1}}
    \sffamily\bfseries\large
    #2
    \vspace*{\stretch{1}}
  \end{flushright}
  \rule{\linewidth}{.7ex}

  \vspace*{\stretch{3}}
  \begin{center}
    \Large Dissertation\\
    \Large an der #4\\
    \Large der Ludwig--Maximilians--Universit\"at\\
    \Large M\"unchen\\
    \vspace*{\stretch{1}}
    \Large vorgelegt von\\
    \Large #2\\
    \Large aus #3\\
    \vspace*{\stretch{2}}
    \Large M\"unchen, den #6
  \end{center}

  \newpage
  \thispagestyle{empty}

\begin{large}
  \noindent This thesis is based on the author's work published
  in \cite{Cadamuro:2010cz,Cadamuro:2011fd,Arias:2012mb} conducted from October 2009 until March 2012 at
  the Max-Planck-Institut f\"ur Physik
  (Werner-Heisenberg-Institut), M\"unchen, under the supervision of
  Dr.~Georg G.~Raffelt and Dr.~Javier Redondo.
\end{large} 

  \vspace*{\stretch{1}}

  \begin{flushleft}
    \large Erstgutachter:  #7 \\[1mm]
    \large Zweitgutachter: #8 \\[1mm]
    \large Tag der m\"undlichen Pr\"ufung: #9\\
  \end{flushleft}

  \cleardoublepage
}

\renewcommand\({\left(}
\renewcommand\){\right)}
\renewcommand\[{\left[}
\renewcommand\]{\right]}

\newcommand{\parte}{{\ttfamily PArthENoPE}}
\newcommand{\be}{\begin{equation}}
\newcommand{\ee}{\end{equation}}
\newcommand{\bea}{\begin{eqnarray}}
\newcommand{\eea}{\end{eqnarray}}
\definecolor{durbeer}{rgb}{1,0,0.3}

\def\bra{\langle}
\def\ket{\rangle}

\def\degree{g}
\def\h{\xi}
\newcommand{\alt}{\mbox{\;\raisebox{.3ex}
  {$<$}$\!\!\!\!\!$\raisebox{-.9ex}{$\sim$}\;}}
\newcommand{\agt}{\mbox{\;\raisebox{.3ex}
  {$>$}$\!\!\!\!\!$\raisebox{-.9ex}{$\sim$}}\;}

\begin{document}

   \frontmatter

   \LMUTitle
      {\Huge Cosmological limits on axions and axion-like particles}  
      {Davide Cadamuro}                       
      {Treviso, Italien}                             
      {Fakult\"at f\"ur Physik}                         
      {M\"unchen 2012}                          
      {31 August 2012}                            
      {PD Dr.~Georg Raffelt}                          
      {Prof.~Dr.~Stefan Hofmann}                         
      {12 Oktober 2012}                         

\renewcommand\contentsname{Table of Contents} 
\chapter*{Zusammenfassung}

Das Axion ist ein pseudo-Nambu-Goldstone Boson. Es tritt in Erscheinung
nach dem spontanen Bruch der Peccei-Quinn Symmetrie, die als L\"osung des
starken CP-Problems vorgeschlagen wurde. Andere pseudo-Nambu-Goldstone
Bosonen, postuliert in einigen Erweiterungen des Standardmodells, werden
``Axion-Like Particles'' (ALPs) genannt, wenn sie bestimmte Eigenschaften mit
dem Axion teilen, insbesondere die Kopplung an zwei Photonen. Bis jetzt
waren alle Suchen nach A\-xio\-nen und ALPs erfolglos. Dies bedeutet, dass
deren Kopplungen extrem schwach sein m\"ussen. Allerdings k\"onnen Axionen
und ALPs einige beobachtbare astrophysikalische und kosmologische Auswirkungen
haben, anhand derer man den Parameterraum dieser Teilchen einschr\"anken
kann.

Wir konzentrieren uns auf Schranken aus der Kosmologie, die ein ideales
Feld f\"ur die Untersuchung von Axionen und ALPs darstellt. Insbesondere
untersuchen wir als erstes die M\"oglichkeit einer Axion- und
ALP-Population, die w\"ahrend der fr\"uhesten Augenblicke des Universums
entstanden ist. Die Bedeutung dieser Analy\-se r\"uhrt daher, dass Axionen
und ALPs wegen ihrer schwachen Wechselwirkung und der besonderen
Produktionsmechanismen ideale Kandidaten f\"ur die dunkle Materie sind.
Schlie{\ss}lich betrachten wir die Folgen des Zerfalls dieser Teilchen f\"ur
bestimmte kosmologische Observablen, n\"amlich f\"ur das Photonenspektrum
von Galaxien, f\"ur den kosmoschen Mikrowellenhintergrund, f\"ur die
effektive Zahl an Neutrinos und die urspr\"ungliche H\"aufigkeit der
Elemente. Unsere Schranken stellen den striktesten Test eines fr\"uhen
Zerfalls von Axionen und ALPs dar und schlie{\ss}en einen Teil des
Parameterraums von ALPs aus, der ansonsten experimentell nur schwer
zug\"anglich ist.

\chapter*{Abstract}

The axion is a pseudo-Nambu-Goldstone boson.
It appears after the spontaneous breaking of the Peccei-Quinn symmetry, which was proposed to solve the strong-CP problem.
Other pseudo-Nambu-Goldstone bosons, postulated in some extensions of the standard model of particle physics, are called axion-like particles (ALPs) if they share certain characteristics with the axion, in particular a coupling to two photons.
Thus far, axion and ALP searches have been unsuccessful, indicating that their couplings have to be extremely weak.
However, axions and ALPs could be responsible for some observable effects in astrophysics and cosmology, which can also be exploited to constrain the parameter space of these particles.

We focus on limits coming from cosmology, which is an optimal field for studying axions and ALPs.
In particular, we first investigate the possibility of a primordial population of axions and ALPs arising during the earliest epochs of the universe.
The importance of this analysis lies on the fact that axions and ALPs are ideal dark matter candidates because of their faint interactions and their peculiar production mechanisms.
Finally, we consider the consequences of the decay of such a population on specific cosmological observables, namely the photon spectrum of galaxies, the cosmic microwave background, the effective number of neutrino species, and the abundance of primordial elements.
Our bounds constitute the most stringent probes of early decays and exclude a part of the ALP parameter space that is otherwise very difficult to test experimentally.

\newpage
\mbox{}
\newpage
\vspace*{\stretch{1}}
\begin{flushright}
\textit{To my wife Sara}
\end{flushright}
\vspace*{\stretch{1}}
\newpage
\mbox{}
\newpage
\pagestyle{fancy}

  \tableofcontents
   \markboth{Table of Contents}{Table of Contents}



  \mainmatter\setcounter{page}{1}

\chapter*{Preface}
\addcontentsline{toc}{chapter}{Preface}
\markboth{Preface}{Preface}
Day after day, cosmology has become more useful as a tool for particle physics. 
In its earliest epoch, the universe provided the environment with the highest known energy, much higher than those presently tested in colliders. 
We still miss a particle physics description of such high energy scales, therefore many open cosmological questions are waiting for a satisfying answer in terms of microphysics.
Every proposal of physics beyond the standard model of particle physics (SM) must deal with cosmology, notably providing mechanisms for inflation and baryogenesis and candidates for dark matter and dark energy.
All these issues provide further motivations to reach a complete picture of particle theory.
 
On the other hand, through the study of those phases of cosmic evolution, when both the microphysics and cosmological descriptions are well settled, physicists can put severe constraints on physics beyond the SM.
Big-bang nucleosynthesis (BBN), the cosmic microwave background (CMB) and large scale structure (LSS) give precise indications about what happened when the universe was older than a second.  
Moreover, in the last few years the vastly increasing amount and quality of data has propelled cosmology to its precision era.
New particle physics models have to face these broad data sets and must be consistent with them.
Cosmology is therefore a crucial testing ground for particle physics.

But cosmology does not play only the killjoy role.
In particular, because of the intrinsic \emph{long exposure time} which is peculiar of cosmological phenomena, together with the \emph{high luminosity} provided by the universe itself, cosmology gives the possibility of testing very weakly interacting particles otherwise not accessible to high energy particle physics experiments. 
Axion and axion-like particles (ALPs) are among these species, and they are the main characters of this dissertation. 
Very stringent limits on their mass and coupling constants come from observations of high-intensity astrophysical settings like globular clusters or the sun. 
Cosmology also plays an important role in the quest for finding them, and, as we will see, the bounds that it provides are complementary to those provided by astrophysics.
 
In the first chapter, we introduce the QCD axion and ALPs, providing some theoretical motivation.
We also review briefly the present status of the exploration of their parameter space through non-cosmological approaches. 
We already mentioned the astrophysical observations, and here we describe the main experimental approaches to axion and ALP physics.
We conclude this chapter with some hints for the presence of such particles, supported by astrophysical clues.

The primordial production of a population of axions and ALPs is the topic of chapter~\ref{chap:DM}. 
This is the basic question to scrutinise before even to start the discussion about the bounds cosmology could provide: without a relic population there can be no cosmological limits.
Moreover, establishing a relic population of axions or ALPs is an intriguing topic in its own right, as it could provide the solution to the dark matter mystery.
The analysis of the stability of these particles is therefore the optimal way to conclude this chapter, and to introduce the core question of this dissertation.

If axions and ALPs are unstable, they decay into photons.
Their decay products could influence the evolution of the universe and the cosmological observables.
In chapter \ref{chap:photons} we treat the limits coming from the \emph{late} decays of axions and ALPs.
These are obtained studying how the decay affects the CMB and considering the possibility of directly detecting the photons emitted in the spectrum of galaxies and the extragalactic photon background.

At earlier times, when the universe was hot and dense enough to rapidly lose the direct imprint of the decay products via thermal scattering, the decay would have a more subtle and indirect influence. 
In chapter~\ref{chap:dilution} we notice that the decay, which injects a large amount of photons in the primordial plasma, can effectively dilute baryons and neutrinos, if it happens after about ten milliseconds since the big bang.
The primordial elemental abundances and the amount of radiation during the first minutes of our universe are strongly affected by this dilution and provide further limits on the axion and ALP parameters. 
To conclude this chapter, we reanalyse the constraints we found considering other decay channels besides the photon one. 
Finally, chapter~\ref{chap:conclusion} is devoted to summarise our arguments and to draw our conclusions.

The research on axions and ALPs will be one of the main frontiers of particle physics in the near future, since these pseudoscalars can solve some of the unresolved problems of particle physics. 
Their experimental discovery would be a true milestone along our path to understand Nature.
The technological challenge to reach this aim is pushing our capability of measuring extraordinarily small signals, but a long path remains to be covered.
Astrophysical and cosmological observations provide directions about the particle parameters, offering a guidance in the design of devoted discovery experiments.
The improving quality and quantity of astrophysical and cosmological data gives us confidence on the possibilities the sky offers.
That is why we consider this kind of analysis of fundamental importance in the quest for new physics.

\chapter{Axions and axion-like particles}\label{chap:intro}
\section{The strong-CP problem and the axion} 

Quantum chromodynamics (QCD), the theory of strong interactions, includes a P, T and thus CP violating term~\cite{Peccei:2006as,Kim:1986ax,Cheng:1987gp,Kim:2008hd},
\be \label{thetaterm}
\mathcal{L}_\theta=\bar{\theta}\frac{\alpha_s}{8\pi}\frac{\epsilon_{\mu\nu\rho\sigma}}{2} G^{\rho\sigma}_a G^{\mu\nu}_a\; ,
\ee
where $G^{\rho\sigma}_a$ is the gluon field strength, $\alpha_s$ the fine structure constant for colour interactions and the so called $\theta$ angle, $-\pi<\bar{\theta}\leq\pi$, is the effective parameter controlling CP violation. 
Here, and in the following, we will use natural units with $\hbar=c=k_\mathcal{B}=1$, where $k_\mathcal{B}$ is the Boltzmann constant.
It is common practice to write the dual tensor $\epsilon_{\mu\nu\rho\sigma} G^{\rho\sigma}_a/2$ simply as $\tilde{G}_{a\mu\nu}$, thus we will simply write $G\tilde{G}$ for the trace in the Lagrangian \eqref{thetaterm}.
The sources of this $\theta$-term are the Adler-Bell-Jackiw anomaly of the axial current in QCD~\cite{'tHooft:1976up,'tHooft:1976fv} and the topology of the QCD vacuum~\cite{Callan:1976je}. 
There are no theoretical hints about which value between $-\pi$ and $+\pi$ the parameter $\bar{\theta}$ could choose, so we just expect it to be an $\mathcal{O}(1)$ quantity.

The gluon field $A^\mu_a$ transforms like a 4-vector under C, P and T.
We can define the coloured electric and magnetic fields as
\begin{subequations}\label{eq:EBcol}
\begin{align}
E_a^k\equiv\ & G^{0k}_a=\partial^0 A^k_a-\partial^k A^0_a+g_s f_{abc}A^0_b A^k_c\; ,\\
B_a^k\equiv\ & \epsilon^{ijk} G_{ij\,a}=\epsilon^{ijk}\(\partial_i A_{j\,a}-\partial^j A_{i\,a}+g_s f_{abc}A_{i\,b} A_{j\,c}\)\; ,
\end{align}
\end{subequations}
and they transform like their electromagnetic counterparts under C, P and T.
In the definitions \eqref{eq:EBcol}, $g_s$ is the strong interaction coupling constant and $f_{abc}$ are the $SU(3)$ structure constants. 
The colour indices are $abc$, while $ijk$ are spatial indices.
Because of the properties of the completely antisymmetric tensor $\epsilon_{\mu\nu\rho\sigma}$, the Lagrangian~\eqref{thetaterm} can be written as $G\tilde{G}=-4\vec{B}_a\cdot\vec{E}_a$.
The scalar product of a polar vector and an axial vector violates P, T and thus CP, once CPT is taken for granted.

Whether CP is a good symmetry for QCD is not a fundamental quastion but a phenomenological one.
At low energy the Lagrangian~\eqref{thetaterm} induces electric dipole moments in baryons, which have not yet been observed.
In particular, many measurements have been performed on the neutron one.
Calculations predict the neutron electric dipole moment to be $d_n\sim10^{-16}\bar{\theta}\,e\,{\rm cm}$~\cite{Baluni:1978rf,Crewther:1979pi}, where $e$ is the electron electric charge. 
Currently, the best experimental limit is $d_n<0.29\times10^{-25}\,e\,{\rm cm}$~\cite{Baker:2006ts}, which translates into
\be
|\bar{\theta}|\lesssim10^{-10}\; .
\ee
The essentials of the strong-CP problem are all here: we were not expecting such an extremely small value for $\bar{\theta}$.

Roberto Peccei and Helen Quinn proposed an elegant solution to this puzzle, introducing a global chiral $U(1)$ symmetry, called the \emph{Peccei-Quinn symmetry} $U(1)_{\rm PQ}$~\cite{Peccei:1977hh,Peccei:1977ur}. 
This symmetry is spontaneously broken, and its Nambu-Goldstone boson (NGB) $a$, the celebrated \emph{axion}~\cite{Weinberg:1977ma,Wilczek:1977pj}, couples to gluons because the $U(1)_{\rm PQ}$ symmetry is violated by the colour anomaly.
Thus, another term involving $G\tilde{G}$ enters in the QCD Lagrangian,
\be\label{axionlagr1}
\mathcal{L}_{aG\tilde{G}}=-\frac{a(x)}{f_a}\frac{\alpha_s}{8\pi}G\tilde{G}\; ,
\ee
where $f_a$ is the order parameter associated with the breaking of $U(1)_{\rm PQ}$. 
For the moment we ignore the subtleties related to the definition of $f_a$.
At energies below the confinement scale of colour interactions, $\Lambda_{\rm QCD}\simeq 1$ GeV, gluons and quarks have to be integrated out, and QCD is described by an effective chiral Lagrangian.
The terms \eqref{thetaterm} and \eqref{axionlagr1}, together with the anomalous contribution of the $U(1)_A$ symmetry of the effective chiral QCD Lagrangian, become an effective potential that can be parametrised as
\be\label{eq:potential}
V(a)\simeq \Lambda_{\rm QCD}^4\[1-\cos\(\frac{a}{f_a}-\frac{\eta^\prime}{f_\eta}-\bar{\theta}\)\]\; .
\ee
Here, $\eta^\prime$ is the pseudo-NGB (PNGB) of the anomalous $U(1)_A$, and $f_\eta$ is a parameter with dimensions of a mass.
The periodicity of the potential is due to the topological and instantonic nature of the $G\tilde{G}$ Lagrangian. 
The configuration of minimum energy for the potential \eqref{eq:potential} is realised by the CP-conserving linear combination $\langle a \rangle=\(\bar{\theta}-{\langle\eta^\prime\rangle}/{f_\pi}\)f_a$, as required by the Vafa-Witten theorem~\cite{Vafa:1984xg}.
Since the $a$ field is massless, it can align itself with a null energy cost to the CP-conserving point: \emph{the strong-CP problem is solved}, independently of the value of $f_a$, $\bar{\theta}$ and~$\langle\eta^\prime\rangle$.

Because of the effective potential~\eqref{eq:potential}, the mass of the eigenstate $\eta^\prime+a\,f_\eta /f_a$ is $m_{\eta^\prime}\sim\Lambda_{\rm QCD}^2/f_\eta$, which solves Weinberg's $U(1)$ problem\footnote{
An approximate $U(1)$ axial-vector current would require the presence of a pseudoscalar boson with mass smaller than $\sqrt{3}m_\pi$, which is not observed~\cite{Weinberg:1975ui}. 
This fact was considered a problem before realising that the axial current is violated by colour anomaly~\cite{'tHooft:1976up,'tHooft:1976fv}.
Thus the $\theta$-term in the QCD Lagrangian solves the $U(1)$ problem, and this provides significance to the $\theta$-vacuum.
Of course, the presence of the axion is not required to solve the $U(1)$ problem, but it helps to recover the observed CP conservation.}.
The orthogonal combination, which mainly consists of axion, mixes with the other scalar mesons in the particle spectrum and acquires mass~\cite{Weinberg:1996kr}. 
It is this combination which is usually called axion, and whose phenomenology is studied.
In first approximation --- considering only up and down quark contributions --- the axion mass is~\cite{Bardeen:1977bd}
\be\label{amass}
m_a=\frac{m_\pi f_\pi}{f_a}\frac{\sqrt{m_u m_d}}{m_u + m_d}\simeq 6\ {\rm eV} \(\frac{10^6 \ {\rm GeV}}{f_a} \)\; ,
\ee
where $m_\pi=135$ MeV is the neutral pion mass and $f_\pi=92$ MeV is the pion decay constant.
The errors in the measurements of the light quark mass ratio establish the range $m_u/m_d=0.3$--$0.6$ for this quantity~\cite{Nakamura:2010zzi}, which translates into a 10\% uncertainty on the axion mass, $m_a=5.23$--$6.01$ eV for $f_a=10^6$~GeV.
The preferred value $m_u/m_d=0.56$ gives us the result~\eqref{amass} after rounding it to 10\% accuracy.

In the original Peccei-Quinn model, the additional chiral symmetry was imposed on the SM through two Higgs doublets, linking the spontaneous breaking of $U(1)_{\rm PQ}$ to the electro-weak symmetry breaking, thus $f_a\sim E_{\rm EW}= 246$~GeV.
However, this first attempt to solve the strong-CP problem through the Peccei-Quinn symmetry was quickly ruled out, because the axion did not show up in experimental data~\cite{Ellis:1978my}.

A way to save the Peccei-Quinn mechanism, bypassing the experimental limits, is to raise the parameter $f_a$ by several orders of magnitude.
The scale $f_a$ suppresses both the axion mass and couplings, a fact that will become more clear shortly.
For the moment it is sufficient to notice that since the axion is usually introduced as a phase in the Higgs sector, it needs to be normalised by a constant with dimension of energy.
It is therefore the dimensionless ratio $\(a/f_a\)$ that always appears in effective Lagrangians for axions.
Increasing $f_a$ automatically lowers the axion couplings to SM particles, leaving untouched the validity of the solution to the strong-CP problem, as we have seen before. 
There are several implementations of this idea, which are called \emph{invisible axion} models.
The first of them was the Kim-Shifman-Vainshtein-Zakharov (KSVZ) axion model~\cite{Kim:1979if,Shifman:1979if}, which we discuss here in more detail for its simplicity. 
A Dirac quark field $Q$ --- a colour triplet in the fundamental representation with no bare mass --- and a SM gauge group singlet Higgs-scalar $S$ are added to the standard model.
The Lagrangian we need is 
\begin{align}\label{LKim}
\mathcal{L}_{\rm KSVZ}=&-\frac{1}{4}G_a^{\mu\nu}G_{a\mu\nu} + \bar{\theta}\frac{\alpha_s}{8\pi}G\tilde{G} + i\bar{Q} \gamma^\mu\partial_\mu Q + g_s A_a^\mu \bar{Q} \gamma_\mu \lambda_a Q  \nonumber \\ 
&-y\( Q^\dagger_L S Q_R + Q^\dagger_R S^* Q_L \)+\frac{1}{2}\partial_\mu S^*\partial^\mu S - \lambda\(S^* S-v_a^2\)^2\; ,
\end{align}
where $\gamma^\mu$ are the Dirac matrices, $g_s$ is the strong interaction coupling constant, $A_a^\mu$ the gluon field, $\lambda_a$ are the Gell-Mann matrices, $Q_{L(R)}$ is the left (right) handed projection of the Dirac spinor, $y$ the Yukawa coupling, $\lambda>0$ and $v_a\gg E_{\rm EW}$ are the parameters of the Mexican hat potential, the last one with dimensions of energy. 
For the moment we do not take into account the interaction of $Q$ with the electroweak gauge fields. 
Under the global chiral $U(1)_{\rm PQ}$, the fields transform as
\begin{subequations}\label{PQtrans}
\begin{align}
Q_L &\rightarrow Q_L e^{i {\alpha}/{2}}\\
Q_R &\rightarrow Q_R e^{-i {\alpha}/{2}}\\ 
S &\rightarrow S e^{i \alpha}\; , 
\end{align}
\end{subequations}
leaving the Lagrangian $\mathcal{L}_{\rm KSVZ}$ invariant at the classical level. 

\begin{figure}[tbp] 
   \centering
   \includegraphics[width=7cm]{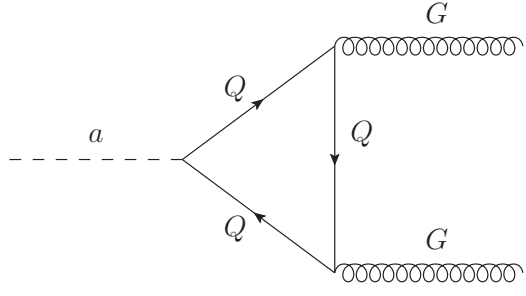}   
   \caption{Triangle loop diagram for the effective axion-gluon interaction of equation~\eqref{deltaL}.}
   \label{fig:triangle}
\end{figure} 

It is convenient to express the scalar Higgs singlet in its polar form, 
\be
S(x)=\rho(x)\exp\({i\frac{a(x)}{v_a}}\) \; .
\ee
At energies lower than $v_a$, $S$ rolls toward the minimum of the Mexican hat potential, and we can make the substitutions $\rho\rightarrow v_a$. 
If we limit our physics consi\-derations to energies lower than the singlet mass, we can keep only the phase field $a$. 
This field changes as $a\rightarrow a+\alpha v_a$ under PQ transformations~\eqref{PQtrans}. 
However, because of the invariance of the Lagrangian~\eqref{LKim}, the energy cost of this shift is null, and therefore $a$ is the NGB for $U(1)_{\rm PQ}$.
Under this approximation, the Lagrangian is now
\begin{align}\label{LKim2}
\mathcal{L}^\prime_{\rm KSVZ}=&-\frac{1}{4}G_a^{\mu\nu}G_{a\mu\nu} + \bar{\theta}\frac{\alpha_s}{8\pi}G\tilde{G}  + i\bar{Q} \gamma^\mu\partial_\mu Q + g_s G_a^\mu \bar{Q} \gamma_\mu \lambda_a Q  \nonumber \\ 
&-y v_a\( Q^\dagger_L e^{i\frac{a}{v_a}} Q_R + Q^\dagger_R e^{-i\frac{a}{v_a}} Q_L \)+\frac{1}{2}\partial_\mu a \partial^\mu a \; .
\end{align}
The spontaneous breaking of the Peccei-Quinn symmetry produces an effective mass term $m_Q=y v_a$ for the quark field, once the phase terms between the left and right spinors are removed.
To achieve this, we chiral rotate the quark field, 
\be
Q_L \rightarrow Q_L \exp\({i\frac{a}{2 v_a}}\)\;,\qquad Q_R \rightarrow Q_R \exp\({-i\frac{a}{2 v_a}}\)\; .
\ee
This transformation adds to the Lagrangian~\eqref{LKim2} the terms
\be\label{deltaL}
\delta\mathcal{L}=-\frac{\partial_\mu a}{2 v_a}\bar{Q}\gamma_5\gamma^\mu Q - \frac{\alpha_s}{8\pi}\frac{a}{v_a} G\tilde{G}\; ,
\ee
where the last term is the contribution of the colour anomaly, originating from the triangle diagram of figure \ref{fig:triangle}. Dealing with energy scales much lower than $m_Q$, the quark field can be integrated out, and the remaining terms of the Lagrangian are
\be\label{axionlagr2}
\mathcal{L}_{\rm KSVZ}=-\frac{1}{4}G_a^{\mu\nu}G_{a\mu\nu} + \(\bar{\theta}-\frac{a}{v_a}\)\frac{\alpha_s}{8\pi}G\tilde{G}+\frac{1}{2}\partial_\mu a \partial^\mu a\; ,
\ee
among which there is the $\mathcal{L}_{aG\tilde{G}}$ term of equation \eqref{axionlagr1}. 

If many heavy quark fields $Q$ are introduced in this model, each of them would produce a contribution like equation \eqref{deltaL}. 
Under a PQ-transformation, these heavy quarks transform if a PQ charge $q^i_{\rm PQ}$ is assigned to them, i.e.~$Q^j_L\rightarrow Q^j_L\exp(i q^j_{\rm PQ}\alpha/2)$ and $Q^j_R\rightarrow Q^j_R\exp(-i q^j_{\rm PQ}\alpha/2)$ if $S \rightarrow S e^{i \alpha}$.
This leaves the Lagrangian invariant under the condition $q^j_{\rm PQ}=1$.
In the effective Lagrangian we define the \emph{Peccei-Quinn scale} or \emph{axion decay constant} to be
\be\label{eq:PQscale}
f_a\equiv\frac{v_a}{\mathcal{N}}\; ,
\ee
where $\mathcal{N}=\sum_j q^j_{\rm PQ}$ counts the number of species that are PQ-charged, and the form of $\mathcal{L}_{aG\tilde{G}}$ of equation \eqref{axionlagr1} is restored.

A triangle diagram, like that of figure~\ref{fig:triangle}, can create other anomalous contributions to equation \eqref{deltaL}, if the heavy quark fields are also coupled to electroweak gauge bosons. 
In the low-energy regime, this means the presence of a two-photon coupling for the axion,
\be\label{2phot}
\mathcal{L}_{a\gamma\gamma}=-\frac{g_{a\gamma\gamma}}{4}a F\tilde{F}\; ,
\ee
$F^{\mu \nu}$ being the photon field strength and $\tilde{F}_{\mu \nu}$ its dual. 
Anyway, because of axion-meson mixing, the two-photon coupling arises even if the heavy quark field is completely decoupled from the $SU(2)\times U(1)$ gauge sector. 
The coupling constant $g_{a\gamma\gamma}$ is actually the sum of two contributions~\cite{Kim:1986ax},
\be \label{2photcoupling}
g_{a\gamma\gamma}=\frac{\alpha}{2\pi f_a}\(\frac{1}{\mathcal{N}}\sum_{i=1}^{N} q^i_{\rm PQ} \(q^i_{\rm EM}\)^2-\frac{2}{3}\frac{4m_d+m_u}{m_d+m_u} \)\equiv\frac{\alpha}{2\pi f_a}C_\gamma\; ,
\ee
where $\alpha$ is the fine structure constant.
The first term is the electromagnetic anomalous contribution of the heavy quarks, $q^i_{\rm PQ\ (EM)}$ being the Peccei-Quinn (electromagnetic) charge of the $i$-th $Q$ field, and the second one is produced by the axion-meson mixing. 
This model dependence can be parametrised with the dimensionless quantity $C_\gamma$.
In the literature, the KSVZ axion is usually defined to have no electromagnetic anomaly and thus to have $|C_\gamma|\simeq1.95$ for $m_u/m_d=0.56$.
Because of the uncertain up and down quark mass ratio, the parameter $|C_\gamma|$ could be anywhere between 1.92 and 2.20.

Another popular axion model was proposed by Zhitnitsky~\cite{Zhitnitsky:1980tq} and by Dine, Fischler and Srednicki~\cite{Dine:1981rt} and is called the DFSZ axion. 
Again, the axion is related to the phase of a Higgs-like scalar singlet $S$ which is added to a two Higgs doublet extension of the SM.
The scalar $S$ is coupled only to the Higgs doublets $H_u$ and $H_d$ in the Higgs potential, 
\begin{align}\label{DFSZpotential}
V(H_u,H_d,\,&S)=\  \lambda_u\({H_u}^\dagger H_u-v_u^2\)^2+ \lambda_d\({H_d}^\dagger H_d-v_d^2\)^2+\lambda\(S^* S-v_a^2\)^2\nonumber \\
&+\chi \({H_u}^\dagger H_u\)\( {H_d}^\dagger H_d\)+ \zeta \({H_u}^\dagger H_d\)\( {H_d}^\dagger H_u\) \nonumber \\
&+\[\gamma_u \({H_u}^\dagger H_u\)+\gamma_d \({H_d}^\dagger H_d\)\]S^* S+\gamma\[\({H_u}^\dagger H_d\)S^2 + {\rm h.c.}\] \; ,
\end{align}
where $\lambda_u$, $\lambda_d$, $\lambda$, $\chi$, $\zeta$, $\gamma_u$, $\gamma_d$ and $\gamma$ are real dimensionless parameters of the potential.
This potential is invariant under the PQ-transformation
\be\label{PQtransDFSZ}
H_u\rightarrow H_u e^{i X_u}\, ,\quad H_d\rightarrow H_d e^{i X_d}\, ,\quad S\rightarrow S e^{i \(X_u-X_d\)/2}\, .
\ee
The model can equally work if the last term of the potential is $\gamma\[\(H_u^\dagger H_d\)S + {\rm h.c.}\]$, but this time $\gamma$ has to have mass dimensions and the PQ-transformations have to be adapted.
The Yukawa Lagrangian is
\be
\mathcal{L}_Y=y_u \epsilon_{ab}\bar{Q}_L^a H_u^{\dagger b} u_R +y_d \bar{Q}_L H_d d_R+y_d \bar{L}_L H_d l_R+{\rm h.c.}\; ,
\ee
where $\epsilon$ is the $2\times2$ antisymmetric matrix and, in this case, $Q_L$ and $L_L$ are the left-handed $SU(2)$ quark and lepton doublets of the SM. 
The right-handed charged lepton component is $l_R$, and the Yukawa couplings are the $y$s.
The vacuum expectation values (VEV) of the Higgs fields can be written as
\be\label{yukawaDFSZ}
\langle H_u\rangle=\frac{1}{\sqrt{2}}\(\begin{array}{c} 0 \\ v_u  \end{array}\)\, , \quad \langle H_d\rangle=\frac{1}{\sqrt{2}}\(\begin{array}{c} 0 \\ v_d  \end{array}\)\, , \quad \langle S\rangle= v_a \; .
\ee
The two Higgs doublets have four degrees of freedom each, and three of them are $SU(2)$ phases, while $S$ has got just a radial excitation and a phase, like in the KSVZ model.
For phenomenological reasons we want $v_a\gg v=\sqrt{v_u^2+v_d^2}\sim E_{\rm EW}$.
Once $U(1)_{\rm PQ}$ is broken, the radial part of $S$ freezes on its VEV, while its phase degree of freedom remains free.
At energies below the EW phase transition, the two Higgs doublets can be written in the unitary gauge deleting from the theory the redundant massless degrees of freedom, which are ``eaten'' by the gauge bosons.
At this point, the neutral orthogonal combination of degrees of freedom which survives mixes with the phase of the $S$ boson due to the last term of the potential \eqref{DFSZpotential}. 
In this mixing, one combination is massive and is the Higgs boson, while the orthogonal massless NGB is the axion $a$.
Redefining the Higgs doublets through transformations like \eqref{PQtransDFSZ} to reabsorb the axion phase in the last term of the potential \eqref{DFSZpotential}, makes $a$ reappear in the Yukawa terms \eqref{yukawaDFSZ}. 
Again, the axion phase can be reabsorbed with some chiral rotations of the fermionic fields, producing a set of anomalous terms like equations~\eqref{deltaL} and~\eqref{2phot}, but this time with the SM fermions instead of the heavy quark $Q$.
In the DFSZ model, the derivative couplings in \eqref{deltaL} play a role in the low energy theory, since they involve the light fermions too. 
The strengths of the couplings depend on the PQ charges $X_u$ and $X_d$, and are therefore model dependent.

We presently do not know which are the features of the high energy model which gives origin to the axion.
However, we can adhere to the proposal of Georgi, Kaplan and Randall~\cite{Georgi:1986df}, and write a generic low energy effective theory.
In the range between $E_{\rm EW}$ and $\Lambda_{\rm QCD}$, the Lagrangian is
\be\label{TheLagr}
\mathcal{L}_a=\frac{1}{2}\partial_\mu a\partial^\mu a - \frac{a}{f_a}\frac{\alpha_s}{8\pi}G\tilde{G} -\frac{g_{a\gamma\gamma}}{4}a F\tilde{F}-\frac{\partial_\mu a}{2 f_a}\sum_{f}C_f\bar{\psi_f}\gamma_5\gamma^\mu \psi_f\; .
\ee
All the model dependencies are hidden in the coupling coefficients $g_{a\gamma\gamma}$ and $C_f$, where the index $f$ runs over all the SM fermions. 
The definition of  the axion-photon coupling constant $g_{a\gamma\gamma}$ is provided by equation~\eqref{2photcoupling}.
In particular, the KSVZ model predicts $C_f=0$ for all the leptons and the ordinary quarks at tree level.
In the DFSZ model, the coupling coefficient to electrons is $C_e=\cos^2 \(\beta\)/3$, while the couplings to the up and down quarks are respectively $C_u=\sin^2 \(\beta\)/3$ and $C_d=\cos^2 \(\beta\)/3$, where $\beta$ is the ratio of the vacuum expectation values of the two $H_u$ and $H_d$ fields~\cite{Nakamura:2010zzi}.
Below $\Lambda_{\rm QCD}$, gluons and quarks confine so we have to write an effective Lagrangian including the couplings to nucleons and mesons,
\be\label{TheLagrQCD}
\mathcal{L}_a=\frac{1}{2}\partial_\mu a\partial^\mu a - m_a^2 f_a^2 \[1-\cos\!\(\frac{a}{f_a}\)\] -\frac{g_{a\gamma\gamma}}{4}a F\tilde{F}-\frac{\partial_\mu a}{2 f_a}\sum_{f}C_f\bar{\psi_f}\gamma_5\gamma^\mu \psi_f+\mathcal{L}_{a\pi}\; .
\ee
The potential is the same of equation~\eqref{eq:potential} up to a reparametrisation.
This time, the sum over $f$ covers the light SM leptons and the nucleons.
The coupling coefficients for axion-proton and axion-neutron interactions are $C_p=-0.55$ and $C_n=0.14$ for $m_u/m_d=0.3$ and $C_p=-0.37$ and $C_n=-0.05$ for $m_u/m_d=0.6$ in the KSVZ model~\cite{Nakamura:2010zzi}.
The DFSZ axion has couplings to nucleons of the same order which depend also on $\beta$~\cite{Nakamura:2010zzi}.  
Among the interactions with mesons, only the pion-axion one is interesting for our purposes.
Therefore in Lagrangran~\eqref{TheLagrQCD} we express $\mathcal{L}_{a\pi}$ as~\cite{Nakamura:2010zzi}
\be\label{apion}
\mathcal{L}_{a\pi}=\frac{\partial_\mu a}{f_a}\frac{C_\pi}{f_\pi}\(\pi^0\pi^+\partial^\mu\pi^- + \pi^0\pi^-\partial^\mu\pi^+ -2 \pi^+\pi^-\partial^\mu\pi^0\)\; ,
\ee
where $C_\pi$ is again a model dependent constant.

\section[Enlarging the parameter space: axion-like particles]{Enlarging the parameter space:\\ axion-like particles} 

As explained in the previous section, several axion models exist, each of them sol\-ving the strong CP problem, but providing different couplings to the SM particles.  
Moreover, the PQ-mechanism works for every value of $f_a$, and the clues about the characteristic PQ-scale are only speculative.
The experimental search for the axion requires therefore not only to explore many orders of magnitude in $f_a$ or $m_a$, but even to scan the model dependencies held in the coupling coefficients of Lagrangians~\eqref{TheLagr} and~\eqref{TheLagrQCD}.

After the seminal paper by Sikivie~\cite{Sikivie:1983ip}, the most relevant axion direct searches try to exploit the two-photon coupling. 
However, as we will show in the next section, the discovery task is tough, the axion being very weekly coupled for the allowed range of $f_a$.
Anyway, this kind of experiments could find in principle any kind of pseudoscalar particle $\phi$ coupled to photons through a term like
\be
\mathcal{L} = - \frac{1}{4} g_\phi \phi F_{\mu \nu} \tilde{F}^{\mu\nu}\; , 
\label{ALPphotoncoupling}
\ee
which mimics the axion-photon interaction~\eqref{2phot}~\cite{Masso:1995tw,Masso:1997ru}.
Therefore, the interest on such hypothetical particles that could couple to two photons has risen in the last years, and the name axion-like particle (ALP) has been coined for them.
These ALPs can in principle also couple to other particles besides photons, however we will always deal with the interaction Lagrangian~\eqref{ALPphotoncoupling}, as it is typically the most important one for the phenomenology at low energies.
We will comment about other possible couplings at the end of chapter \ref{chap:dilution}.

ALPs are even more interesting on the theoretical side, as they can arise in the low energy spectrum of many extensions of the SM.
An ALP can appear in a theory as a PNGB of a continuous global symmetry. 
Examples of these symmetries are related to particle flavour~\cite{Wilczek:1982rv}, lepton number~\cite{Chikashige:1980qk,Chikashige:1980ui} or the $R$-symmetry in supersymmetry~\cite{Chun:1991xm,Nelson:1993nf}.
When a continuous global symmetry is spontaneously broken, massless NGBs appear in the low energy theory as phases of the high energy degrees of freedom. 
Since phases are dimensionless, the canonically normalised theory at low energies always involves the combination $\phi/f_\phi$, where $\phi$ is the NGB field and $f_\phi$ is a scale close to the spontaneous symmetry breaking (SSB) scale. 
The ALP could be for instance related to the generation of right-handed neutrino masses, and hence have a decay constant at an intermediate scale, like $f_\phi\sim 10^{10}$--$10^{12}$~GeV; alternatively it could be associated with a grand-unification theory (GUT), and have a decay constant at the corresponding scale $f_\phi\sim 10^{15}$~GeV.
From the no-hair theorem, we know up to some extent that black-hole dynamics violates global symmetry conservation.
Therefore unbroken global symmetries can not exist in theories with gravity, and we should have PNGBs instead of NGBs.
There are many possibilities for breaking the shift symmetry besides gravity effects, explici\-tly or spontaneously, perturbatively or non-perturbatively.

Moreover, the observation that in string theory ALPs appear in all compactifications has raised even more attention to them~\cite{Svrcek:2006yi}.
These so-called string axions share the NGB properties (having a shift symmetry and being periodic) but with the natural size of $f_\phi$ being the string scale.

The fact that the ALP acquires a mass implies that in the model Lagrangian a potential has to be included.
Taking inspiration from the axion case, the ALP potential can typically be parametrized as
\begin{equation}
\label{ALPpotential}
V(\phi) = m^2_\phi f^2_\phi \left[ 1- \cos\!\( \frac{\phi}{f_\phi}\) \right] \; .
\end{equation}
If the dynamics explicitly breaking the associated global continuous symmetry have a characteristic scale $\Lambda$, the ALP mass is parametrically small, since it is suppressed by powers of $ \Lambda/f_\phi$.
The phenomenology of the SM requires $\Lambda$ to be related to physics beyond the electroweak scale, i.e.~$\Lambda\gtrsim$ TeV (which implies $m_\phi \gg m_a$ for  $f_\phi=f_a$), or to belong to a hidden sector. 
We have here no preconceptions regarding ALP mass, for it depends on the unknown ratio of two unconstrained energy scales.

The dimensionful coupling parameter $g_\phi$ in equation~\eqref{ALPphotoncoupling} can be parametrised as 
\be
\label{genericgvsf}
g_\phi \equiv \frac{\alpha}{2\pi}\frac{ C_{\gamma}}{f_\phi} \; .  
\ee
In the simplest case $ C_{\gamma}$ is an integer, but this is not true in general when the ALP mixes, either kinetically or via symmetry-breaking effects with other ALPs or with pseudoscalar mesons. 
For string axions, the coupling to photons is related via a loop factor to either the string scale or the Planck scale, or it could be even weaker, so an ALP with a large coupling would restrict the string scale to be low. 
For string and field theoretical models the most interesting values are therefore $g_\phi \sim 10^{-11}$--$10^{-15}$~GeV$^{-1}$, $\sim 10^{-19}$~GeV$^{-1}$ and $\sim 10^{-21}$~GeV$^{-1}$ corresponding to intermediate, GUT or Planck scales.

We will adhere to the phenomenological approach to leave the ALP parameters $m_\phi$ and $g_\phi$ free to span many orders of magnitude, exploring the consequences of the presence of these particles in order to limit the ALP parameter space, or to find hints of their existence.

\section{A compendium of limits} 

\begin{figure}[tbp] 
   \centering
   \includegraphics[width=14cm]{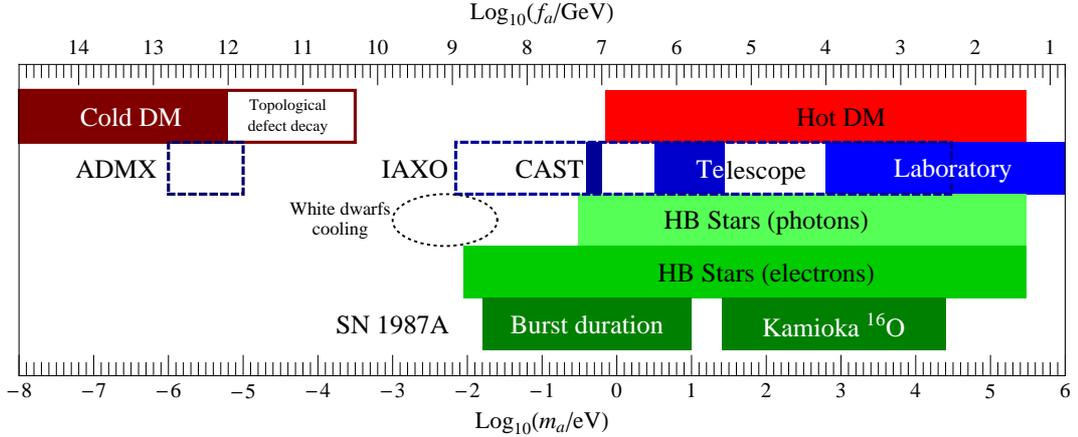}   
   \caption{Axion limits. The red bounds in the first line come from cosmological considerations. The blue ones in the second line are due to direct measurements. The green limits refer to astrophysical arguments related to stellar evolution. These bounds are described in more detail in the text. The dotted ellipse suggest the values of $m_a$ that could help fitting recent white dwarf cooling data. }
   \label{fig:alimits}
\end{figure}

The search for axions and ALPs has not yet been successful.
Up to now we have only indications about where these scalars are not and some hints about where they could hide.
The present bounds on the axion mass, and consequently on the PQ-scale, are plotted in figure~\ref{fig:alimits}.
These bounds are indicative, as the couplings can change according to the axion model, but they give a picture of the situation, especially considering $\mathcal{O}(1)$ coupling coefficients.
The red and brown bounds in the first line, labelled \textbf{Cold DM}, \textbf{Topological defect decay}  and \textbf{Hot DM} are the bounds provided by cosmology and we will thoroughly discuss them in the next chapters.
The blue bounds in the second line come from direct measurements of astrophysical and cosmological quantities and from experiments, while the green ones are related to astrophysics.

As we have already mentioned, very soon after it was proposed the axion was ruled out in the keV--MeV mass range. 
Measurements of heavy quarkonium state decays or of nuclear de-excitations and beam dump or reactor experiments have found no evidence for an axion coupled to fermions or nucleons up to $f_a\gtrsim10^4$~GeV, which in term of mass is $m_a\lesssim0.6$~keV \cite{Kim:2008hd}. 
Pure particle physics experiments can only test values $f_a$ in a rather low energy range.
The bound they put dies out when $f_a$ is high enough to fade the production and detection probabilities through the suppression of the axion couplings.
In figure~\ref{fig:alimits} this bound is labelled \textbf{Laboratory}. 

The \textbf{Telescope} region, $m_a=3$--$27$~eV and $f_a=2.3\times 10^{5}$--$2.1\times 10^{6}$~GeV, is excluded by the non-observation of photons that could be related to the relic axion decay $a\rightarrow\gamma\gamma$ in the spectrum of galaxies and in the extragalactic background light~\cite{Bershady:1990sw,Ressell:1991zv,Overduin:2004sz,Grin:2006aw}.
The Axion Dark Matter eXperiment (ADMX) has provided some constraints in the region $m_a=1.9$--$3.5\ \mu$eV, which translates into $f_a=1.8\times 10^{12}$--$3.3\times 10^{12}$~GeV, after the sensitivity to test the axion was almost reached~\cite{Asztalos:2003px}. 
This collaboration is looking for axion dark matter (DM) using an \emph{haloscope}.
We will describe this instrument in the next chapter.
The plans of the ADMX collaboration are to scan the mass range enclosed in the dashed region called \textbf{ADMX} in figure \ref{fig:alimits} after some upgrades~\cite{Lyapustin:2011pb,Asztalos:2011bm,CarosiTalk}. 
The two latter limits are closely related to cosmology and to the axion being a DM component, we will therefore refer again to them in the following chapters.

\begin{figure}[tbp] 
   \centering
   \includegraphics[width=7cm]{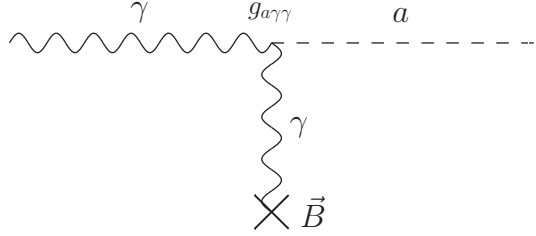}   
   \caption{Diagram of the Primakoff effect.}
   \label{fig:Prima}
\end{figure}

Stars represent a prolific environment for the production of light and weakly-coupled particles, as we have learned for example in the neutrino case~\cite{Raffelt:1996wa}.
The sun is surely the brightest axion source in the sky.
A photon can convert into an axion if it interacts with an external magnetic or electric field by means of the two-photon coupling of equation \eqref{2phot} as in figure \ref{fig:Prima} \cite{Raffelt:1987im}. 
This is the so called Primakoff effect, which was first proposed for the creation of mesons in the electric field of nuclei~\cite{Halprin:1966zz}. 
Therefore, the high concentration of thermal photons, together with the strong electromagnetic fields of the stellar plasma, makes the sun, and stars in general, a rich soil for axion and ALP production.
The total axion luminosity, calculated using the standard solar model, is $L_a \sim (g_{a\gamma\gamma}/10^{-10}\ {\rm GeV^{-1}})^2 10^{-3} L_\odot$, where $L_\odot=3.90\times 10^{25}$ W is the solar luminosity in photons~\cite{Raffelt:2006cw}.
To detect the flux of axions, several solar axion telescopes, like SUMICO~\cite{Inoue:2008zp} and CAST~\cite{Arik:2011rx}, have been built.
These \emph{helioscopes} are essentially vacuum pipes.
They are permeated by a strong magnetic field to exploit the inverse Primakoff effect to convert axions back into photons~\cite{Sikivie:1983ip}.
Axions can enter the telescope because of their very weak interaction with matter, and successively be detected once they have  oscillated into photons. 
Among the helioscopes, CAST currently gives the strongest constraints: its results exclude the part of the axion parameter space, $m_a=0.39$--$0.64$~eV and $f_a=9.8\times10^6$--$1.6\times10^7$~GeV, labelled \textbf{CAST}.
Recently, a new proposal for an axion helioscope has appeared, the International AXion Observatory (IAXO)~\cite{Irastorza:2012qf}. 
The hope is to improve the sensitivity to $g_{a\gamma\gamma}$ of at least one order of magnitude with respect to CAST and therefore, in the most optimistic scenario, to explore the area labelled \textbf{IAXO} which is enclosed by the blue dashed line.  

Waiting for IAXO and its results, the best upper limits on the axion mass in figure~\ref{fig:alimits} come from stellar evolution arguments.
A globular cluster is a gravitationally bound system of stars.
The stars belonging to a globular cluster have approximately the same age and they differ only in their initial mass. 
Since the more massive a star is, the faster it evolves, a globular cluster gives the possibility to study a broad sample of stellar evolution stages and to estimate how long each phase lasts.
In particular, if axions are produced inside a star and escape, they provide an additional cooling channel, besides the photon and neutrino ones. 
If there are more efficient energy release channels, the nuclear fuel consumption has to be faster, and thus the ageing quicker.
Counting the stars in each evolution stage inside a globular cluster permits us to study how fast the fuel consumption is and therefore to put bounds on the production of axions in stellar cores.
The best constraints come from the stars which have reached the helium burning phase, which are called horizontal branch (HB) stars because of the position they occupy in the Hertzsprung-Russel diagram.
The non-standard energy loss prolongs the red giant (RG) phase and shortens the HB one~\cite{Raffelt:1996wa}.
Counting the RG and the HB stars in globular clusters and comparing the two numbers it is possible to evaluate the axion production rate in stars, and to obtain the two \textbf{HB Stars} bounds in figure~\ref{fig:alimits}.
In particular, if the axion is directly coupled to the electron, i.e.~$C_e$ is $\mathcal{O}(1)$, it has a significant production channel more which is reflected in the broader exclusion bound.
HB stars have a typical core temperature of $T\sim10^8$ K $\sim 10$~keV. 
The thermal distribution of photons, averaged over the large volume of the star, still includes many $\gamma$s that are energetic enough to efficiently produce axions if their mass is not $m_a\gtrsim 300$~keV, which is where the HB bounds stop.

Also supernova explosions (SN) are used to put limits on axions. 
Stars with 6--8~$M_\odot$ mass or more reach the ultimate phase of the processing of nuclear fuel, creating an iron nucleus. 
Iron has the largest binding energy per nucleon and therefore cannot be efficiently burnt inside a star.
It does not contribute to produce the radiation pressure necessary to contrast the gravitational pull and to maintain the hydrostatic equilibrium. 
If the iron core reaches a critical mass, it collapses under its own weight. 
Once the nuclear density is reached the collapse stops and the bounce produces a shock wave that expels the outer layers in a core-collapse SN explosion.
In the collapse, electrons are jammed inside protons forcing inverse $\beta$-decays $e^-+p^+\rightarrow n+\nu_e$: a neutron star forms and lots of neutrinos are created.
The density of matter in a SN core is so high, that even neutrinos remain trapped and it takes some time before they can diffuse out~\cite{Raffelt:1996wa}.
After SN~1987A, 24 neutrino events were measured above the background in about 10~s. 
Their distribution in energy and time agrees well with the standard picture for type II SNe.
A particle with a weaker matter interaction than the neutrino would provide a more efficient energy dissipation channel than the standard ones.
If this is the case, the neutrino burst duration would have been shorter than what was measured.
In such a high nucleon density environment, axions would be produced by virtue of their nucleon coupling in reactions like $N+N\rightarrow N+N+a$. 
In figure~\ref{fig:alimits}, the upper bound on the axion mass of \textbf{SN 1987A}, $m_a<16$~meV, thus $f_a>4\times 10^8$~GeV, comes from this burst duration argument.
Lighter axions, and thus less coupled ones, would not be efficiently produced and thus they would have not significantly affected the timing of SN 1987A neutrino events~\cite{Raffelt:2006cw}.
On the other side, if axions couple much more strongly to matter, they could be trapped inside the core, and the neutrino burst would have suffered little or no modification. 
This is why the SN~1987A bound stops at low $f_a$. 
However, in this case some axions are emitted and it would have been possible to detect them in the Kamiokande II experiment thanks to the nuclear reaction $a+^{16}\!\!{\rm O}^{} \rightarrow ^{16}\!\!{\rm O}^*\rightarrow ^{16}\!\!{\rm O}+\gamma$.
The region labelled \textbf{Kamioka~$\mathbf{^{16}}$O} is excluded by the non-observation of these events~\cite{Engel:1990zd}.
The SN~1987A bounds are very uncertain and have to be taken with a grain of salt. 
They are particularly interesting if the axion has not a direct coupling with the electron, for the stronger limits coming from HB stars are not valid in this case.

\begin{figure}[tb] 
   \centering
   \includegraphics[width=14cm]{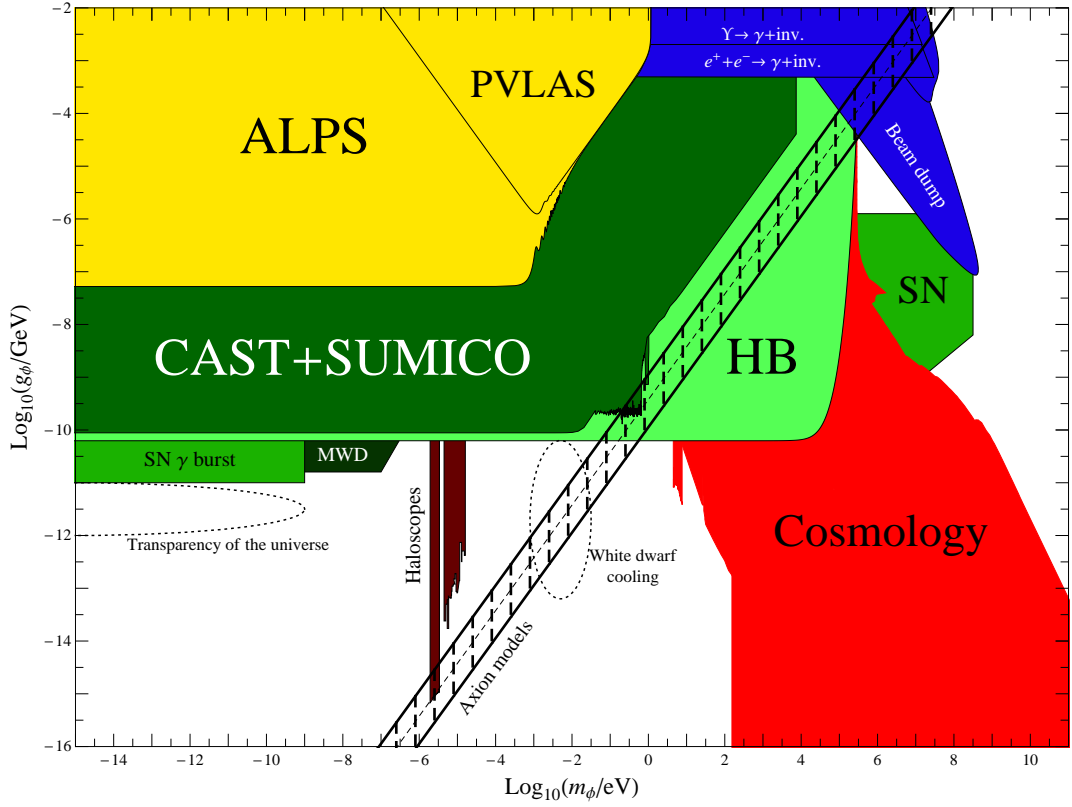}   
   \caption{ALP limits. The blue regions are constrained by particle physics experiments. In green are the astrophysics related limits, in particular CAST+SUMICO are helioscope measurements. The laser experiments exclude the yellow region. Haloscopes have explored the brown region and in red are the bounds provided by cosmology. 
The dotted ellipses suggest the values of the parameter space which should characterise a pseudoscalar particle in order to solve some astrophysical conundrums.
The band of axion models includes values of $C_\gamma$ between 0.6 and 6.}
   \label{fig:alplimits}
\end{figure} 

Most of the bounds just described are also valid in the ALP case, as they are directly constraining $g_{a\gamma\gamma}$ and thus $g_\phi$.
Moreover, those based on measurements of the coupling to electrons or nucleons can also be used to constrain ALPs, once they are translated to the effective ALP-fermion coupling mediated by photons.
Because of their larger parameter space, ALPs have a richer phenomenology and more constraining experiments and observations than the axion.
Figure~\ref{fig:alplimits}, where we collected the most relevant ALP bounds, provides a panoramic view of the ALP parameter space, in comparison with the black-dashed band which represents the axion models, whose central value is the KSVZ axion and the shading covers the interval $0.6\lesssim C_\gamma\lesssim6$.

The bounds related to cosmology are plotted in red and they will be explained in the next chapters, together with the brown constraints that come from the search for ALP DM with ADMX and other haloscopes.
In blue are the bounds coming from particle physics experiments, in green astrophysical observations and arguments.
The yellow area is constrained by laser experiments dedicated to the search for ALPs.

Particle physics experiments, as in the axion case, can set constraints only for quite strongly coupled particles and they pile up in the upper part of figure~\ref{fig:alplimits}.
They are very useful anyway, for they can test larger mass than astrophysics.
In figure~\ref{fig:alplimits} we draw in blue the exclusion bounds due to $\Upsilon$ decay and positro\-nium annihilation into invisible channels, labelled respectively $\boldsymbol{\Upsilon\rightarrow\gamma+{\rm inv}}$ and $\boldsymbol{e^+ e^-\rightarrow\gamma+{\rm inv}}$.
The beam dump experiments performed in SLAC exclude the patch labelled \textbf{Beam dump}~\cite{Riordan:1987aw,Bjorken:1988as}.

Astrophysics plays a central role in constraining the ALP case. The HB bound on $g_{a\gamma\gamma}$ described before is directly  applicable to ALP photon coupling, and it is plotted in figure~\ref{fig:alplimits} in light green and labelled \textbf{HB}.
The SN 1987A gives some constraints too. 
The first one, labelled \textbf{SN}, is given by the duration of the measured neutrino pulse, as in the axion case, and was derived in~\cite{Masso:1995tw}. 
The second bound related to SN 1987A is labelled \textbf{SN $\boldsymbol{\gamma}$ burst} and it is related to the transparency of the dense SN core to the ALP propagation. 
If an ALP exists in this region of the parameter space, it would be produced during the core collapse and it would subsequently escape from it.
Then, the propagating ALPs can oscillate into high energy photons interacting with the galactic magnetic field, and finally detected on earth.
Since no $\gamma$-ray pulse was measured in correspondence of SN 1987A, the \textbf{SN $\boldsymbol{\gamma}$ burst} region can be excluded~\cite{Jaeckel:2010ni}.
However the SN bounds can be considered rather weak, as they rely on an insufficient understanding of the SN dynamics and, in the cases just discussed, of the ALP emission from a nuclear-density environment.

The CAST helioscope constrains a large part of ALP parameter space.
The Japanese experiment SUMICO gives also a bound in a small part of the parameter space not constrained by CAST~\cite{Inoue:2008zp}. 
Their limits are both plotted in dull green in figure~\ref{fig:alplimits} and labelled \textbf{CAST+SUMICO}.
It is interesting to notice in this picture how it is actually difficult to reach the sensitivity to constrain the axion.
In particular, just a corner of the bound penetrates the axion model stripe, which implies the slimness of the CAST bound in figure~\ref{fig:alimits}.

A light shining through the wall (LSW) experiment is the table-top version of an helioscope, in the sense that in this case the source is not the sun but a high intensity laser.
The experiment consists of a pipe divided in two by an opaque barrier, the wall, in a production section and a detection one.
In both of them a high vacuum has been created and an intense magnetic field imposed.
The beam of photons emitted by the laser passes through the magnetic field before ending its run against the wall, converting some of the photons of the beam into ALPs via the Primakoff effect.
Because of their feeble coupling, ALPs can cross the wall passing into the detection section, where they have to be converted back into photons interacting with the magnetic field before being finally detected~\cite{Redondo:2010dp}.
The yellow patch labelled \textbf{ALPS} in figure~\ref{fig:alplimits} shows the constraints of the latest LSW, the Any Light Particle Search (ALPS)~\cite{Ehret:2010mh}.

Measuring changes in polarization is a different approach to laser experiments \cite{Maiani:1986md,Raffelt:1987im}.
Again, the laser beam passes through an intense magnetic field. 
The component of the vector potential parallel to the magnetic field effectively mixes with the pseudoscalar field and it is partially absorbed and retarded.
This causes a light beam, linearly polarised at a given angle with the magnetic field, to rotate a small amount because of the absorptive process and to gain a small ellipticity due to the dispersion effect.
Therefore, if pseudoscalar particles coupled to two photons exist, a magnetic field induces respectively \emph{dichroism} and \emph{birefringence} on the magnetised vacuum. 
Both effects are proportional to the square of the magnetic field and of the photon coupling $g_\phi$.
Moreover, the acquired ellipticity is proportional to $m_\phi^2$~\cite{Battesti:2007um}.
The QED background processes are virtual pair creation for birefringence \cite{Schwinger:1951nm} and photon splitting for dichroism \cite{Adler:1971wn}. 
Both of them are very suppressed, especially the photon splitting, thus measuring significant magneto-optical properties of the vacuum would be a signal of the existence of particles coupled to two photons.
The best constraints on these phenomena are provided by the Polarizzazione del Vuoto con LASer (PVLAS) experiment, whose results are shown in figure~\ref{fig:alplimits} in the yellow zone labelled \textbf{PVLAS}~\cite{Zavattini:2007ee}. 
In particular, this refers to the birefringence measurements, for the dichroism ones are completely superseded by ALPS results.

Finally, the \textbf{MWD} region is constrained by similar consideration, but powered by the high energy phenomena which sometimes characterise the astrophysical environments.
In the past decades many white dwarfs with very strong magnetic fields --- up to $10^{10}$~G --- were discovered. 
We call these objects magnetised white dwarfs (MWD).
The polarisation of light coming from MWD is strongly influenced by the magnetic field.
A typical fraction of 5\% of the light is circularly polarised, as required to electromagnetic waves propagating in a magnetised atmosphere, while the linearly polarised fraction is something less but of the same order, and could be explained by photon-ALP conversion.
If ALPs with $m_\phi\lesssim 10^{-6}$~eV and $g_\phi\gtrsim 10^{-11}$~GeV$^{-1}$ exist, the linearly polarised fraction of light would be larger than 5\% for MWDs with $B=10^9$~G, and consequently this region has to be excluded~\cite{Gill:2011yp}.
The bound can improve by more than an order of magnitude for the coupling to photons --- up to $g_\phi\sim10^{-12}$~GeV$^{-1}$ --- if the data about the MWD with the strongest magnetic field, $B\sim10^{10}$~G, are confirmed~\cite{Gill:2011yp}.

\section{Where could axions and ALPs hide?}

Until now each axion and ALP search has been unfruitful, providing only exclusion bounds.
However, besides the phenomenological need to have an axion in order to solve the strong CP problem, there are some observational hints about which regions of the parameter space could hide an axion or an ALP.
We draw dashed ellipses in figures \ref{fig:alimits} and \ref{fig:alplimits} to highlight the regions of parameter space suggested by these observations.

Recently, the measurement of TeV photons from very far sources --- namely some active galactic nuclei (AGN) --- has puzzled astrophysicists~\cite{Aharonian:2005gh,Mazin:2007pn}. 
Very high energy photons should inelastically scatter with the background light producing $e^+e^-$ pairs.
They should be rapidly absorbed by the intergalactic medium and we should observe none of them if coming from very far sources.
Of course, it could be that the spectrum of the source is much harder than we expect or that we overestimate the amount of background photons in the medium.
However, it could also be that these TeV photons, interacting with the extragalactic magnetic field, oscillate into ALPs, then escape the absorption and finally, once converted back into the photon form, are detected by our telescopes~\cite{DeAngelis:2007dy,Hooper:2007bq,Hochmuth:2007hk}.
To solve this transparency issue, this mechanism needs an ALP with very low mass, $m_\phi\sim10^{-12}$--$10^{-9}$, and coupling greater than $g_\phi\sim10^{-12}$ GeV$^{-1}$~\cite{Jaeckel:2010ni}. 
Of course, much about these considerations depends on the assumptions on the source and on the extragalactic magnetic field.
Especially the strength of the latter has to be assumed very close to the upper limit to have an appreciable effect.
To avoid this problem, one can assume the photon-ALP conversion to happen in the regions around the source and in our galaxy, where the magnetic fields are better known~\cite{Simet:2007sa,Horns:2012kw}.
We have nevertheless to notice that if the constraint coming from MWD is solid, not too much space is left for ALPs to solve the transparency of the universe problem.
But even if this is the case, the ALP-photon conversion affects the polarisation of distant astrophysical sources and, once some more information about the coherent components on the intergalactic magnetic field are obtained, it will be possible to extract some useful limits from AGNs~\cite{Bassan:2010ya,Horns:2012pp} and other celestial objects, like quasars~\cite{Payez:2012vf}.

Astrophysics provides another very interesting clue related to the evolution of white dwarf stars. 
It seems that if the axion has a direct coupling to electrons and a decay constant $f_a\sim 10^9$ GeV, it provides an additional energy-loss channel that permits to obtain a cooling rate that better fits the white dwarf luminosity function than the standard one~\cite{Isern:2008nt}.
The selected mass range is in the meV range and $g_{a\gamma\gamma}\sim 10^{-12}\ {\rm GeV}^{-1}$.
The hadronic axion would also help in fitting the data, but in this case a stronger value for $g_{a\gamma\gamma}$ is required to  perturbatively produce an electron coupling of the required strength. 

Finally, axions and ALPs are also perfect dark matter candidates.
We will devote a large section of chapter \ref{chap:DM} to this topic.

After the presentation made in this chapter, we will deal from now on with the cosmological bounds.
Since many of the topics that will be treated involve both axions and ALPs, we will refer to them using the term \emph{pseudoscalars} when they are on the same level.
We also want to underline that in the following discussion $g_{a\gamma\gamma}$ and $g_\phi$ are phenomenologically equivalent.

\chapter{Establishing an axion or ALP relic population}\label{chap:DM}
In order to obtain information about the axion and its ALP relatives from cosmological considerations it is first of all necessary to understand if a primordial population of pseudoscalars can be established.
Several mechanisms can achieve this task and they will be described in sections~\ref{sec:nonthermal} and~\ref{sec:thermal}.
Sections~\ref{sec:excess} and~\ref{sec:decay} deal with the ways a pseudoscalar population can disappear from the cosmic plasma.

In the rest of the dissertation we will deal with a flat Robertson-Walker metric
\be
ds^2=dt^2-R^2(t)\(dr^2+r^2d\vartheta^2+r^2 \sin^2 \vartheta^2 d\varphi^2\)\; ,
\ee
where $R$ is the cosmic scale factor, which has length dimensions, $t$ is the time coordinate, $r$ is the dimensionless radial comoving coordinate, and $\(\vartheta,\varphi\)$ are the dimensionless comoving angular coordinates.
The cosmic scale factor is growing in time, representing the expansion of the universe. 
A useful way of measuring the expansion is through the redshift $z$, which measures the ratio between the wavelength $\lambda_e$ of a light signal emitted at time $t_e$ and the wavelength $\lambda_d$ of the same signal detected at time $t_d$, and it is defined to be
\be
1+z=\frac{\lambda_d}{\lambda_e}=\frac{R(t_d)}{R(t_e)}\; .
\ee 
The redshift of a signal measured today is also a practical way to refer to cosmological time scales.
From the present temperature of the CMB, $T_0=2.73\ {\rm K}=2.35\times 10^{-13}$~GeV, the temperature at a given redshift is easily obtained with the formula $T=T_0(1+z)$ if we assume no heating of the thermal bath.

\begin{figure}[tb] 
   \includegraphics[width=100mm]{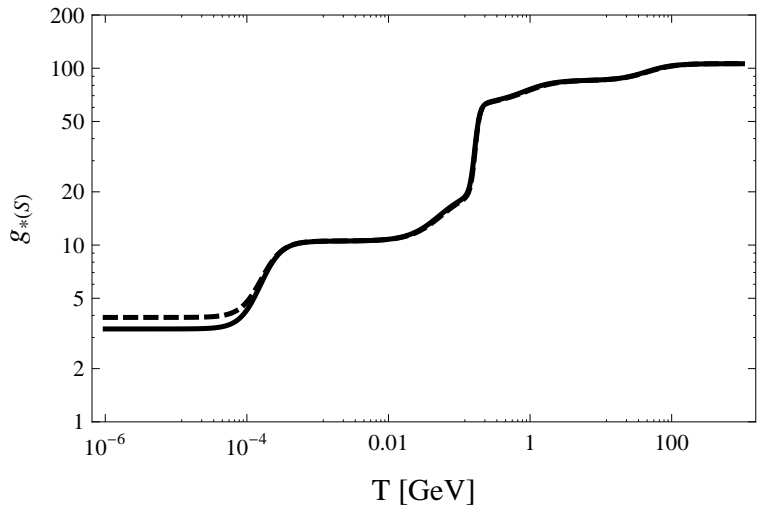}   
   \centering
   \caption{The number of relativistic internal degrees of freedom $g_*$ (solid) and the number of relativistic entropy degrees of freedom $g_{*S}$ (dashed) as functions of temperature.}
   \label{fig:gigi}
\end{figure}

The expansion rate is defined to be
\be
H(t)\equiv \frac{\dot{R}(t)}{R(t)}\; 
\ee
where the dot stands for the time derivative.
The Friedmann equation links $H$ with the energy density of the universe $\rho$,
\be\label{eq:exprate}
H^2=\frac{8\pi}{3}\frac{\rho}{m_{\rm Pl}^2}\; ,
\ee
the Planck mass being $m_{\rm Pl}=1.2211\times 10^{19}$ GeV.
In the radiation dominated universe, the energy density is
\be
\rho=\frac{\pi^2}{30}g_*(T)T^4\; ,
\ee
which depends on the temperature $T$, and the equation~\eqref{eq:exprate} becomes
\be\label{eq:exprateRAD}
H\simeq1.66 g_*(T)^{1/2}\frac{T^2}{m_{\rm Pl}}\; .
\ee
We will refer very often to this form for $H$.
The $T$ dependent quantity $g_*$ is the number of relativistic internal degrees of freedom, which is plotted in figure~\ref{fig:gigi}.
Its definition is
\be
g_*(T)=\sum_{i={\rm bosons}}g_i\(\frac{T_i}{T}\)^4+\frac{7}{8}\sum_{i={\rm fermions}}g_i\(\frac{T_i}{T}\)^4\; ,
\ee
where the indices $i$ run over the bosons and fermions with temperature $T_i$ and $g_i$ internal degrees of freedom, which are relativistic when the photon temperature is~$T$.

The critical energy density 
\be
\rho_c=3 \frac{\(H_0 m_{\rm Pl}\)^2}{8\pi}=10.5 \,h^{2}\,\frac{{\rm keV}}{{\rm cm}^3}\; ,
\ee
where $H_0= 100\, h\ {\rm km\,s^{-1}\, Mpc^{-1}}$ is the present value of the expansion rate and $h\simeq 0.7$ is its present-day normalized value, defines the energy density of a flat universe expanding at $H_0$ rate.
It can be used as a unit of measure for the energy density of the different constituents of the universe.
Thus, defining the present ratios $\Omega_r=\rho_r/\rho_c$, $\Omega_m=\rho_m/\rho_c$, and $\Omega_\Lambda=\rho_\Lambda/\rho_c$ respectively for radiation, matter and vacuum energy, the expansion rate can be conveniently expressed as a function of the redshift with
\be
H(z)=H_0\sqrt{\Omega_r(1+z)^4+\Omega_m(1+z)^3+\Omega_\Lambda}\; ,
\ee  
if there are no reactions converting one energy form into the other.

The entropy density of the radiation dominated universe is
\be
s=\frac{2\pi^2}{45}g_{*S}(T)T^3\; ,
\ee  
and $g_{*S}$ are the number of relativistic entropy degrees of freedom,
\be
g_{*S}(T)=\sum_{i={\rm bosons}}g_i\(\frac{T_i}{T}\)^3+\frac{7}{8}\sum_{i={\rm fermions}}g_i\(\frac{T_i}{T}\)^3\; ,
\ee
which is also plotted in figure~\ref{fig:gigi}.
For all the usual cosmological quantities, like $g_*$ and $g_{*S}$, we follow the definitions of ``The Early Universe'' by Kolb and Turner~\cite{Kolb:1990vq}.
We also define comoving quantities scaling out the effect of the universe expansion.
Therefore, comoving quantities change only in force of physical processes which are not the stretching of space-time.
Some examples are the comoving entropy $S=sR^3$, the comoving number density $N=nR^3$ and comoving momentum $K=kR$.

\section{Production mechanisms I: non-thermal relics}\label{sec:nonthermal} 

Soon after the invisible axion proposal, it was recognised that this new particle is a perfect DM candidate.
The spontaneous breaking of the PQ-symmetry and the consequent phase-transition, which shifts the vacuum of the theory to the minimum of the potential~\eqref{eq:potential}, provides the axion with an efficient non-thermal production mechanism~\cite{Dine:1982ah,Abbott:1982af,Preskill:1982cy}.  
This \emph{realignment mechanism} was later proposed also for the non-thermal productions of string axions~\cite{Arvanitaki:2009fg,Acharya:2010zx,Marsh:2011gr,Higaki:2011me}, general ALPs~\cite{Masso:2004cv} and hidden photons~\cite{Nelson:2011sf}.
More recently, the realignment mechanism for hidden photons and ALPs has been reviewed in~\cite{Arias:2012mb}. 
Here it is also shown that once produced, a population of very light non-thermal dark matter particles is extremely difficult to reabsorb in the primordial plasma. 

\subsection{Realignment mechanism}

In the realignment mechanism a field, which in the early universe can take a random initial state, rolls down towards the minimum of the potential.
Once it has reached the bottom, it overshoots the minimum and starts to oscillate around it.
If the quanta of the field are cosmologically stable, these oscillations behave as a cold dark matter fluid.
Their energy density in fact is diluted by the expansion of the universe as $\rho\propto R^{-3}$.

\begin{figure}[tb] 
   \includegraphics[width=80mm]{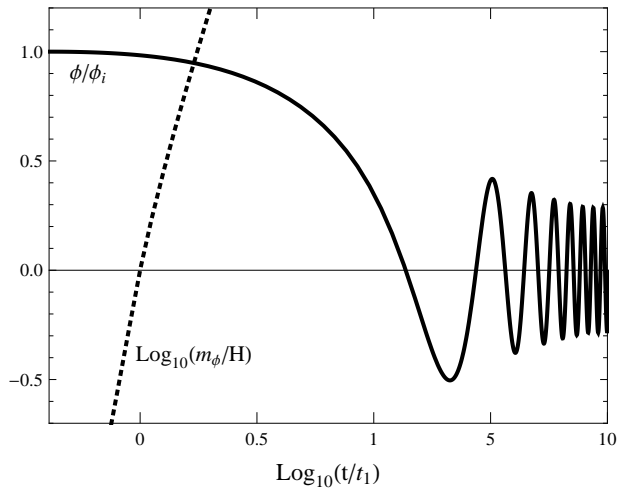}   
   \centering
   \caption{Numerical solution of equation \eqref{eq:condensev} (solid) and the evolution of $\log_{10}(m_\phi/H)$ (dashed) for the values used in the equation.}
   \label{fig:CDMsolution}
\end{figure}

The simplest example is that of a scalar field $\phi$ of mass $m_\phi$ with Lagrangian
\be 
\mathcal L= \frac{1}{2}\partial_\mu \phi \partial^\mu \phi-\frac{1}{2}m_{\phi}^2\phi^2 +{\cal L}_I \; ,  
\ee
where ${\cal L}_I$ encodes interactions of the scalar field with itself and the rest of particles in the primordial bath.
In general, the mass receives thermal corrections from ${\cal L}_I$ which might be crucial, thus $m_{\phi}=m_{\phi}(t)$ should be understood. 
In each causally connected patch of the universe the scalar field has an initial value, $\phi_i$.
If inflation already happened, $\phi$ is uniformly equal to $\phi_i$ in the whole observable universe.
The equation of motion in the expanding universe for the homogeneous component of $\phi$ --- called the \emph{zero mode} --- is obtained neglecting the gradient effects, 
\begin{equation}
\label{eq:condensev}
\ddot{\phi}+3H\dot{\phi}+m_{\phi}^{2}\phi=0 \; .
\end{equation}  
Its solution can be separated into two regimes. 
In a first epoch, $3H\gg m_{\phi}$, so $\phi$ is an overdamped oscillator and gets frozen, $\dot \phi=0 $. 
At a later time, $t_1$, characterized by $3H(t_1)=m_{\phi}(t_1)\equiv m_1$, the damping becomes undercritical and the field can roll down the potential and starts to oscillate. During this epoch, the mass term is the leading scale in the equation and the solution can be found in the WKB approximation,
\be
\label{eq:WKB}
\phi\simeq \phi_i \(\frac{m_1 R_1^3}{m_{\phi} R^3}\)^{1/2} \cos\(\int_{t_1}^t m_{\phi}\, dt\) ,  
\ee
where $\phi(t_1)\sim \phi_i$, since up to $t_1$ the field evolution is frozen. 
Note that to obtain this solution only the definition of $H=\dot{R}/R$ has been taken into account, not its actual time dependence, and so it is valid for radiation, matter, and vacuum energy dominated phases of the universe and their transitions. 
Figure \ref{fig:CDMsolution} shows a numerical solution of equation~\eqref{eq:condensev}.

The approximate solution \eqref{eq:WKB} corresponds to fast oscillations with a slow amplitude decay. 
Defining this amplitude ${\cal A}(t)=\phi_i(m_1 R_1^3/m_{\phi} R^3)^{1/2}$ and the phase $\alpha(t)=\int^t m_{\phi}(t)dt$, 
the energy density of the scalar field is
\be
\rho_{\phi} =  \frac{1}{2}\dot \phi^2+\frac{1}{2}m^2_{\phi}\phi^2 = \frac{1}{2}m^2_{\phi}{\cal A}^2 + ...\; ,
\ee
where the dots stand for terms involving derivatives of ${\cal A}$, which by assumption are much smaller than $m_{\phi}$, because $m_{\phi}\gg H$ in this regime. 
The pressure is then 
\be
p_{\phi}    = \frac{1}{2}\dot \phi^2-\frac{1}{2}m_{\phi}^2\phi^2 =-\frac{1}{2}m_{\phi}^2{\cal A}^2\cos\(2\alpha\)-{\cal A}\dot{\cal A} m_{\phi} \sin\(2\alpha\)+\dot {\cal A}^2\cos^2\(\alpha\)\; .
\ee
When the field just starts to oscillate the equation of state is a non-trivial and strongly time dependent function.
However, at much later times, $t\gg t_{1}$, the oscillations in the pressure occur at time scales $1/m_{\phi}$, much faster than the cosmological evolution.
We can therefore take an average over these oscillations, and the pressure is then
\begin{equation}
\langle p_{\phi} \rangle   = \langle\dot {\cal A}^2\cos^2\(\alpha\)\rangle=\frac{1}{2}\dot{\cal A}^2\; .
\end{equation}
At leading order in $\dot{\cal A}/({\cal A}m)$, the equation of state is just
\begin{equation}  
w=\langle p\rangle/\langle \rho\rangle \simeq 0 \; ,
\end{equation}
which is exactly that of non-relativistic matter.

It follows from equation~\eqref{eq:WKB} that the energy density in a comoving volume, $\rho_\phi R^3$, is not conserved if the scalar mass changes in time. 
However, the quantity 
\begin{equation}
N_\phi=\frac{\rho_\phi R^3}{m_{\phi}}=\frac{1}{2}m_{1} R^{3}_{1}\phi^{2}_{i} \; ,
\end{equation} 
is constant, and can be interpreted as a comoving number of non-relativistic quanta of mass $m_{\phi}$.
We can use this conservation to compute the energy density today to obtain
\begin{equation}
\label{mos}
\rho_{\phi} (t_0)=m_0 \frac{N_\phi}{R_0^3}\simeq
 \frac{1}{2}m_0 m_1 \phi_i^2\left(\frac{R_1}{R_0}\right)^{3} \;  , 
\end{equation}
where quantities with a $0$-subscript are evaluated at present time.  

\begin{figure}[tb] 
   \includegraphics[width=100mm]{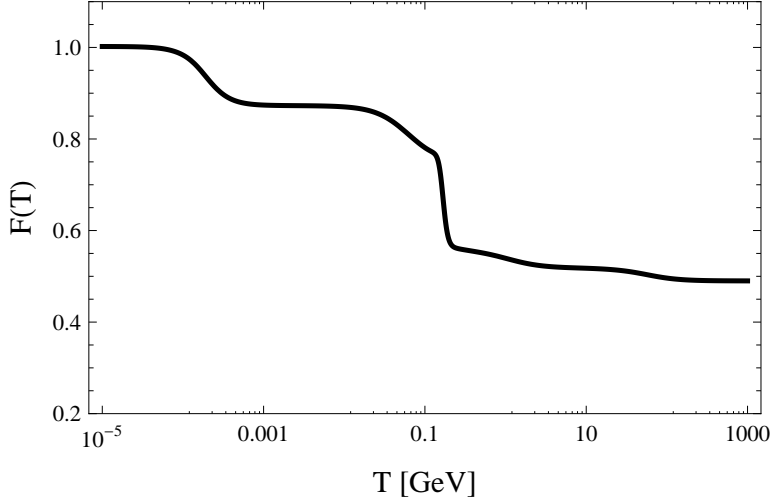}   
   \centering
   \caption{The function $\mathcal{F}(T)$, used in equations~\eqref{eq:CCDMall}.}
   \label{fig:effe}
\end{figure} 

Through the conservation of comoving entropy $S$, the dilution factor $(R_1/R)^3$ in equation~\eqref{mos} can be expressed  as 
\be\label{eq:dilution}
\(\frac{R_1}{R}\)^3=\frac{g_{*S}(T)T^3}{g_{*S}(T_1)T_1^3} \; .
\ee 
Using the expression for the Hubble constant in the radiation dominated era, equation~\eqref{eq:exprateRAD}, and the definition of $T_1$, $3H(T_1)=m_1$, $T_1$ can be written in terms of $m_1$ and the Planck mass.
In this way, equation \eqref{mos} for the matter density from the misalignement mechanism today can then be expressed in a more quantitative way as 
\begin{subequations} \label{eq:CCDMall}
\begin{align}
\label{eq:CCDM}
\rho_{\phi,0}&\simeq \(0.17\, \frac{{\rm keV}}{{\rm cm}^3}\) \sqrt{\frac{m_0}{{\rm eV}}} \sqrt{\frac{m_0}{m_1}}\(\frac{\phi_i}{10^{11}\, {\rm GeV}}\)^2 {\cal F}(T_1) \; ,\\
\label{eq:CCDMb}
\Omega_{\phi}&\simeq \(0.016\, h^{-2}\)\sqrt{\frac{m_0}{{\rm eV}}} \sqrt{\frac{m_0}{m_1}}\(\frac{\phi_i}{10^{11}\, {\rm GeV}}\)^2 {\cal F}(T_1) \; ,
\end{align}
\end{subequations}
where ${\cal F}(T_1)\equiv (g_{*}(T_1)/3.36)^\frac{3}{4}(g_{*S}(T_1)/3.91)^{-1}$ is a smooth function ranging from $1$ to $\sim 0.3$ in the interval $T_1\in (T_0, 200 {\rm GeV})$, and it is plotted in figure~\ref{fig:effe}.
The quantity $\Omega_\phi=\rho_{\phi,0}/\rho_c$ is the fraction of energy density of the universe in ALPs.
The abundance is most sensitive to the initial amplitude of the oscillations, being proportional to $\phi_i^2$, and to a lesser degree to the present mass value $m_0$. 
The factor $1/\sqrt{m_1}$ reflects the damping of the oscillations in the expanding universe:  the later the oscillations start, i.e.~the smaller $T_1$ and therefore $H_1$ and $m_1$, the less damped they are for a given $m_0$. 

If we compare the above estimate with the DM density measured by WMAP and other large scale structure probes~\cite{Komatsu:2010fb},
\be
\label{eq:observedDM}
\rho_{\rm CDM} = 1.17\, \frac{{\rm keV}}{{\rm cm}^3}\; ,  \qquad \Omega_{\rm CDM} = 0.11\,h^{-2}\; , 
\ee
it seems that only very large values of $\phi_i$ can provide a $\rho_{\phi,0}$ that accounts for all the dark matter. 
However, a relatively small $\phi_i$ could be compensated by a small $m_1\ll m_{0}$.  

It is now necessary to provide a model for the thermal mass of the pseudoscalar particle to compare the estimate~\eqref{eq:CCDM} with the observed abundance~\eqref{eq:observedDM}.
For the potential~\eqref{ALPpotential}, the ALP satisfies the equation of motion~\eqref{eq:condensev} as long as the angle $\varphi\equiv\phi/f_\phi$ is small. 
The inaccuracy of the quadratic approximation can be anyway cured by an additional correction factor to \eqref{eq:CCDM}. 
This is normally an ${\cal O}(1)$ factor except if we fine tune the initial condition to $\varphi=\pi$.

\begin{figure}[phtb] 
   \includegraphics[width=120mm]{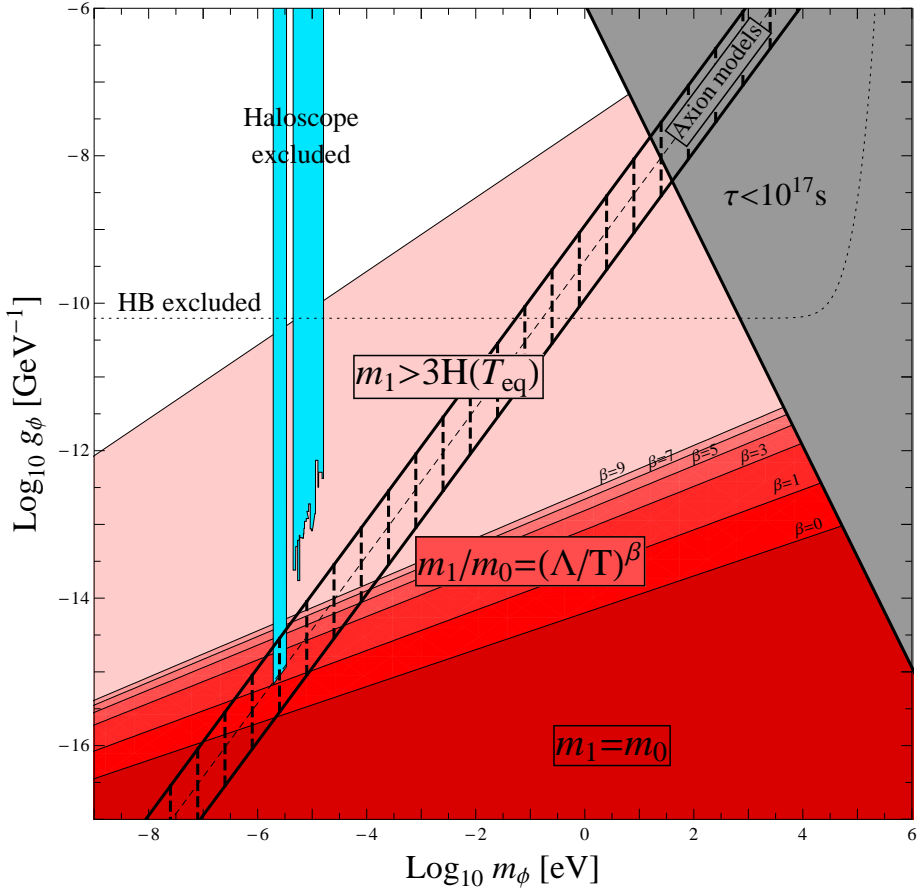}   
   \centering
   \caption{Axion and ALP DM from the realignment mechanism. The axion model band is plotted as in figure~\ref{fig:alplimits}. The regions of the ALP parameter space in which the realignment mechanism provides sufficient --- or too large --- DM abundance are plotted in different red gradations for different thermal mass models. The pink region fulfils the minimum requirement for non-thermal produced ALP to behave like CDM. The grey region cover the part of parameter space in which the ALP lifetime, which is provided by equation~\eqref{eq:lifetime}, is shorter than the age of the universe. The turquoise regions have been tested by haloscopes and no ALPs have been found. As a reference, the HB upper bound is plotted as a dotted line.}
   \label{fig:ALPDM}
\end{figure} 

Figure~\ref{fig:ALPDM} represents the region of the parameter space of ALPs in which the realignament mechanism could provide a sufficient production of DM. 
The allowed regions of ALP dark matter in the $m_\phi$--$g_\phi$ plane in figure~\ref{fig:ALPDM} can be obtained using
\be
\phi_i =\varphi_i \frac{\alpha  C_{\gamma}}{ 2\pi g_\phi}
\ee
with $\varphi_i=\phi_i/f_\phi$, the initial misalignment angle, whose range is restricted to values between $-\pi$ and $+\pi$.
The model dependent factor $ C_{\gamma}$ will from now on be taken to be unity for presentation purposes, but the reader should keep in mind that in principle it can assume a very different value according to the model it comes from. 
To tune the right DM abundance we have used the a priori unknown value of  $\varphi_i$. 
The upper bound on $g_\phi$ reflects the fact that $\varphi_i$ cannot be larger than $\pi$, and thus assumes $\varphi_i\sim \pi$. 
Moving to lower values of $g_\phi$ requires inflation happening after SSB in order to have a homogeneous small value of $\varphi_i$ which is increasingly fine-tuned to zero to avoid over-abundant DM production. 
In this sense the values closest to the boundary, corresponding to the largest values of the photon coupling, can be considered the most natural ones.
We will further comment on the fine-tuning issue later on.

In the simplest realization of an ALP model, the mass receives no thermal corrections, thus $m_\phi$ is constant throughout the universe expansion and the DM yield can be inferred directly from equation \eqref{eq:CCDM} using $m_1=m_0$. 
In figure~\ref{fig:ALPDM} it is the dark red region labelled $\mathbf{m_0=m_1}$.

If the global symmetry associated to the ALP is anomalous, then the mass is provided by the instantonic potential just as the $\eta'$ or the axion acquire their mass thanks to QCD instantons.  
The ALP needs in principle another unbroken $SU(N)$ group, which condenses at a scale $\Lambda$.  
If $T$ is the temperature of the $SU(N)$ sector, then the thermal ALP mass is
\begin{equation}
\label{eq:ALPmassVar}
m_\phi\simeq 
\begin{cases}
\frac{\Lambda'^2}{f_\phi}\equiv m_0&\hbox{for $T\ll \Lambda$}\; ,\\
\\
m_0\(\frac{\Lambda''}{T}\)^\beta&\hbox{for $T\gg \Lambda$}\; .
\end{cases}\ 
\end{equation}  
It makes sense to naively assume $\Lambda'\sim \Lambda''\sim \Lambda$, while the precise relation among these scales depends on the details of the model.  
At temperatures larger than $\Lambda$, electric-screening damps long range correlations in the plasma and thus the instantonic configurations, resulting in a decrease of the ALP mass.
In specific models the exponent $\beta$ can be obtained for instance from instanton calculations, but here it is left as a free parameter.
Assuming the onset of ALP coherent oscillations to happen in the mass suppression regime, it is easy to obtain an expression for $m_0/m_1$ which is the expected enhancement in the DM abundance,
\be
\label{eq:enha1}
\sqrt{\frac{m_0}{m_1}}=\(\frac{\sqrt{m_0 m_{\rm Pl}}}{\Lambda''}\)^\frac{\beta}{\beta+2}\(3\times 1.66\sqrt{g_{*1}}\)^\frac{-\beta}{2\beta+4}
\ee
and the factor that controls the enhancement is
\be
\label{eq:enha2}
\frac{\sqrt{m_0 m_{\rm Pl}}}{\Lambda''}\sim \frac{\Lambda'}{\Lambda''} \sqrt{\frac{m_{\rm Pl}}{f_{\phi}}}\; .
\ee
These models can provide only a moderate enhancement of the DM density with respect to the constant $m_\phi$ case. 
The gained regions for the ALP DM case for values of $\beta=1,3,5,7,9$ can be seen in figure~\ref{fig:ALPDM} from bottom to top --- the lowermost region $m_1=m_0$ corresponds, of course, to $\beta=0$. 
Actually, even considering unrealistically huge values of $\beta$ does not help much, as can be seen from the asymptotic approach of the highest $\beta$ cases. 
This is reflected by the finite limit of equation \eqref{eq:enha1} when $\beta\to \infty$, but it follows from its definition, equation \eqref{eq:ALPmassVar}. 
In the $\beta\to \infty$ limit, $m_\phi$ is a step function of temperature, $m_\phi\propto \Theta(T-\Lambda)$, and the relation $m_0=\Lambda^2/f_\phi$ determines $\Lambda$ from $m_0$ and $f_\phi$. 
Thus each point in the $m_\phi$--$g_\phi$ parameter space has an implicit maximum DM abundance, independent of $\beta$.   
The crucial assumption that leads to these conclusions is that $\Lambda'\sim \Lambda$, because it does not allow to consider arbitrary small values for $\Lambda$ for a given mass. 
Therefore, models in which $\Lambda'\gg \Lambda$ imply generically higher DM abundance and require smaller initial amplitudes $\phi_i$.
Unfortunately, at the moment we can not provide a fully motivated example.

Finally, the minimal requirement for an ALP to behave as CDM is that at latest at matter-radiation equality, at a temperature $T_{\rm eq}\sim 1.3$~eV, the mass attains its current value $m_0$ and therefore the DM density starts to scale truly as $1/R^3$. 
In particular, at this point the field should already have started to oscillate.
This corresponds to a lower limit on $m_1$, $m_1>3 H(T_{\rm eq})=1.8\times 10^{-27}$ eV, which implies an upper 
bound on  $\rho_{\phi,0}$,
\be
\label{eq:ALPDMbound}
\rho_{\phi,0}<  1.17\, \frac{{\rm keV}}{{\rm cm}^3}\ \frac{m_0}{{\rm eV}}\(\frac{\phi_i}{54\, {\rm TeV}}\)^2\; .
\ee
In other words, if we want these particles to be the DM, we need 
\be
(m_0/{\rm eV})(\phi_i/54\,{\rm TeV})^2>1\; ,
\ee  
giving a constraint on the required initial field value as a function of the mass today, which is plotted in figure~\ref{fig:ALPDM} in pink.
Again, we are not able to provide a well motivated model for an ALP that could perform this late realignment production and this limit has to be considered just as the broadest region of ALP parameter space that in principle could provide the right amount of CDM. 

The dark matter generated by the realignment mechanism has interesting properties beyond those of cold dark matter. At the time of their production, particles from the realignment mechanism are semi-relativistic. Their momenta are of the order of the Hubble constant $p\sim H_{1}\ll T_{1}$. 
Accordingly we have today --- outside of gravitational wells --- a velocity distribution with a very narrow width of roughly,
\begin{equation}
\delta v(t)\sim \frac{H_{1}}{m_{1}}\left(\frac{R_{1}}{R_{0}}\right)\ll 1\; .
\end{equation}
Combined with the high number density of particles, $n_{\phi,0}=N_\phi/R_0^3=\rho_{\rm CDM}/m_0$, this narrow distribution typically leads to very high occupation numbers for each quantum state,
\begin{equation}
N_{\rm occupation}\sim \frac{(2\pi)^3}{4\pi/3}\frac{n_{\phi,0}}{m^{3}_{0}\delta v^3}
\sim 10^{42}\left(\frac{m_{1}}{m_{0}}\right)^{3/2}\left(\frac{\rm eV}{m_{0}}\right)^{5/2}\; ,
\end{equation}
where we used $R_0/R_1\sim T_1/T_0\sim \sqrt{m_1 m_{\rm Pl}/T_0}$. 
If the ALP self-interactions are strong enough to achieve thermalisation and ALP-number-conserving, as argued in references~\cite{Sikivie:2009qn,Erken:2011dz} for the case of axions, this high occupation number leads to the formation of a Bose-Einstein condensate. 
This could imprint interesting signatures in cosmological observations, like galactic caustic rings~\cite{Sikivie:2009qn,Erken:2011dz,Duffy:2008dk,Kain:2011pd}.  

The axion realignment production is a particular case of what is described above. 
The $\Lambda$ parameter in this case is $\Lambda_{\rm QCD}$, and the phase transition that turns on the axion mass is the QCD phase transition.
The axion thermal mass was first calculated from the dilute instanton gas approximation~\cite{Gross:1980br} and more recently using the instanton liquid one~\cite{Wantz:2009it}.
The theoretical uncertainties in these calculations make the parameters for the thermal axion mass slightly different.
For example, a simple and recent approximation in the dilute gas approximation, that also agrees very well at high temperature with the liquid instanton one, is~\cite{Wantz:2009it}
\be
m_a(T)=4.1\times10^{-4}\,\frac{\Lambda^2}{f_a}\(\frac{\Lambda}{T}\)^{3.34}\; ,
\ee
for $\Lambda=400$ MeV.
This approximation is valid above $T\sim0.1$ GeV, since below this value the axion mass approaches the zero-temperature value $m_a$ given by equation~\eqref{amass}.
The exact shape of the thermal axion mass determines $t_1$, thus it affects the $m_0/m_1$ ratio in equation~\eqref{eq:CCDM}. 

\subsection{Before or after inflation?}

We do not know when cosmic inflation occurred and if it took place before or after the spontaneous breaking of the PQ-symmetry. 
This is one of the crucial points in calculating the axion energy density from the realignment mechanism. 
If the initial $\phi_i$ was set after inflation, we need an average of $\varphi_i^2$ over all the universe to calculate the amount of axions produced through equation~\eqref{eq:CCDM}, for the current universe includes many patches that were initially causally disconnected. 
However, this is not possible if the universe inflated at $T<f_a$: in this case the resulting axion energy density $\rho_a$ remains $\varphi_i^2$ dependent.
Assuming a flat prior on the value of the initial misalignment angle, $\varphi_i\sim\mathcal{O}(1)$ is the most likely outcome.
If this is the case, it is possible to state that the axion mass should be $m_a\gtrsim6\ \mu$eV, and thus $f_a\lesssim 10^{12}$~GeV, to avoid DM over-production. 
This is the \emph{traditionally} quoted lower bound, first obtained in 1982~\cite{Dine:1982ah,Abbott:1982af,Preskill:1982cy}. 
It is represented in figure~\ref{fig:alimits} with the \textbf{Cold DM} label.
Successive estimates, with a more refined treatment of the average of the squared initial angle or of the thermal axion mass, give more severe bounds but of the same order. 
For the realignment contribution, the recent reference~\cite{Wantz:2009it} gives a limit which is stronger by one order of magnitude.
The traditional $f_a\lesssim 10^{12}$~GeV bound is therefore the most conservative one.

\begin{figure}[htb] 
   \includegraphics[width=120mm]{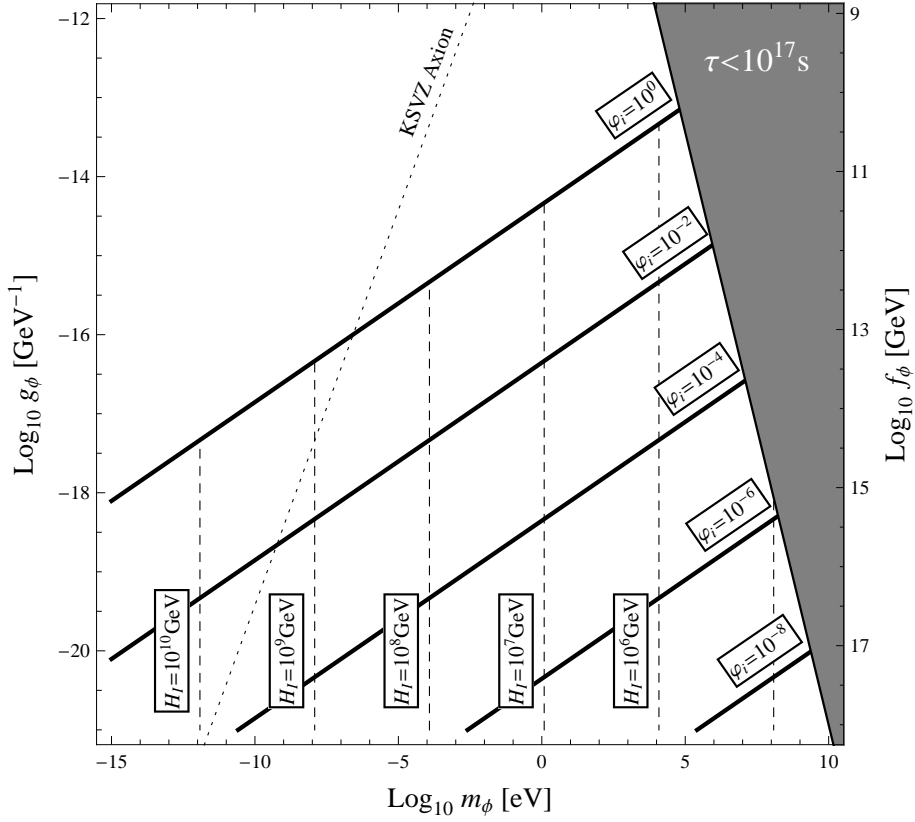}   
   \centering
   \caption{The anthropic window. 
The solid lines represent the isocontours of the values of $\varphi_i$ that provide the whole cold DM density, according to equation~\eqref{eq:CCDMb}, taking $m_1=m_0$ and $\mathcal{F}(T)=1$. 
Dashed lines give the isocontours of the maximum value of $H_I$ allowed by the limit~\eqref{eq:isoHI}, calculated using the same value of $\varphi_i$ as before.
As a reference, the KSVZ axion parameters are plotted as a dotted line.
In the grey region ALPs are cosmologically unstable.}
   \label{fig:anthropic}
\end{figure} 

It seems that this cold DM bound could be easily avoided in the case inflation happened after the spontaneous symmetry breaking, if we accept a fine-tuned very small value for $\varphi_i$. 
In figure~\ref{fig:anthropic} are plotted the isocontours of $\varphi_i$ that provide the whole cold DM density, if in equation~\eqref{eq:CCDMb} we take $m_1=m_0$ and $\mathcal{F}(T)=1$.
However, $\varphi_i$ has a minimum value due to the quantum fluctuations generated in the pseudoscalar field during inflation. 
These unavoidable inhomogeneities in $\varphi_i$ are of order $\delta \varphi_i \sim H_I/(2\pi f_\phi)$, where $H_I$ is the value of the expansion rate at the end of inflation. 
This precludes fine-tuning of the universe average of $\phi_i$ below $H_I/2\pi$, and sets a minimum DM abundance for a specified value of $H_I$. 
Since the $\phi$ field is effectively massless during inflation in this scenario, these inhomogeneities correspond to isocurvature perturbations of the gravitational potential. 
This has been discussed extensively in the literature in the context of axions and of string axions, see for example~\cite{Hertzberg:2008wr,Hamann:2009yf,Acharya:2010zx}. 

The WMAP7 observations of the primordial density fluctuations set very stringent constraints on isocurvature perturbations from which one can obtain an upper bound on $H_I$, assuming a given $f_\phi$. 
WMAP measures~\cite{Komatsu:2010fb}
\be
\alpha \equiv \frac{\bra |S^2| \ket}{\bra |S^2| \ket + \bra |R^2| \ket}< 0.077\; ,
\ee
where $\bra |S^2| \ket \approx \frac{H_I^2}{\pi^2 \phi_i^2}$ is the isocurvature power spectrum, and $\bra |R^2| \ket$ the adiabatic one, which is generated by the inflaton or by other fields. 
At the pivot scale $k_0 = 0.002\ \mathrm{Mpc}^{-1}$, WMAP finds $\bra |R^2| \ket = 2.42 \times 10^{-9}$, which gives the bound
\be\label{eq:isoHI}
H_I \lesssim 4 \times 10^{-5} \varphi_i f_\phi \; .
\ee
The lower bound for the reheating temperature is $T_{\rm RH}>4$~MeV~\cite{Hannestad:2004px}. 
Linking naively the two quantities gives $H_I\approx T_{\rm RH}^2/m_{\rm Pl}>1.3\times10^{-24}$~GeV. 
In the light of equation~\eqref{eq:isoHI}, it is rather a weak bound by itself.
However, in the most common inflation models, $H_I$ is preferred to be much larger, to allow mechanisms that require very high initial temperatures of the universe --- like leptogenesis, which needs $T>10^9$~GeV~\cite{Buchmuller:2004nz}.
The isocontours of the maximum $H_I$ allowed by limit~\eqref{eq:isoHI} are plotted in figure~\ref{fig:anthropic}, where the value of $\varphi_i$ that provides the whole DM density is used.
A given value of $H_I$ excludes the region of the parameter space that lies on the right of its isocontour.
Then, along the $H_I$ isocontour $f_\phi$ selects the $\varphi_i$ that is provided by primordial fluctuations, while on the left half-plane the initial misalignment allowed is larger and thus less fine-tuned.

In the case of the axion, the values of $f_a$ that satisfy the constraint~\eqref{eq:isoHI} for large $H_I$ are above the GUT scale, which seems very reasonable from the model building point of view: they form the so-called \emph{anthrophic window}\footnote{
The adjective anthropic refers to the fact that the existence of the human species requires the universe to fulfil some conditions about its age and composition, which too abundant axion DM would not meet.}. 
Considering that the axion was introduced to avoid fine-tuning to solve the strong-CP problem, the legitimacy of the choice of a very small $\varphi_i$ to justify an axion scale linked to $\Lambda_{\rm GUT}$ or the string scale is questionable~\cite{Mack:2009hv}.  
However, the anthropic point of view can find a vindication regarding the DM density and other significant physical quantities simply as observational data, from which it is possible to extract the value of the initial misalignment with an a posteriori analysis~\cite{Tegmark:2005dy}. 
This procedure is legitimate because of the intrinsic randomness of $\varphi_i$, and converts the fine-tuning issue into an observational problem.
 
On the other hand, if the SSB took place after inflation the $\phi_i$ picked up random values in different causally disconnected regions of the universe and the anthropic window loses its significance.
If this is the case, also the gradient effects should be taken into account.
At larger scales, the DM density averages to a constant value corresponding to 
$\langle \phi_i^2\rangle \sim \pi^2 f^2_\phi /3 $, bearing the mentioned ${\cal O}(1)$ correction due to the non-harmonic behaviour of large initial phases. 
The initial size of the domains of different $\phi_i$ cannot be larger than 
\begin{equation}
L_{i, {\rm dom}}\sim \frac{1}{H_{\rm SSB}} \sim \frac{m_{\rm Pl}}{f_\phi^2\sqrt{g_*(f_\phi)}}\; .
\end{equation}   
Non-linear effects, due to the attractive self-interaction caused by higher order terms in the expansion of the potential~\eqref{ALPpotential}, drive the overabundances to form peculiar DM clumps that are called \emph{miniclusters}~\cite{Hogan:1988mp,Kolb:1993zz,Kolb:1993hw,Zurek:2006sy}. 
These act like seeds enhancing the successive gravitational clumping that leads to structure formation. 
The minicluster mass is set by the dark matter mass inside the Hubble horizon $d_H=H^{-1}$ when the self-interaction freezes-out, i.e.~$M_{\rm mc}\sim \rho_{\phi}(T_{\lambda})d_H(T_{\lambda})^3$ for the freeze-out temperature $T_{\lambda}$. 
Long-range interactions will be exponentially suppressed at distances longer than $1/m_{\phi}$ so we can expect $T_{\lambda}$ to be of the order  $T_1$, with at most a logarithmic dependence on other parameters. 
This is indeed the case for QCD axions, for which the miniclustering is quenched soon after the QCD phase transition that turns on the potential~\eqref{ALPpotential}~\cite{Sikivie:2009qn} giving $M_{\rm mc}\sim 10^{-12}M_{\odot}$, where $M_{\odot}=1.989\times10^{33}\ {\rm g}$  is the solar mass, and a radius $R_{\rm mc}\sim10^{11}$~cm~\cite{Kolb:1995bu}. 
In the case of ALPs, $M_{\rm mc}$ can be larger if  the mass is lighter. 
The authors of~\cite{Zurek:2006sy} pointed out that the present data on the CDM power spectrum constrain $M_{\rm mc}\lesssim 4 \times 10^3 M_{\odot} $ which translates into a lower bound 
in temperature $T_{\lambda}>2\times 10^{-5} {\rm \ GeV}$ and in the ALP mass $m_{\phi}>H(T=2\times 10^{-5} {\rm \ GeV})\sim 10^{-20}$~eV. 
If some of these miniclusters survive the tidal disruption during structure formation they should be observable in forthcoming lensing experiments~\cite{Zurek:2006sy,Kolb:1995bu}. 

\subsection{Contribution of topological defects}

During the spontaneous symmetry breaking of a global symmetry topological defects such as cosmic strings and domain walls are formed~\cite{Vilenkin:1982ks}. 
Strings form because of the breaking of a $U(1)$ symmetry. 
They have a thickness $\delta \sim 1/f_\phi$  and typical sizes of the order of the horizon, $L \sim t$. 
As strings enter into the horizon they can rapidly reconnect, form loops and decay into PNGBs and we have to consider also their contribution to the axion and ALP energy density.

Axions resulting from string decay are known to contribute significantly to their cold DM density, but the exact amount is subject to a long-standing controversy~\cite{Kolb:1990vq}. 
The debate is focused around the axion emission spectrum. 
Some authors argue that the decay proceeds very fast, with an emission spectrum $1/k$ with high and low energy cutoff of order respectively $1/\delta$ and $1/L$.
In this case the contribution to $\rho_\phi$ is of the same order as the realignment mechanism~\cite{Harari:1987ht,Hagmann:1990mj}.
Others put forward that the string decays happen after many oscillations, with a radiation spectrum peaked around $2\pi/L$, which enhances the contribution to cold DM by a multiplicative factor of $\log (L/\delta)\sim \log (f_a/m_a)\sim {\cal O}(60)$~\cite{Davis:1985pt,Davis:1986xc,Vilenkin:1986ku,Davis:1989nj,Dabholkar:1989ju,Battye:1993jv,Battye:1994au}.   

Once the axion potential builds up at the QCD phase transition, also domain walls build up. 
If the axion potential has only one minimum the domain walls can still efficiently decay into axions --- since on their two sides there is the same vacuum --- leading to a third axion population. 
If different exactly degenerate vacua exist\footnote{From the definition~\eqref{eq:PQscale} of $f_a$, the axion potential~\eqref{eq:potential} has exactly $\mathcal{N}$ equivalent vacua.} the domain walls are persistent. 
The energy stored in a domain wall is huge, and can very easily run in conflict with observations. 
An option to overcome this problem requires to assume a small explicit breaking of the Peccei-Quinn symmetry, which breaks the degeneracy and select one among the possible vacua, otherwise it has to be $\mathcal{N}=1$~\cite{Sikivie:1982qv}.
It has been recently pointed out that the axions emitted by decaying domain walls constitute the dominant population in the slow string decay scenario~\cite{Hiramatsu:2012gg}.
Summing all the contributions to axion non-thermal DM in this scenario would constrain the axion decay constant to be $
f_a\lesssim (1.2$--$2.3) \times 10^{10}\ {\rm GeV}$, two orders of magnitude lower than the traditional axion CDM bound~\cite{Hiramatsu:2012gg}.
We plotted this bound in figure~\ref{fig:alimits}, only enclosed by a line because of its uncertainty: in the case the SSB would have happened before inflation, the topological defects, being diluted away, would provide a negligible contribution and thus the traditional value is again the one to be considered.
The same type of behaviour is expected for ALPs with characteristics similar to those of the axion, i.e.~ALPs whose mass is generated at a late phase transition due to a hidden sector which becomes strongly interacting. 
In this case we should keep in mind the controversy of the string decay spectrum and assume an uncertainty of order $\log(f_\phi/m_\phi)$ in the DM abundance. 
The domain wall problem can in principle be solved by strong enough explicit breaking if $\mathcal{N}>1$.
Again, the contribution to ALP energy density from wall decay should mimic the axion case in the $\mathcal{N}=1$ case.

\subsection{Detection of pseudoscalar cold dark matter}  

The axion and the ALP are optimal cold DM matter candidates, but only if they are cosmologically stable, which depends on their mass and couplings.
The two-photon coupling provides a decay channel, and the consequent lifetime is 
\be\label{eq:lifetime}
\tau=\frac{64 \pi}{g^2_\phi m^3_\phi}\simeq 1.3 \times 10^{25} {\rm s}\ \(\frac{10^{-10} \, {\rm GeV}^{-1}}{g_\phi}\)^2 \(\frac{{\rm eV}}{m_\phi}\)^3\; .
\ee
Of course, if the pseudoscalar is massive enough other decay channels are kinema\-tically allowed.
In figure~\ref{fig:ALPDM} the grey area hides the part of the parameter space in which the ALP and the axion have a lifetime shorter than the age of the universe, which is approximately $10^{17}$ s.
In the rest of the dissertation we will very frequently deal with $\tau$, probably the most important parameter for our cosmological considerations.

Axion and ALP cold DM is currently searched with resonant cavities called \emph{haloscopes}~\cite{Sikivie:1983ip}. 
These are high quality resonant microwave cavities permeated by a strong and static magnetic field.
The cavity is tuned to convert the pseudoscalars in the DM halo into photons exploiting the two-photon coupling. 
The signal is a monochromatic peak at the frequency of the mass of the particle, broadened by the virial distribution of particle velocity in the galactic potential~\cite{Nakamura:2010zzi,Carosi:2007uc}.

A haloscope provides a power of pseudoscalars converted to photons equal to 
\be \label{eq:halopower}
W=g_\phi^2 V B^2 \frac{\rho_\phi}{m_\phi} C \min\(Q^c,Q^\phi\)\; ,
\ee
where $B$ is the magnetic field, $V$ the cavity volume, $C$ the form factor for the cavity mode, and $Q^c$ and $Q^\phi$ are respectively the quality factors for the cavity and for the pseudoscalar signal, which is basically the ratio of the energy of the signal over its spread in energy~\cite{Carosi:2007uc}.
A technological challenge in building such apparatus is to balance the quality factor of the cavity. 
For the moment, the highest $Q^c$ that can be obtained is still about an order of magnitude less than quality factor of the expected signal.
Of course, a very high magnetic field is preferable.
The geometry of the cavity selects the range of masses to which the instrument is sensitive and also influences the power \eqref{eq:halopower} through the volume factor. 
The design of the device has therefore to find a compromise between the two, considering that the geometry of the cavity has to be 
optimised to measure the most probable range for pseudoscalar cold DM, which in the axion case means very suppressed couplings. 
The cavity moreover has to be tunable, in order to span several values of $m_\phi$.
To explore a broad mass range takes a long time, for just one mass value per time can be tested.
Finally, since the signal provided by pseudoscalar DM would be rather feeble, many techniques to suppress the noise in the apparatus are applied, in order to enhance the signal to noise ratio.

The first haloscope was the Rochester-Brookheaven-Fermilab detector~\cite{Wuensch:1989sa}, followed by the University of Florida experiment~\cite{Hagmann:1990tj}, but both of them had not enough sensitivity to test the axion. 
Only a few years later ADMX was almost able to reach the sensitivity to test the axion as DM in the $1.9$--$3.5\ \mu$eV range~\cite{Asztalos:2003px}.
ALP cold DM was tested and not found in the turquoise regions labelled \textbf{Haloscope excluded} in figure~\ref{fig:ALPDM}.
In this mass range, for $g_\phi\gtrsim10^{-10}\ {\rm GeV}^{-1}$ this is not surprising because we do not expect any ALP condensate.
The haloscope tested regions are also plotted in figure~\ref{fig:alplimits} and labelled \textbf{Haloscopes}.
A new experiment, called ADMX-HF, is planned to operate at Yale University: the plan is to scan the ALP parameter space in the range around 10--100 $\mu$eV, with enough sensitivity to reach the axion band~\cite{CarosiTalk}.

\section{Production mechanisms II: thermal relics}\label{sec:thermal} 

We learned in the previous section that the higher the ALP/axion decay constant, the more abundant would be the final yield, it being proportional to $\phi_i^2\propto f_\phi^2$, see equation \eqref{eq:CCDM}.
If however $f_\phi$ is not high enough to guarantee a sufficient non-thermal production, the ALP or the axion can be nevertheless a DM component as it would be more strongly coupled and more prone to thermal production. 

A population of particles with $\degree$ internal degrees of freedom --- which counts the polarisation and the particle-antiparticle states --- and mass $m$ in \emph{thermal equilibrium} has number and energy densities and pressure given respectively by
\begin{align}
n_{\rm eq}&=\frac{\degree}{2\pi^2}\int_m^\infty \frac{\sqrt{E^2-m^2}}{\exp\(E/T\)\pm1}E\,dE \; ,\\
\rho_{\rm eq}&=\frac{\degree}{2\pi^2}\int_m^\infty \frac{\sqrt{E^2-m^2}}{\exp\(E/T\)\pm1}E^2\,dE \; , \\
p_{\rm eq}&=\frac{\degree}{6\pi^2}\int_m^\infty \frac{\(E^2-m^2\)^{3/2}}{\exp\(E/T\)\pm1}\,dE\; ,
\end{align}
if the temperature is $T$.
According to the fermionic or bosonic nature of the particle, the sign $+$ or $-$ holds in the previous formulae.
In the case of negligible chemical potential, the previous equations in the relativistic limit $T\gg m$ for bosons are 
\begin{align}
n_{\rm eq}&=\frac{\zeta(3)}{\pi^2}\degree\, T^3\; ,\label{eq:nrelbos}\\
\rho_{\rm eq}&=\frac{\pi^2}{30}\degree\, T^4 \; , \label{eq:rhorelbos}\\
p_{\rm eq}&=\frac{\rho}{3}\; ,
\end{align}
while for fermions are
\begin{align}
n_{\rm eq}&=\frac{3}{4}\frac{\zeta(3)}{\pi^2}\degree\, T^3\; ,\label{eq:nrelferm}\\
\rho_{\rm eq}&=\frac{7}{8}\frac{\pi^2}{30}\degree\, T^4 \; , \label{eq:rhorelferm}\\
p_{\rm eq}&=\frac{\rho}{3}\; .
\end{align}
In the non-relativistic limit, $T\ll m$, the quantum statistics makes no difference and we have
\begin{align}
n_{\rm eq}&=\degree \(\frac{m\, T}{2\pi}\)^\frac{3}{2}\exp\({-\frac{m}{T}}\)\; ,\label{eq:nnonrel}\\
\rho_{\rm eq}&=m n \; , \\
p_{\rm eq}&=T n\; .
\end{align}

The dynamics of a particle distribution is governed by the Boltzmann equation.
The Boltzmann equation links the total time derivative of a distribution on the left hand side to the microphysics of particle interactions that lies on the right hand side.
The effects of the universe expansion are included on the left hand side.
If $\sigma_i$ is the cross-section for the scattering of the particle under examination with the $i$-species in the bath, the Boltzmann equation for number density can be written as
\be\label{eq:boltzmanngeneral}
\dot{n}+3 H n=-\sum_i\bra \sigma_i v \ket\(n n^i-n_{\rm eq}n^i_{\rm eq}\)\; ,
\ee
assuming that the interaction products rapidly thermalise, which allows us to write the term $n_{\rm eq}n^i_{\rm eq}$ on the right-hand side~\cite{Gondolo:1990dk}. 
The thermal average $\bra \sigma_i v \ket$ is obtained integrating in momentum the matrix element for the interaction process multiplied by the distribution functions of the particles involved.

Our aim is to calculate the rate at which an axion or ALP population can arise from the scattering of SM particles in the thermal bath during the early phases of the universe.
The axion and ALP self-interaction can be neglected, being suppressed by $f_\phi^{-4}$. 
Then, we can consider only the interactions with particle species in thermal equilibrium.
We will specify better which interactions with which particles in the following. 
Under this assumption, $n^i$ is always $n^i_{\rm eq}$ and the Boltzmann equation simplifies as 
\be\label{eq:boltzmann}
\dot{n}+3 H n=-\Gamma\(n-n_{\rm eq}\)\; ,
\ee
where $\Gamma=\sum_i \Gamma_i$ is the total interaction rate and the $i$-th partial one is defined to be $\Gamma_i=\bra \sigma_i v \ket n^i_{\rm eq}$.
From equation \eqref{eq:boltzmann} we infer that if $\Gamma$ is much bigger than the expansion rate $H$ the number density is forced to follow the equilibrium one.
If the right hand side becomes negligible, $n$ is only diluted by the expansion of the universe, and changes through the factor provided by equation~\eqref{eq:dilution} 
\be\label{eq:afterfo}
n(t)=n(t_*) \(\frac{R(t_*)}{R(t)}\)^3\; .
\ee
We can consider $\Gamma/H>1$ as a rule of thumb to define the thermal equilibrium condition, while if $\Gamma/H<1$ the species under examination is decoupled.
Both $\Gamma$ and $H$ are temperature dependent.
When a particle species loses thermal contact with the bath because all its interactions \emph{freeze out}, it is said to \emph{decouple}, and this moment is characterised by the temperature $T_{\rm fo}$ which is the solution of the equation $H(T_{\rm fo})=\Gamma(T_{\rm fo})$.
At any later time between the freezing out and the decay, the pseudoscalar density can be obtained in first approximation thanks to equation~\eqref{eq:afterfo} and the dilution factor~\eqref{eq:dilution}
\be
\label{ALPyield}
n_\phi(T) = n^{\rm eq}_\phi(T_{\rm fo})\frac{g_{*S}(T)T^3}{g_{*S}(T_{\rm fo})T_{\rm fo}^3}\; .
\ee
Assuming the particle content of the standard model, there is a minimum pseudoscalar yield which is given by the value of 
$g_{*}(T_{\rm fo}>E_{\rm EW})$
\be \label{nphionngamma}
\frac{n_\phi}{n_\gamma}\geq \frac{1}{2}\frac{g_{*S}(T)}{106.75}\simeq 0.005 g_{*}(T)\; .
\ee

Our simple criterion about $\Gamma/H$ is sufficient to give a first approximation of the axion thermal history according to the axion scale $f_a$.
The key events take place during the radiation domination era, and therefore $H$ can be expressed through equation \eqref{eq:exprateRAD}.
It is conceptually convenient to divide the thermal history of the universe into three epochs, classifying them through the main axion interaction at work.

\begin{itemize}
\item $f_a\gtrsim T\gtrsim \Lambda_{\rm QCD}$ \\
We learned in the previous section that when the universe temperature falls below $f_a$ the PQ-symmetry is spontaneously broken and the axion, still massless, pops up. 
If $f_a$ is not too high, the axion coloured interactions depicted in figure \ref{fig:colour} are efficient. 
The interaction rate is $\Gamma\sim (\alpha_s^3/f_a^2) n_{\rm col}\propto (\alpha_s^3/f_a^2) T^3 $, where $n_{\rm col}$ is the number density of coloured particles.
A more precise estimate of $\Gamma$ is given in~\cite{Graf:2010tv}.
The ratio $\Gamma/H\propto T$. 
Thus at high $T$ the axions are in equilibrium --- if the required temperature was ever achieved --- and they decouple when $T$ becomes too small.
To decouple during this phase, the axion must have a decay constant in the range $f_a\simeq 10^8$--$10^{10}$~GeV, thus a mass around the 1--10 meV.
If this is the case, the fraction of the universe energy density in thermal axions is negligible, $\Omega_a h^2\sim10^{-6}$--$10^{-4}$~\cite{Graf:2010tv}.

\begin{figure}[htb] 
   \centering
   \subfloat
   {\includegraphics[width=60mm]{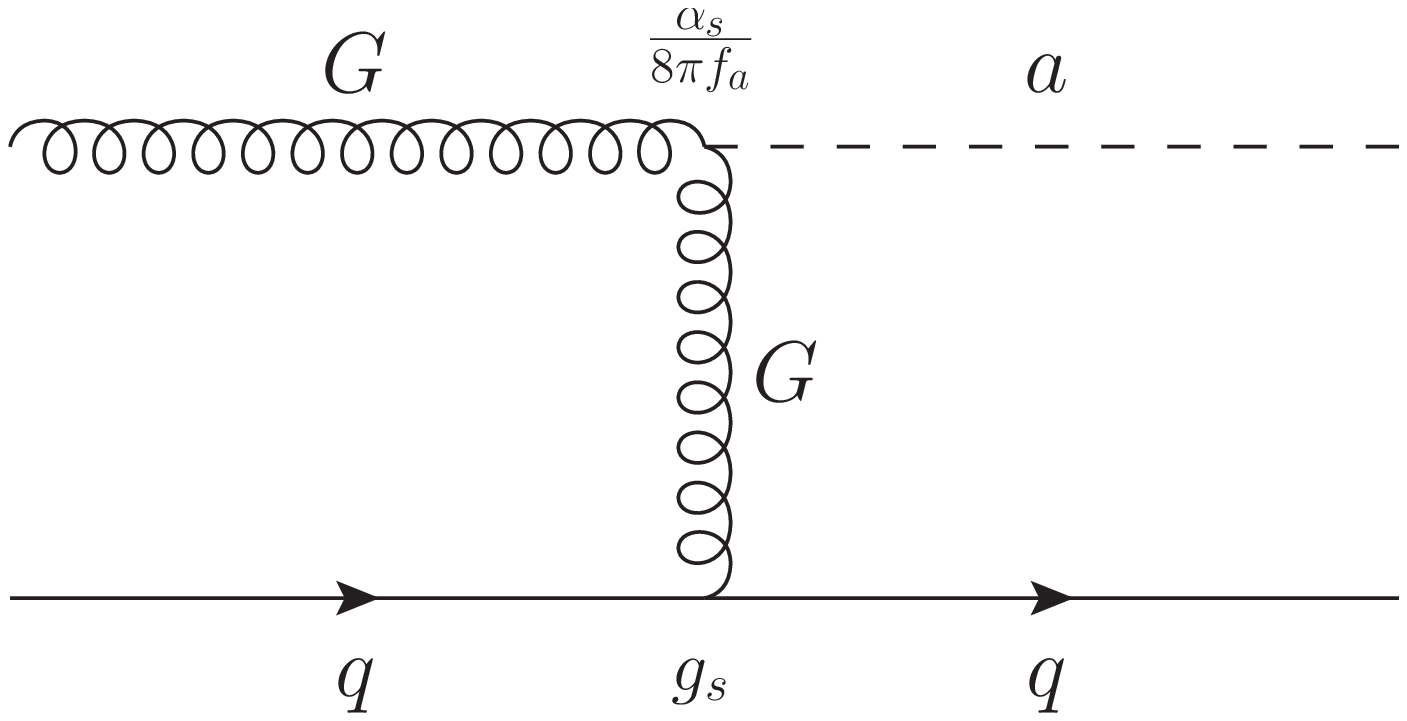}   }
	\qquad
   \subfloat
   {\includegraphics[width=60mm]{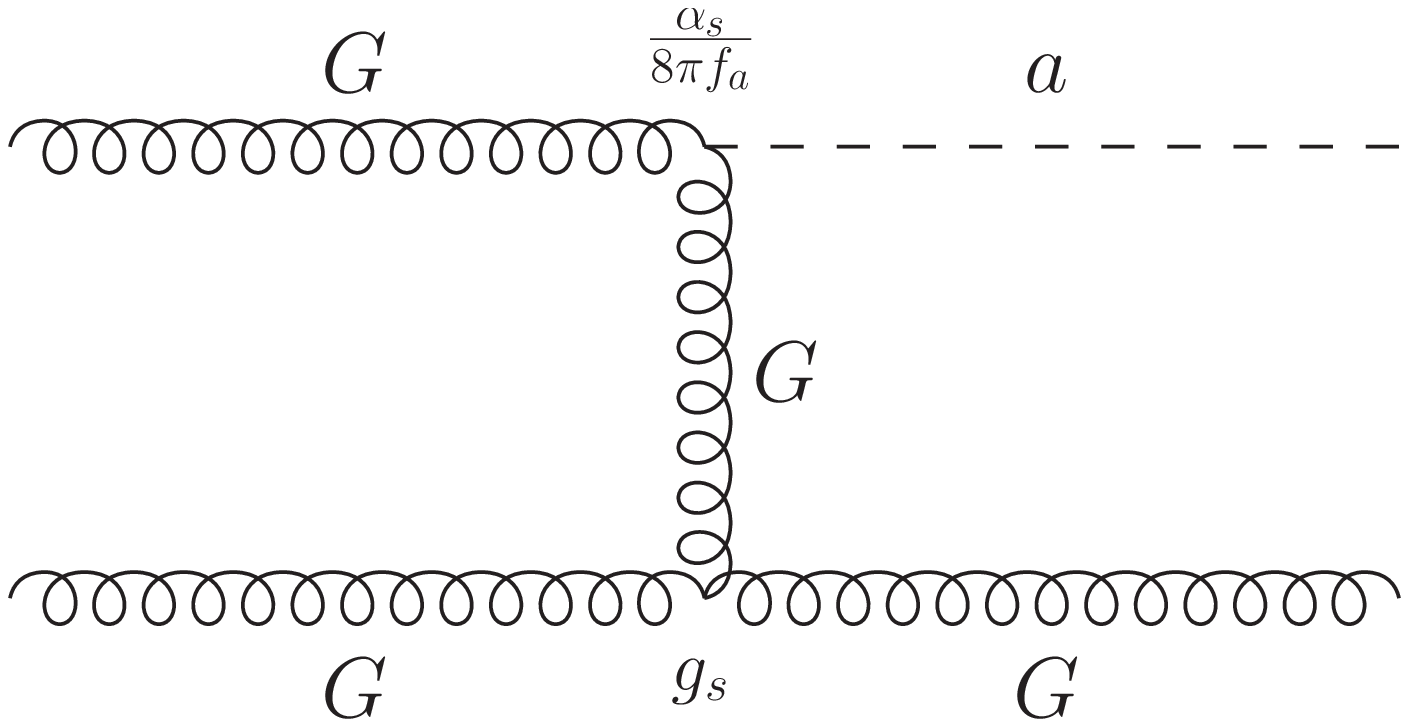}   }\\
 \subfloat
   {\includegraphics[width=60mm]{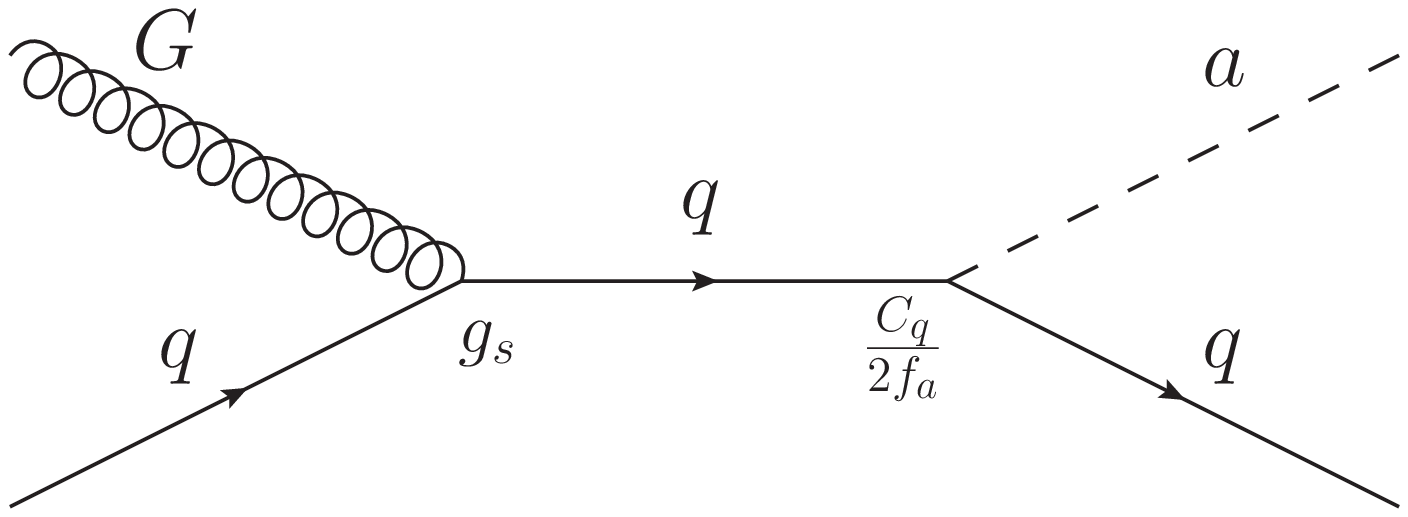}   }
   \caption{Relevant axion interactions for $T>\Lambda_{QCD}$.}
   \label{fig:colour}
\end{figure} 

\item $\Lambda_{\rm QCD}\gtrsim T\gtrsim m_\pi$ \\
At $T\sim\Lambda_{\rm QCD}$ gluons and quarks confine. 
At this stage, if not yet decoupled, axions can still interact with pions thanks to the effective Lagrangian~\eqref{apion} (figure~\ref{fig:pion}), with $\Gamma\sim T^5/(f_\pi f_a)^2$~\cite{Hannestad:2005df}.
If the axion decay constant is $f_a\simeq 10^4$--$10^7$ GeV, which in terms of mass is $m_a\sim0.01$--$100$ eV, axions decouple at this stage. 
The axion yield in this case is larger than in the previous one, and in principle it can be abundant enough to influence the evolution of the universe.
The formation of LSSs like those we observe today requires the percentage of relativistic matter to be not too high, otherwise free streaming light particles would prevent the gravitational binding of the smallest clumps. 
Together with the CMB measurements, LSS constrains the axion mass $m_a$ to be smaller $0.7$ eV  and thus $f_a > 9\times 10^{6}$~GeV~\cite{Hannestad:2005df,Hannestad:2010yi}.
In figure \ref{fig:alimits} this upper bound is labelled as \textbf{Hot DM}.
This bounds applies to a small range of axion masses, for its validity ceases if the axion is cosmologically unstable. 
In this case the cosmological effects of the axion needs other tools to be analysed, which will be discussed in the following chapters.

\begin{figure}[htb] 
   \centering
   \includegraphics[width=55mm]{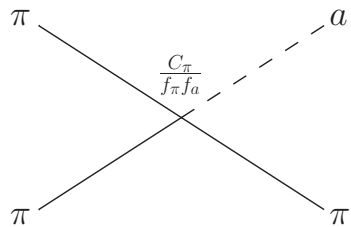}   
   \caption{Axion-pion interaction relevant for $\Lambda_{\rm QCD}>T>m_\pi$.}
   \label{fig:pion}
\end{figure} 

\item $T \lesssim m_\pi$ \\
Once the pions become non-relativistic, their abundance is exponentially suppressed, as it can be seen from equation~\eqref{eq:nnonrel}, and the axion-pion interaction freezes out.
Axions are only left to interact electromagnetically with photons and charged particles.
The interaction of axions during this phase is less efficient in keeping them in equilibrium, not only because the electromagnetic force is weaker than the strong one, but also because the interacting species are just a few and very diluted by the expansion. 
Figure~\ref{fig:electron} shows how a charged particle interacts with an axion via the Primakoff effect, which remains active until the $e^+e^-$ annihilation at $T\sim m_e$. 
For this interaction, $\Gamma_{a\gamma\gamma}\sim\alpha g^2_{a\gamma\gamma}n_e$.
The axion decay is also an electromagnetic interaction, with $\Gamma=1/\tau$. 
In particular, this process is characterised by a \emph{freezing in} temperature, since $\tau$ has no $T$ dependence and thus the ratio $\Gamma/H$ grows with time. 
Axions with a mass larger than $\mathcal{O}(10)$ keV never freeze out, because for them $H\sim1/\tau$ before the Primakoff interaction with electrons decouples~\cite{Cadamuro:2010cz}.
In this case, the thermal equilibrium condition implies their disappearance from the bath when they becomes non-relativistic.

\begin{figure}[htb] 
   \centering
   \includegraphics[width=70mm]{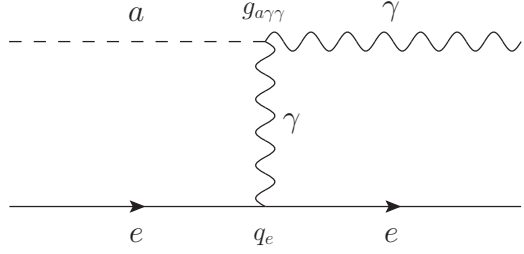}   
   \caption{Relevant axion interaction for $T<m_\pi$.}
   \label{fig:electron}
\end{figure} 

\end{itemize}

\begin{figure}[htb] 
   \centering
   {\includegraphics[width=60mm]{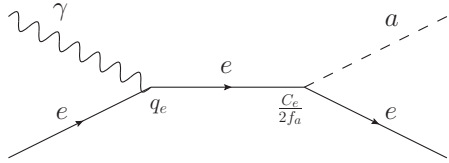}   }
   \caption{Additional axion-electron interaction, which is relevant in the DFSZ model for $T<m_\pi$.}
   \label{fig:compton}
\end{figure}

Of course, if the axion has direct coupling to other SM particles, like the electron as in figure~\ref{fig:compton}, the final yield would be higher, because, according to equation~\eqref{eq:boltzmann}, a higher total interaction rate $\Gamma$ means a later decoupling.
In the ALP case, we are mainly interested in the interactions involving the two-photon coupling~\eqref{ALPphotoncoupling}.
Additional interactions would delay the decoupling and increase the final abundance.

As already mentioned, the cosmological stability is the second key factor to obtain limits on the parameters of pseudoscalars from cosmological observables.
If the axion mass is larger than 20 eV, we can see from equation~\eqref{eq:lifetime} that its lifetime is shorter than the age of the universe.
Considering our schematic representation of axion thermal production as a function of $f_a$, it seems worthwhile to give some more details about the last processes, for they determine the abundance of cosmologically unstable axions. 
Moreover, the generalisation to the ALP case is straightforward.

The rate of Primakoff production due to scattering on relativistic electrons was computed in~\cite{Bolz:2000fu} to be 
\be
\Gamma_{\rm P} = \frac{\alpha\,{g_{a\gamma\gamma}}^2}{12}T^3\[\log\(\frac{T^2}{m_\gamma^2}\)+0.8194\]\; , 
\ee
where ${m_\gamma}^2= 2\pi\alpha T^2/3$ is the squared plasmon mass in a relativistic electron-positron plasma and $T$ the temperature.
Taking the rate to be proportional to the number density of electrons, $n_e=3\zeta(3)T^3/\pi^2$, we can generalise to a multicomponent plasma, using instead the effective number density of charged particles 
\be
n_q=\sum_i {q^i_{\rm EM}}^2 n_i \equiv \frac{\zeta(3)}{\pi^2}g_q(T) T^3\; .
\ee
The parameter $g_q(T)$ represents the effective number of relativistic charged degrees of freedom.
The plasmon mass has also to be corrected by a factor $m_\gamma\propto g_q^{1/2}$.
The Primakoff interaction rate becomes
\be
\Gamma_{\rm P} \simeq \frac{\alpha\, {g_{a\gamma\gamma}}^2}{12}\frac{\pi^2\,n_q }{3\zeta(3)}\[\log
\(\frac{T^2}{m_\gamma^2}\)+0.8194\] \; .
\ee 

The freezing out temperature for the Primakoff interaction is easily calculated from $\Gamma_{\rm P}/H=1$, which gives
\be\label{eq:T_fo}
T_{\rm fo}\simeq \frac{11}{\alpha\,{g_{a\gamma\gamma}}^2\, m_{\rm Pl}}\frac{\sqrt{g_*}}{g_q}\simeq
123 \frac{\sqrt{g_*}}{g_q} 
\(\frac{10^{-9}\, {\rm GeV}^{-1}}{g_{a\gamma\gamma}}\)^2 {\rm GeV}\; . 
\ee
In the ALP case, for values of $g_\phi \lesssim 2 \times 10^{-9} \ {\rm GeV}^{-1}$, interactions freeze out at temperatures above the electroweak scale, where the particle content of the plasma is somewhat speculative. 
For instance in the minimal supersymmetric standard model scenario, above the supersymmetry breaking energy scale we have
$g_*=228.75$, while the SM alone provides only 106.75 relativistic degrees of freedom.
For $g_\phi \lesssim  10^{-17} \ {\rm GeV}^{-1}$ we require a freeze out temperature above the Planck scale, which is most likely meaningless.

Cosmologically stable ALPs must not exceed the measured abundance of DM, therefore
\be
\Omega_\phi h^2 =\frac{\rho_\phi}{\rho_c}h^2=\frac{m_\phi\,n_{\phi}}{\rho_c}h^2<\Omega_{\rm DM}h^2=0.11\; .
\ee
Using~\eqref{ALPyield}, we see that ALPs with a mass $m_\phi=154$ eV would account for all the dark matter of the universe, thus larger masses are excluded.
The discovery of new degrees of freedom (dof) above the EW scale would relax this bound, which is linearly sensitive to $g_*(T_{\rm fo})$, by a factor $(106.75+\, {\rm new\ dof })/106.75$. 
This means that $\mathcal{O}(100)$ of them are needed for a sizeable change.

\section{Absorption of overabundant pseudoscalars}\label{sec:excess}

The prototypical Boltzmann equation \eqref{eq:boltzmann} shows that if an interaction rate is effective, the particle number density tends to its equilibrium value.
Therefore, in case that the pseudoscalars are more abundant than those required by the equilibrium condition because of the realignment production, the excess tends to be absorbed in the thermal bath if the particles thermalise with it.
However, in the range of parameters in which the realignment plays an important role, the Primakoff interaction is rather inefficient in thermalising the pseudoscalar condensate.
Moreover, the process $\phi+e^{\pm}\rightarrow \gamma+e^{\pm}$ is exponentially suppressed at high temperature because the energy of the incoming electron $E$ has to be high enough to produce a photon whose plasmon mass $m_\gamma$ is of the order of the temperature.
This threshold implies $2m_\phi E > m_\gamma^2 $, and thus $E\gtrsim\alpha T^2/m_\phi$, which is disfavoured especially in the case of small $m_\phi$. 
This threshold could be overcome in a three-body interaction with a photon more in the initial state. 
However also this process is suppressed at high temperature. 
Since the $\tilde{F}F$ term is a total derivative, the two-photon interaction involves a derivative of the pseudoscalar field whose only component for the condensate is $\partial_0\phi\sim m_\phi \phi$.
Again, for small-mass pseudoscalars the interaction is suppressed by powers of $m_\phi/T$, because all absorption amplitudes for the zero mode are proportional to $m_\phi$ and thus the probability to $m_\phi^2$. 

A more subtle effect involves the presence of primordial magnetic fields, and could in principle absorb the excess of pseudoscalar DM pointed out in the previous section.
We have mentioned that, in presence of a magnetic field $\vec{B}_{\rm ext}$, a pseudoscalar particle and the component of the electromagnetic vector potential parallel to $\vec{B}_{\rm ext}$ mixes, because of the two-photon coupling \eqref{ALPphotoncoupling}.
The two fields can therefore oscillate into one another.
The mixing allows a pseudoscalar surrounded by a magnetic field to have effectively the same interactions as the photon, but with amplitude suppressed by the mixing parameter $\chi_{\rm eff}$ which is usually small.
These photon-like interactions should therefore be added to our thermalisation considerations, besides the Primakoff effect.
Plasma effects complicate further the problem, because the mixing parameter can be substantially enhanced during its thermal evolution.
More details about the magnetic field induced evaporations and the derivation of the related quantities are provided in appendix \ref{app:axphotmix}. 

The effective mixing parameter in the primordial plasma can be approximated by
\be \label{eq:chieff}
\chi_{\rm eff}^2\simeq\frac{ \(g_\phi B_{\rm ext} \omega\)^2}{ \({m_\gamma}^2-{m_\phi}^2\)^2+\(\omega D\)^2}\; .
\ee
Here, $B_{\rm ext}(T)$ is the modulus of the magnetic field, and $D(\omega,T)$ is the photon damping factor, which depends on the photon wavelength $\omega$ and the temperature.

Magnetic seeds could be generated during cosmological phase transitions~\cite{Grasso:2000wj}, thus plausible temperatures for the birth of a primordial magnetic field could be $\Lambda_{\rm QCD}$ or $E_{\rm EW}$.
We call $T_B$ this temperature.
Because of equipartition of energy --- and dimensional analysis --- we assume for primordial magnetic fields $B\sim T^2$.
Then, the magnetic flux conservation implies that $B=B_i(R_B/R)^2\sim B_i(T/T_B)^2$ if $B_i$ is the initial strength of the field~\cite{Ahonen:1995ky}, suppressing the mixing \eqref{eq:chieff} at low temperatures.

If the damping factor dominates the mixing parameter~\eqref{eq:chieff}, the magnetic evaporation affects the pseudoscalar condensate if
\be
\(\frac{g_\phi}{10^{-10}\ {\rm GeV}^{-1}}\)^2\(\frac{B_0}{ {\rm nG}}\)^2\(\frac{106}{g_*(T_B)}\)^{1/2}\frac{T_B}{10^{9}\ {\rm GeV}}\gtrsim1\; ,
\ee
where $B_0$ is the present value of the magnetic field.
An axion cold DM condensate would not be affected by the magnetic field~\cite{Ahonen:1995ky}, since in this case $g_{a\gamma\gamma}\ll 10^{-10}\ \mbox{GeV}^{-1}$ and $B_0$ in the nG range is very close to the upper limit for a primordial magnetic field~\cite{Kahniashvili:2010wm,Paoletti:2010rx}.

\begin{figure}[tbp]
   \centering
   \includegraphics[width=8cm]{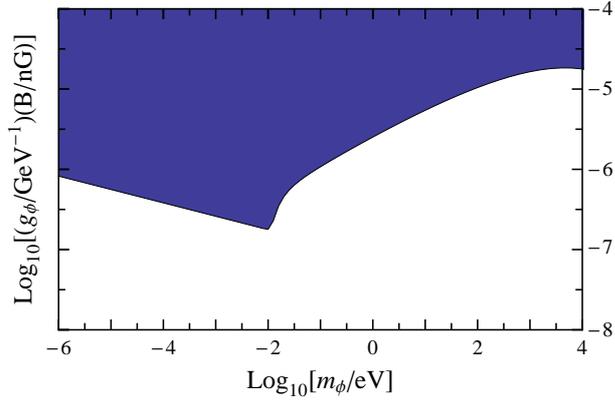}
   \caption{Region of the parameter space that would be affected by the resonant pseudoscalar-photon mixing, leading to the evaporation of the DM condensate, in case of primordial magnetic field of present strength $B$.}
   \label{fig:Bevaporation}
\end{figure}

In the lower temperature range, when the plasmon mass $m_\gamma$ is more important than the damping factor, there are three different regimes for $\chi_{\rm eff}$.
We are mostly interested in the resonant regime, defined by $m_\gamma(T) \sim m_\phi$, during which the mixing is enhanced.
In the resonant case, the magnetic field induces the evaporation of an ALP condensate if
\be
\frac{\pi}{2}\frac{\(g_\phi B_{\rm res}\)^2}{r m_\phi H_{\rm res}}\gtrsim 1
\ee
where $r=d \log\({m_\gamma}^2\)/d \log(T)$ and the subscript res denotes quantities evaluated at the time of the resonance $t_{\rm res}$.
The region of the ALP parameter space where the resonant enhancement can in principle affect the primordial population is plotted in figure \ref{fig:Bevaporation}, as a function of $m_\phi$ and the product $g_\phi B$, where $B$ is the today value of the primordial magnetic field.
The two different behaviours in the curve are due to the resonance happening during the two different regimes of $m_\gamma$, which depends on the plasma electrons being relativistic or not.

If we consider primordial magnetic fields originating at the electro-weak phase transition, then in order to obtain effects on the ALP population in a region not constrained by HB stars we would need for the field an average intensity of $B\sim3\ \mu$G at present time. Cosmological observables presently exclude effective values of the intergalactic magnetic field larger than few~nG~\cite{Kahniashvili:2010wm,Paoletti:2010rx}, which leaves not too much space for the resonant magnetic field evaporation, at least in the ALP unconstrained region.

\section{Pseudoscalar decay}\label{sec:decay}

The freezing in of the decay makes the pseudoscalars regain thermal contact with the bath.
The decay temperature $T_{\rm d}$ can be obtained from $\tau H=1$, so we need to know the energy density of the universe in order to calculate $H$.
If pseudoscalars do not dominate the universe energy budget when they decay, $T_{\rm d}$ is
\be
\label{decayT}
T_{\rm d} \simeq 
\frac{0.6}{g^{1/4}_*(T_{\rm d})}\(\frac{g_\phi}{10^{-7}\, {\rm GeV}^{-1}}\)
\(\frac{m_\phi}{\rm  MeV}\)^{3/2} {\rm MeV}\; . 
\ee
If instead the pseudoscalar energy density does dominate, which happens when $\phi$ is very non-relativistic and far out of equilibrium, the decay temperature is instead
\be
\label{decayT2}
T_{\rm d} \simeq 
\frac{0.7}{\[g_*(T_{\rm d})/g_*(T_{\rm fo})\]^{1/3}}\(\frac{g_\phi}{10^{-7}\, {\rm GeV}^{-1}}\)^{4/3} \;
\(\frac{m_\phi}{\rm  MeV}\)^{5/3} {\rm MeV}\; . 
\ee 
This is typically larger than the previous case, since a matter dominated universe expands more slowly than a radiation dominated one. 
Since the universe becomes radiation dominated after the decay, the temperature in equation \eqref{decayT} gives the correct order of magnitude for the reheating temperature.

If the decay temperature $T_{\rm d}$ is smaller than $3m_\phi$, the pseudoscalar reaches thermal contact with the bath being non-relativistic.
From equation \eqref{eq:nnonrel} and \eqref{ALPyield}, we know that $n_{\rm eq}(T_{\rm d})\ll n_{\phi}(T_{\rm d})$, and thus the Boltzmann equation \eqref{eq:boltzmann} becomes
\be
\dot{n}+3 H n=- \frac{n}{\tau}\; .
\ee
The solution is easily found to be $n(t)=n(t_{\rm d})\(R_{\rm d}/R\)^3 \exp\({-(t-t_{\rm d})/\tau}\)$, where $t_{\rm d}$ and $R_{\rm d}$ are respectively the time and the scale factor when $T=T_{\rm d}$.
The number density rapidly decreases because of the decay, and two photons per decaying pseudoscalar are created.

On the other side, if $T_{\rm d}\gtrsim 3m_\phi$, $n_{\rm eq}(T_{\rm d})$ can not be neglected, and the ALP can regain the thermal abundance --- in this case it would be given by equation \eqref{eq:nrelbos}, and so higher than $n_\phi(T_{\rm d})$ --- thanks to the inverse decay process, $\gamma\gamma\rightarrow \phi$.
The rate for this process is 
\begin{equation}
\Gamma_{\gamma\gamma\to \phi} \simeq
\frac{1}{\tau}\frac{m_\phi^2-4 m_\gamma^2}{m_\phi^2}
\left\langle\frac{m_\phi}{\omega}\right\rangle\,,
\end{equation}
where $\left\langle {m_\phi/\omega}\right\rangle$ is the thermally averaged time dilatation factor, being $\omega\geq m_\phi$ the energy of the outcoming pseudoscalar. 
Decay or inverse decay is only kinematically allowed if $m_\phi<2m_{\gamma}$.
For $T\gg m_e$, we have $m_\gamma\sim T$ while for $T\ll m_e$ we have $m_\gamma \ll T$, thus the decay/inverse decay channels open up not far from $m_\phi\sim {\rm max}\{T,m_e\}$.
In this case, some photons are subtracted from the thermal bath to make the pseudoscalars regain the equilibrium distribution. 
The pseudoscalar population then follows the equilibrium distribution, and becomes Boltzmann suppressed, basically disappearing from the bath, when the temperature drops below $m_\phi$. 
This \emph{equilibrium decay} of relativistic pseudoscalars, together with the previous case of \emph{non-equilibrium decay} of non-relativistic particles, should be well kept in mind as they will be a key point to understand the topics of chapter~\ref{chap:dilution}.

\begin{figure}[tp] 
   \centering
   \includegraphics[width=90mm]{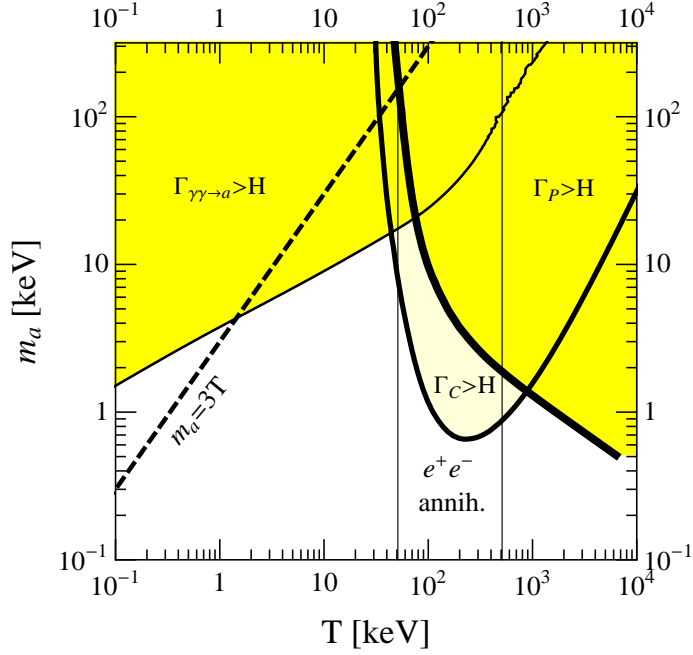}   
\caption{Axion decoupling and recoupling ($C_\gamma=1.9$, in the light yellow region it is also $C_e=1/6$).
   Thick solid line:~freeze-out of Primakoff process. 
   Medium solid line:~coupling and freeze-out of the Compton process.
   Thin solid line:~recoupling of inverse decay. In the yellow shaded region, axions are in thermal equilibrium. The light yellow region is relevant only if the Compton process is effective. 
The dashed line denotes $m_a=3T$, on the left of it axions are non-relativistic. 
    The vertical lines delimit the $e^+e^-$ annihilation epoch.}
   \label{fig:axionTd}
\end{figure}
\begin{figure}[tbp]
   \centering
   \includegraphics[width=90mm]{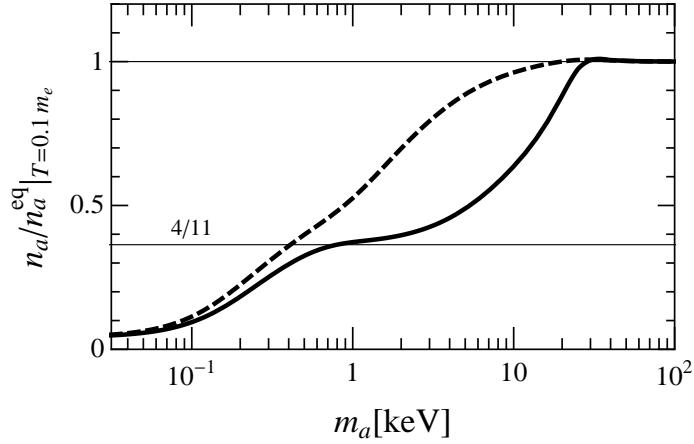}
   \caption{Axion number density $n_a$ after
   $e^+e^-$ annihilation from
   numerically solving the Boltzmann equation
   until $T=m_e/10$. The equilibrium density
   $n_a^{\rm eq}$ is defined in terms of the photon temperature.
   Solid line: only Primakoff process ($C_\gamma=1.9$). Dashed line: Primakoff and Compton ($C_\gamma=1.9$, $C_e=1/6$).}
   \label{fig:axionabu}
\end{figure}

Focusing on the axion case, a direct coupling with the electron can be relevant or not according to the model.
Since $\Gamma_{\rm P}$, $\tau$ and eventually the Compton interaction rate $\Gamma_{\rm C}\sim\alpha C_{e}^2 n_e/{f_a}^2$ are unambiguously determined as a function of $f_a$ or $m_a$ once a model is chosen, it is possible to plot the freezing out and freezing in temperatures, like in figure~\ref{fig:axionTd}, where the parameters chosen are $C_\gamma=1.9$ and $C_e=1/6$. 
The yellow area represents the range in which the axion interactions are efficient, while below the dashed line the axion is relativistic.
Therefore, the inverse decay is active in the yellow area delimited by the dashed line and the thin solid one, along which $\Gamma_{\gamma\gamma\rightarrow a}=H$.
From this figure it is clear that the lower is $f_a$, the higher $m_a$, and thus the later the axion decouples, in particular above $10$--$20$ keV this never happens.
In figure~\ref{fig:axionabu} we plot the ratio between the axion number density and its equilibrium value, both at $T= m_e/10$. 
It is obtained solving numerically the set of Boltzmann equations for axions, photons and electrons.
Including the Compton interaction, the resulting yield is higher, for the axion, having an additional interaction channel, decouples later so it its abundance is less diluted by the factor \eqref{eq:dilution}.
The range mostly affected by the further coupling is $m_a=0.5$--10~eV.
We can see in figure~\ref{fig:axionTd} that is where lies the Compton light yellow region, and from this plot we can better appreciate the decrease in the decoupling temperature.
We will deal again with figures~\ref{fig:axionTd} and~\ref{fig:axionabu} in chapter~\ref{chap:dilution}, where the Boltzmann equations will be discussed.

We have shown in this chapter how a population of pseudoscalars can arise and then disappear in the early universe.
Now it is time to develop its consequences.

\chapter{Signals from the sky: relic decay photons}
\chaptermark{Signals from the sky}
\label{chap:photons}
Each pseudoscalar decay produces a couple of photons, as we have seen in the previous chapters.
The fate of these photons depends mainly at which stage of the universe evolution the decay takes place.

The many charged particles in the primordial plasma efficiently scatter the photons propagating among them.
The universe is therefore optically opaque until it becomes cold enough to permit the electrons which survived the $e^{\pm}$ annihilation to combine with nuclei to form neutral atoms.
This event, known as \emph{recombination}, happens rather late in the history of the universe. 
The temperature has to cool down to a value around a couple of orders of magnitude below the binding energy of hydrogen and helium atoms.
This delay trend is very peculiar of combination events in cosmology, as it affects also primordial nucleosynthesis. 
The reason is the overwhelming number of photons compared with the number of baryons.
The \emph{baryon-to-photon ratio} is measured to be~\cite{Komatsu:2010fb}
\be\label{eq:eta}
\eta=\frac{n_b}{n_\gamma}=\(6.23\pm 0.17\)\times10^{-10}\; .
\ee
The number of photons whose energy is above the photo-dissociation threshold therefore can be easily larger than $n_b$. 
The combination event has to wait behind, until this high energy tail of the photon distribution empties because of the cooling.

The recombination is a crucial event for our understanding of the history of the universe. 
Just after this epoch the \emph{cosmic microwave background} (CMB) is released. 
Most of the information we have about the early universe is extrapolated from what was imprinted on the CMB around the recombination time.

In this chapter we address the effects of the decay photons injected after the universe becomes transparent or immediately before.
In the first case they can freely propagate and in principle be detected, while in the other they should leave an imprint on the CMB. 
We will analyse also the effect of the pseudoscalar decay on the ionization history of the universe.

\section{Spectral distortions of the cosmic microwave background}

The photon distribution is maintained in thermal equilibrium because of Compton scattering, double Compton scattering and bremsstrahlung interactions with the residual electrons. 
This means that if a perturbing event happens, the photon distribution can evolve again to the equilibrium condition, but only if these interactions have enough time to operate.
Compton scattering $\gamma+e^- \leftrightarrow \gamma+e^-$ has the fastest rate among these processes, $\Gamma\propto \alpha^2$, but it can only lead to kinetic equilibrium.
Since it conserves the photon number, it can not erase a chemical potential in the distribution. 
Double Compton scattering $\gamma+e^- \leftrightarrow \gamma+\gamma+e^-$ and bremsstrahlung $e^- + p^+ \leftrightarrow e^- + p^+ + \gamma$ are comparatively slower, $\Gamma\propto \alpha^3$. 
They change the number of photons, and thus they permit to achieve the full thermal equilibrium condition, i.e.~a Planckian spectrum~\cite{Hu:1992dc}.

FIRAS measured the CMB spectrum in the range 2--21~cm$^{-1}$, and found it to be is a perfect black-body within very small experimental errors, only an ${\cal O}(10^{-5})$ deviation is allowed by data~\cite{Fixsen:1996nj}.
Thus, any perturbation of the photon equilibrium condition must have been smaller than the measurement uncertainties, or the perturbing event must occur long enough before the bremsstrahlung or double-Compton cease to be effective.

Because of the very small value of $\eta$, the bremsstrahlung process is subdominant with respect to double Compton until $e^+e^-$ annihilation.
The exact value of the freezing out temperature for each photon-electron interaction depends on the wavelength.
Double Compton scattering is effective at all wavelengths until the temperature drops below $T_{\rm DC}\sim750$ eV at $t_{\rm DC}$.
At temperatures lower than $T_{\rm DC}$ double Compton and bremsstrahlung can still produce or absorb photons, but only if their frequency is $\omega\ll T$. 
Compton scattering redistributes photons along the whole spectrum and decouples later, at $T_{\rm C}\sim25$ eV.
Therefore, if a photon injection is not completely reabsorbed before $t_{\rm DC}$, the resulting CMB spectrum follows a Bose-Einstein distribution at high frequency and a Planckian one in the frequency range where double Compton and bremsstrahlung are still active~\cite{Hu:1992dc}.
The residual degeneracy parameter is constrained by FIRAS to be $|\mu|<0.9\times 10^{-4}$~\cite{Fixsen:1996nj}.

After $t_{\rm DC}$, the photons follow a random walk path scattered by the electrons.
Since the average photon energy is very small compared with the electron mass, the electron recoil is negligible and photons simply bounce off in random directions.
Photons emitted at this stage distort the overall photon spectrum in a very peculiar way. 
Electrons rapidly thermalise with the non-thermal population of photons and their effective temperature increases. 
However, as we said, CMB photons cannot gain energy efficiently out of them. 
The CMB spectrum is however slightly influenced by the heated electrons.
They dissipate the exceeding thermal energy pushing few photons towards higher frequency and this imprints a typical pattern on the CMB spectrum. 
This phenomenon is described by the Kompaneets equation~\cite{Zeldovich:1969ff,Peebles:1993ppc}
\be\label{eq:kompannets}
\frac{\partial n_\gamma}{\partial y}=\frac{1}{x^2}\frac{\partial}{\partial x}\[x^4\(\frac{\partial n_\gamma}{\partial x}+n_\gamma+{n_\gamma}^2\)\]\; .
\ee
The variable $x=\omega/T_e$ is the ratio between the photon frequency $\omega$ and the electron temperature $T_e$.  
The parameter $y$ is given by
\be
dy=\frac{T_e(t)-T(t)}{m_e} \sigma_{\rm T}n_e(t) dt\; ,
\ee
where $n_e$ is the electron number density, $T$ the photon temperature and $\sigma_{\rm T}$ is the Thomson cross section.
A stationary solution of equation~\eqref{eq:kompannets} is 
\be
n(x)=\frac{1}{e^{x+\mu}-1}\; ,
\ee
because of the number conserving Compton interaction.
If $T_e\gg T$, the Kompaneets equation simplifies\cite{Zeldovich:1969ff},
\be
\frac{\partial n_\gamma}{\partial y}=\frac{1}{x^2}\frac{\partial}{\partial x}x^4\frac{\partial n_\gamma}{\partial x}\;
\ee
and if the perturbation to the equilibrium is not too big the approximated solution gives
\be
\frac{\delta n_\gamma}{n_\gamma}=y\frac{x\, e^{x}}{e^x-1}\(x\frac{e^x+1}{e^x-1}-4\)\; ,
\ee
whose limiting case are $-2y$ for $x\ll1$ and $yx^2$ for $x\gg1$~\cite{Zeldovich:1969ff,Peebles:1993ppc}.
Computing the exact spectrum requires to calculate the effect of the decay on $T_e$ and nume\-rically integrate equation~\eqref{eq:kompannets}. 
Like in the $\mu$ case, the observational constraint is very limiting, $|y|<1.5\times 10^{-5}$~\cite{Fixsen:1996nj}.
Of course, all of this depends on the rate of photon-electron interactions. 
At temperature around $T_{\rm dec}=0.26$ eV, most electrons and protons are combined into neutral hydrogen, the universe becomes effectively transparent and the CMB decouples.

The photon injection due to pseudoscalar decay is an example of a perturbation strictly constrained by FIRAS data.
The non-thermal contribution to the photon spectrum from decay is a peak centred at energy $\omega\sim m_\phi/2$. 
Assuming a thermal population, from equation~\eqref{nphionngamma} we have
\be
\frac{\rho_\phi}{\rho_\gamma}\gtrsim 0.007 \frac{m_\phi}{T}\; ,
\ee
where we have taken $g_{*S}=3.9$.
The energy injection is always greater than $10^{-5}\rho_\gamma$, since the ratio $m_\phi/T>1$, and thus it is potentially in radical conflict with observations.
To conventionally escape this limit only a small fraction of the energy injection should influence the photon distribution, thus the decay has to be just started when CMB is released or to occur long before recombination.  
However, we will see in the following that there are other cosmological data sets which exclude pseudoscalars able to clear the CMB distortion hurdle through late or early enough decay.
These further limits largely overlap with the CMB distortion bound.
Therefore, here we consider sufficient only a rough estimate of the pseudoscalar parameter space which is excluded by CMB spectral distortions. 
A cosmological scenario with $g_*(T_{\rm fo})\gtrsim {\cal O}(10^{3})$ would permit to evade unconventionally this CMB constraint.
Also the dilution of the pseudoscalar density due to some consistent entropy injection during the epoch in which $\phi$ is decoupled from the thermal bath can in principle relax this bound. 
However, this has to happen much before primordial nucleosynthesis, otherwise it can be constrained by the same arguments which will be applied to pseudoscalars in the next chapter.

In figure~\ref{fig:photons} --- which gives a summary of this chapter --- the light green regions labelled {\bf CMB} $\boldsymbol{\mu}$ and {\bf CMB} $\mathbf{y}$ corresponds to $T_{\rm DC}>T_{\rm d}>T_{\rm C}$ and $T_{\rm C}>T_{\rm d}>T_{\rm dec}$ respectively.
The influence of later decays requires a different treatment and it is discussed in the following sections.
We always assumed that the amount of photons injected is always too large to be thermalised, since we are not considering here the unconventional scenario of pseudoscalar dilution just discussed.
Note that these bounds are somewhat conservative.
The decay is not an instantaneous event, so a significant amount of energy can be released after $T_{\rm d}$.
This is especially true for $m_\phi\gtrsim$ keV, for in this case the pseudoscalar energy density dominates the universe before the decay and the subsequent perturbation of the photon spectrum is huge.

\begin{figure}[tbp]
\begin{center}
   \includegraphics[width=63mm]{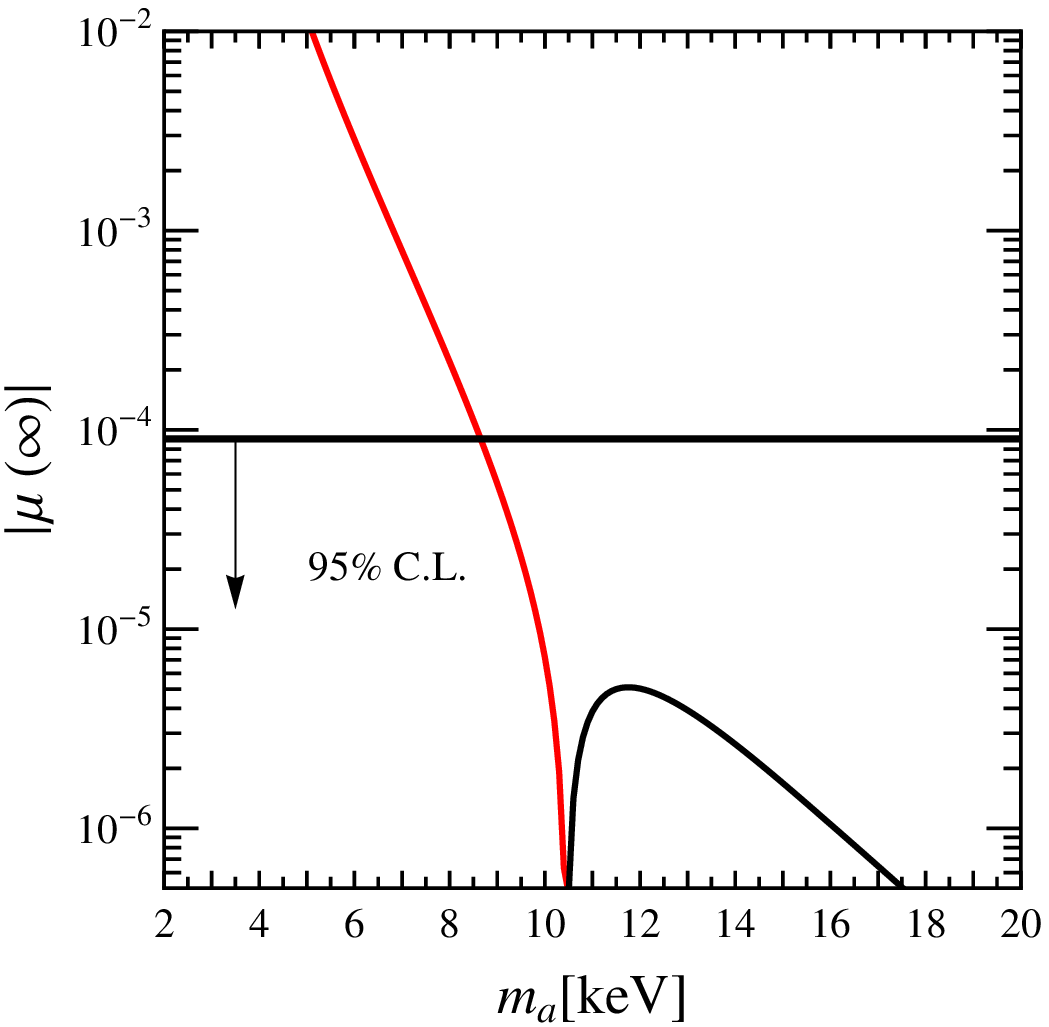}
   \hspace{3mm}
   \includegraphics[width=63mm]{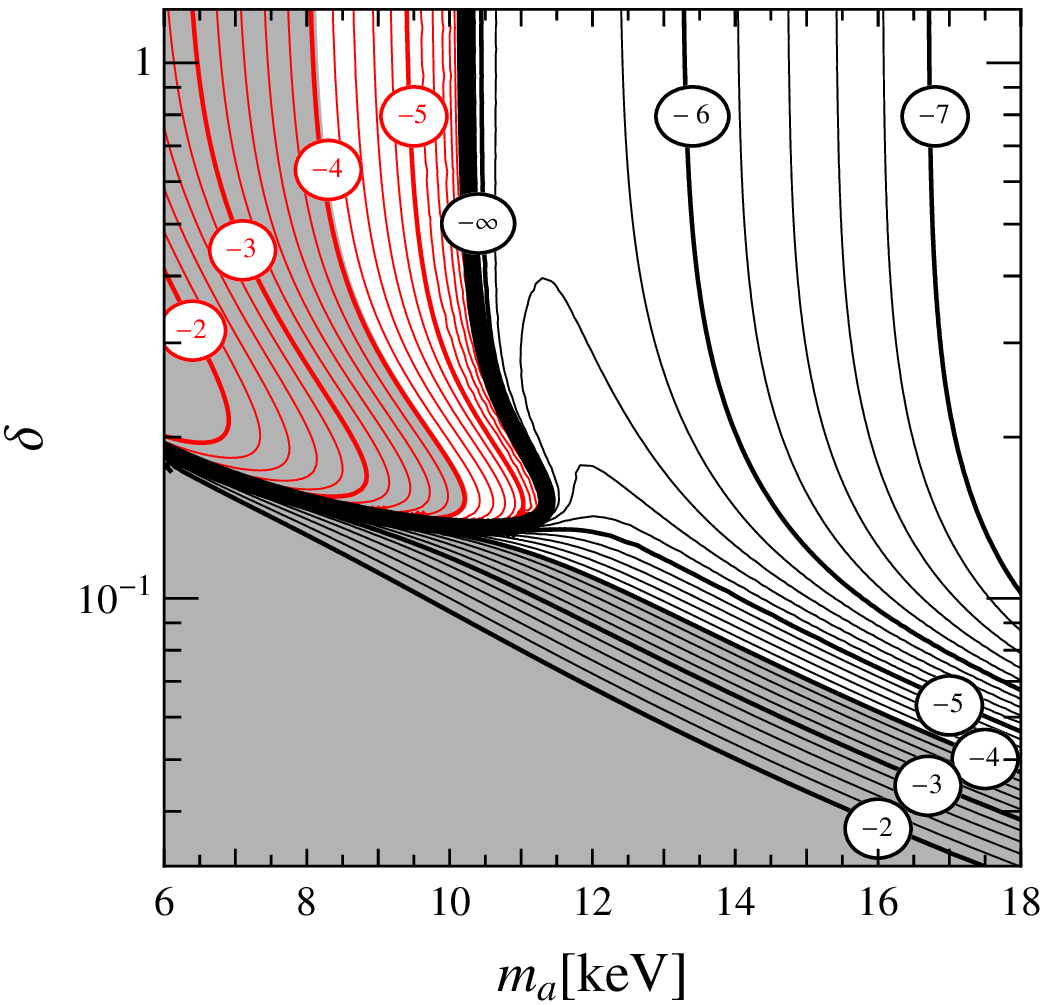}
\end{center}
   \caption{{\it Left:} Photon $\mu$ parameter after axion decay ($\delta=1$).
   The observational upper bound is indicated.
   {\it Right:}
   The contours of $\log_{10}|\mu|$ in the $\delta$--$m_a$ plane where
   black/red corresponds to positive/negative values of $\mu$.
   The thick black line corresponds to the boundary $\mu=0$.
   The shaded area is excluded.}
   \label{fig:axionmu}
\end{figure}

In the axion case, we have numerically solved the differential equation for the evolution of the degeneracy parameter $\mu$ of the photon distribution~\cite{Cadamuro:2010cz}.
Assuming $\mu$ small, its evolution is governed by the equation~\cite{Hu:1992dc},
\begin{equation}
\label{muevolution}
\frac{d \mu}{d t} = \frac{d \mu_a}{dt}-\mu\(\Gamma_{\rm DC}+\Gamma_{\rm B}\) \; ,
\end{equation}
where $\Gamma_{\rm DC}$ and $\Gamma_{\rm B}$ are the double Compton and bremsstrahlung rates for decreasing $\mu$, which are the inverse of the relaxation times for the two processes.
The rate-of-change due to the axion decay photon injection is given by 
\begin{equation} \label{muinject}
\frac{d\, \mu_a}{dt}=-\frac{2}{2.14}\(\frac{3}{\rho_\gamma} \frac{d\, \rho_a}{dt}-\frac{4}{n_\gamma}\frac{d n_a}{dt}\)\; ,
\end{equation}
where $d\,n_a/dt=-n_a/\tau$ and $d\,\rho_a/dt=m_a d\,n_a/dt$.
Our results are shown in fi\-gure~\ref{fig:axionmu}, where we plot the final $\mu$ value as a function of $m_a$ and $\delta\equiv C_\gamma/1.9$ for hadronic axions. 
In the left panel we fixed $C_\gamma=1.9$, the value in the simplest KSVZ model, and in this case we found
\begin{equation}
m_a > 8.7~{\rm keV }~~\hbox{at~~95\%~C.L.}\; .
\end{equation}
This is a robust bound, since $\mu$ is a steep function of $m_a$.
The CMB distortion effect depends sensitively on the axion-photon interaction strength for $\delta<1$ (right panel of figure~\ref{fig:axionmu}). 
Generally the spectral distortions get larger for smaller $C_\gamma$ at a given $m_a$, because if the decay happens later the photon distribution is less protected against distortions.
For large $C_\gamma$, the final $\mu$ changes sign from negative to positive with increasing $m_a$. 
For $\delta<0.1$, $\mu$ is always positive since axions decay non-relativistically, thus the energy injection is more important than photon number, see equation~\eqref{muinject}.
Of course, because of the sign change in $\mu$ some fine-tuned cases exist where the final $\mu$ can be accidentally zero.

\section{Ionization history of the universe}

CMB is released at around $T_{\rm dec}$, whose corresponding redshift is $z_{\rm dec}\sim1100$, when most of the recombination has already taken place.
Before recombination, the number of ultraviolet photons was too high to have long-lived atomic bonds, which were broken just after their formation.

Around $z_{\rm rec}\sim1500$, when the universe temperature was $T\sim 0.35$~eV, helium and hydrogen nuclei start to capture free electrons to form neutral atoms.
Helium recombines before hydrogen because of its higher atomic number and thus larger Coulomb potential.
The universe is still opaque after helium recombination, since He constitutes just around one fourth of the baryonic density, and the CMB has to wait the hydrogen recombination to break free~\cite{Mukhanov:2005sc}.

To exactly compute the ionization history of the universe it is necessary to solve a set of coupled Boltzmann equations for photons, free electrons and atoms in their different energy levels.
The result of this calculation is usually given in terms of the fraction of free electrons, or \emph{ionization fraction} $x_{\rm ion}$, as a function of redshift. 
The recombination of the hydrogen freezes out at $z\sim800$ because of the expansion of the universe, leaving a residual ionization fraction of order $x_{\rm ion}\sim\mathcal{O}(10^{-4})$.
The universe fully ionizes again much later, between redshifts $6$ and $10$, presumably due to ultraviolet emission from the first galaxies. 
The details of the reionization process are still not well understood~\cite{Choudhury:2006nr}. 

The slight imprint that the free electrons leave in the CMB through Thomson scattering, for instance in the polarization, gives us information about the history of recombination and reionization.  
The optical depth for CMB photons is one of the parameters that can be measured from the CMB multipole analysis and it is defined to be
\be
\tau_{\rm opt}(z_1,z_2)=-\int_{z_1}^{z_2}\frac{\sigma_T n_e(z) x_{\rm ion}(z)}{H(z)(1+z)}dz \; ,
\ee
where all the quantities in the integral are expressed as a function of redshift.

The WMAP7 measured $\tau_{\rm opt}$ after decoupling to be $0.088\pm0.015$~\cite{Komatsu:2010fb}. 
A factor $0.04$--$0.05$ of this can be attributed to a fully ionized universe up to about redshift 6, which is supported by the absence of Ly-$\alpha$ features in quasar spectra. 
The origin of the remaining fraction, $\tau_6=\tau_{\rm opt}(6,z_{\rm dec})\simeq 0.04$, is still uncertain and leaves some space for ALP decay.

The photons produced by ALP decay after decoupling are free to propagate since there are almost no free electrons to interact with. 
However, ultraviolet ra\-diation can interact with atoms and photoionize them.
The universe is indeed very opaque to ultraviolet radiation. 
This prevents us from detecting decay photons in the $13.6$--$300$ eV range, but it does not mean that we can not constrain them, since the photoionisation triggers an increasing of the ionization fraction of the universe. 
This argument does not hold for axions because in this mass range they decay before recombination.
If we assume that each decay photon ionizes only one H atom immediately after the ALP decay, which is a first rough approximation, the number of ionizations per unit time can be estimated to be~\cite{Natarajan:2010dc}
\be
\xi(z)\sim \frac{2}{ \tau} \frac{ n_{\phi}(z)}{n_{H}(z)}
\sim 2 \times 10^{-3}\(\frac{m_\phi}{100\ {\rm eV}}\)^3\(\frac{g_\phi}{10^{-13}\ {\rm GeV}^{-1}}\)^2e^{-\frac{2}{3}\frac{1}{H(z)\tau}}\ {\rm Myr}^{-1}\; .
\ee
If we now multiply this quantity by a typical time scale~\cite{Natarajan:2010dc}
\be 
t_H=1/H(z)\sim2.4\ {\rm Myr} \[501/(1+z)\]^{3/2}\; , 
\ee 
where we have neglected $\Omega_r$ and $\Omega_\Lambda$ in $H(z)$, we get a conservative estimate of the degree of ionization induced by the ALP decays up to a certain redshift. 
ALPs with 100 eV mass and $g_\phi\sim10^{-13}$~GeV$^{-1}$ would have produced an ionization comparable with the standard residual value $10^{-4}$ already at high redshifts, $z\sim 500$.
Since the ALP-induced ionization grows in time as $(1+z)^{-3/2}$, it shows a potentially interesting effect. 
We could even think that ALPs close to these parameters may be responsible for the full reionisation of the universe. 
However, the $(1+z)^{-3/2}$ dependence is too soft --- reionization seems to be a much more abrupt process, usually parametrised to be almost a step function in $x_{\rm ion}(z)$ --- and we can only check if ALPs provide a $\tau_6$ compatible with observations. 

\begin{figure}[tbp]
   \includegraphics[width=7cm]{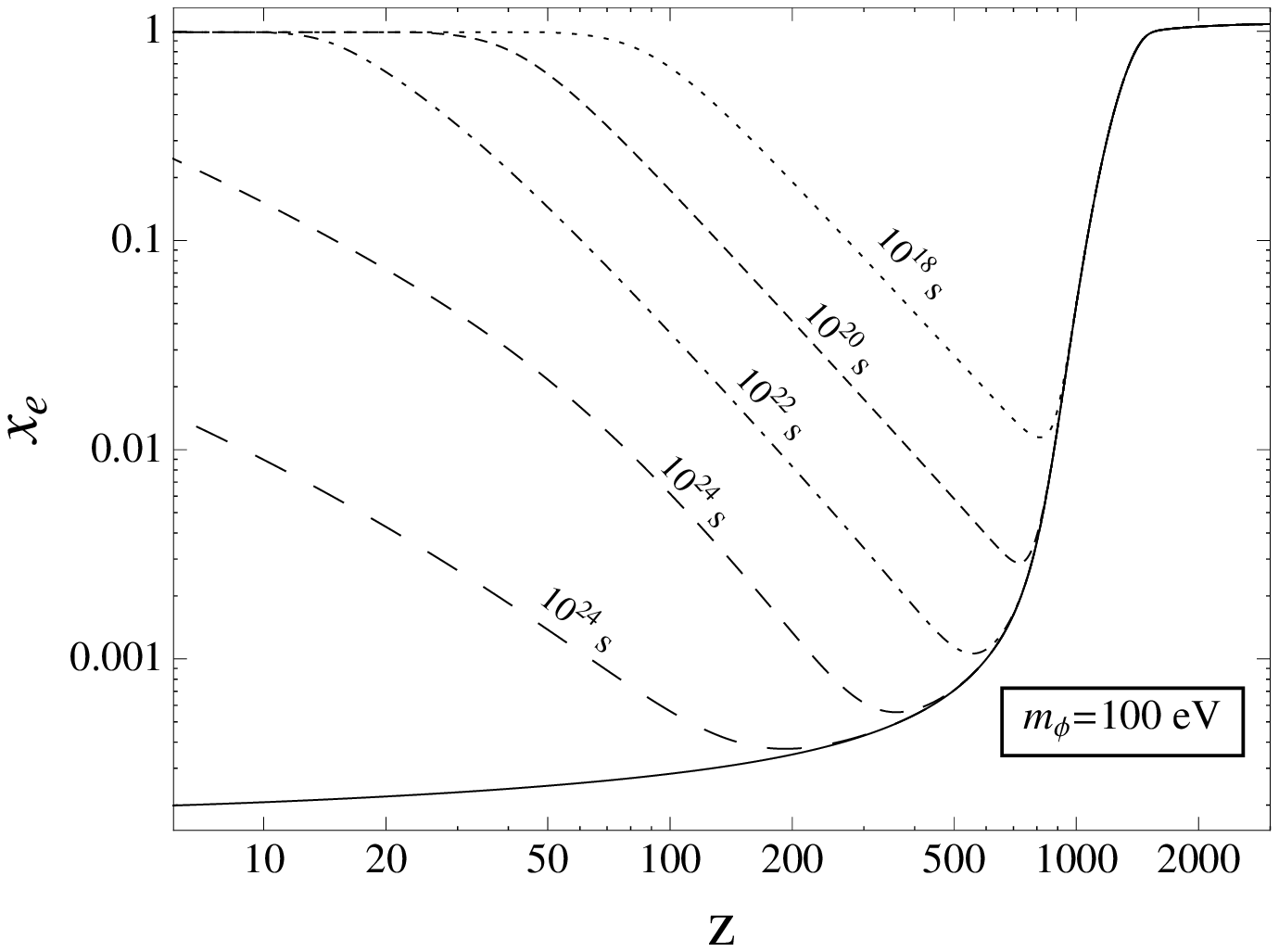}
   \includegraphics[width=7cm]{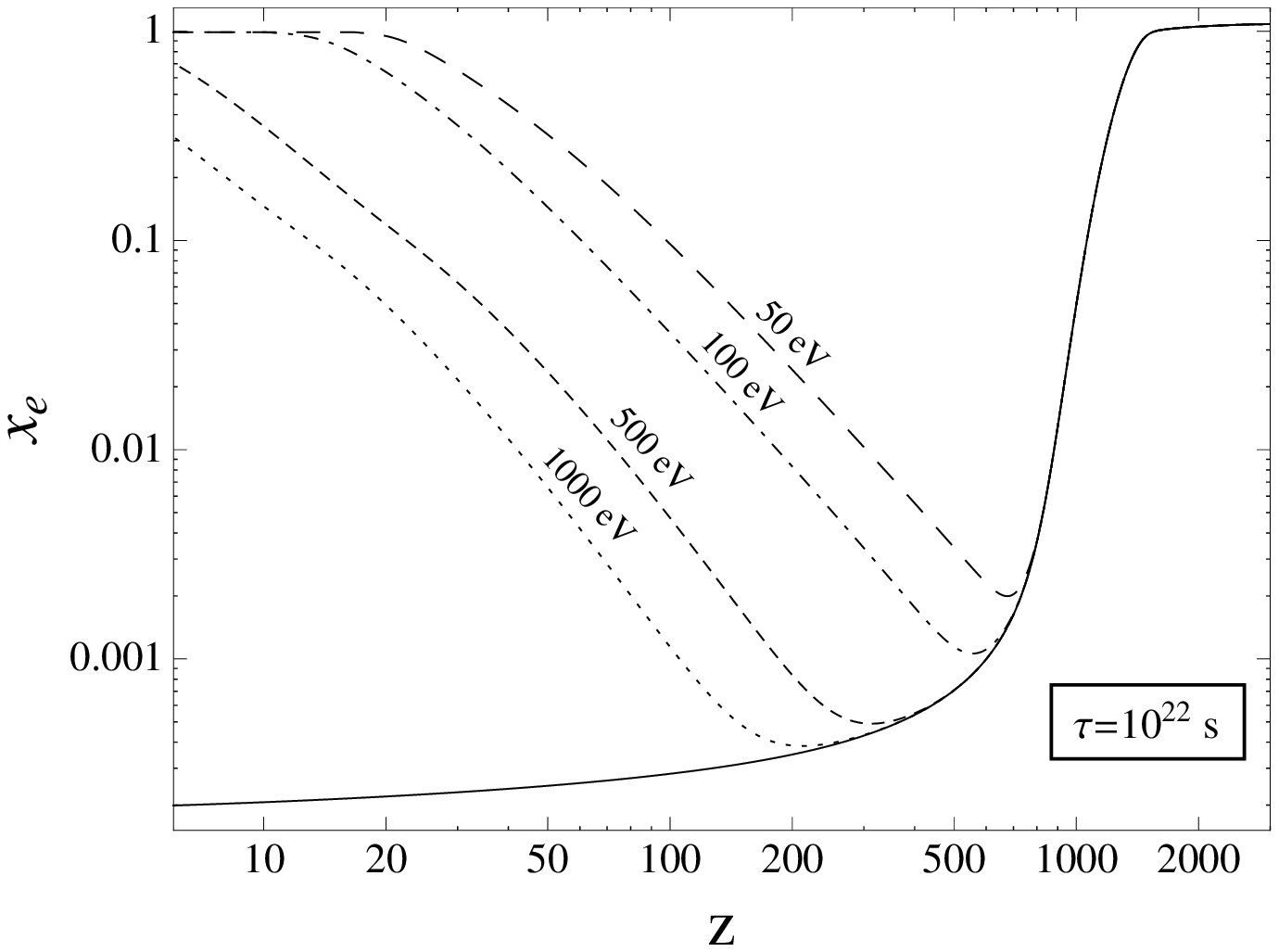}
   \caption{The effect of ALP decay on the ionization fraction, plotted in function of the reshift $z$. The standard case with no decay is plotted with the solid line. {\it Left:} decay of a $100$ eV mass ALP with $\tau=10^{18}$ s, $10^{20}$ s, $10^{22}$ s, $10^{24}$ s and $10^{26}$ s.  
   {\it Right:} decay of a $\tau=10^{22}$ s ALP with $m_\phi=1000$ eV, $500$ eV, $100$ eV and $50$ eV.}
   \label{fig:ionization}
\end{figure}

In order to obtain a more detailed constraint, we have calculated the ionization history of the universe in the decaying ALP cosmology~\cite{Cadamuro:2011fd,Arias:2012mb} by introducing the ALP ionizations in the recombination code RECFAST~\cite{Seager:1999bc}.
In figure~\ref{fig:ionization}, we plot the ionization history of the universe in function of $z$ for several examples of ALP mass and lifetime.
We set the ALP fraction to to account for all the DM, $\Omega_{\rm DM}h^2=0.11$, and made the conservative assumption that each photon emitted during the decay can ionize only one atom.
On the left panel the mass is fixed, $m_\phi=100$ eV, and the lifetime varies. 
Of course, for longer lifetimes the ALP ionising effect appears later.
On the right panel the lifetime is instead fixed to $10^{22}$ s, and the mass varies from 50 eV to 1 keV.
The higher the mass, and consequently the energy $\omega$ of the emitted photons, the less efficient is the ionising effect.
The one-electron atom photoionisation cross section is suppressed for very high energy photons~\cite{bransden2003physics},
\be \label{eq:ph-ion}
\sigma_{\rm ph\mbox{-}ion}\sim\frac{256\pi}{3}\frac{\alpha_{\rm em}}{Z^2}\(\frac{E_{1s}(Z)}{\omega}\)^{7/2}a_0^2\; ,
\ee
where $Z$ is the atomic number, $E_{1s}=13.6 Z^2$ eV the energy of the $1s$ state, $a_0=(\alpha m_e)^{-1}=5.292\times 10^{-9}$ cm is the Bohr radius and $\omega$ the photon energy.

To scan the ALP parameter space, we computed the optical depth in the interval $z=6$--$100$, requiring it to not exceed $\tau_6$. 
We made two different calculations, assuming the ALP thermal abundance in~\cite{Cadamuro:2011fd}, and secondly that ALPs constitute the whole DM in~\cite{Arias:2012mb}. 
Our results are excluding the light green region labelled $x_{\rm ion}$ in figure \ref{fig:photons}, where the thermal origin of ALPs is considered.
A similar result was obtained in~\cite{Arias:2012mb}. 
This bound would increase up to one order of magnitude at the largest masses for which ionization is effective, $m_\phi\lesssim 300$ eV, if we assume optimistically that all the energy of the emitted photons can be converted into ionization.
The ionization history constrains ALP lifetimes much longer than the age of the universe, $\tau\gtrsim10^{24}$ s, which means that  only less than one ALP out of ten millions can decay.
The effect of the decay of a large population of particles has catastrophic effects, but only extremely small perturbations to the standard cosmological scenario are allowed.

\section{Direct detection of relic decay photons}

\begin{figure}[tbp]
\center
   \includegraphics[width=11cm]{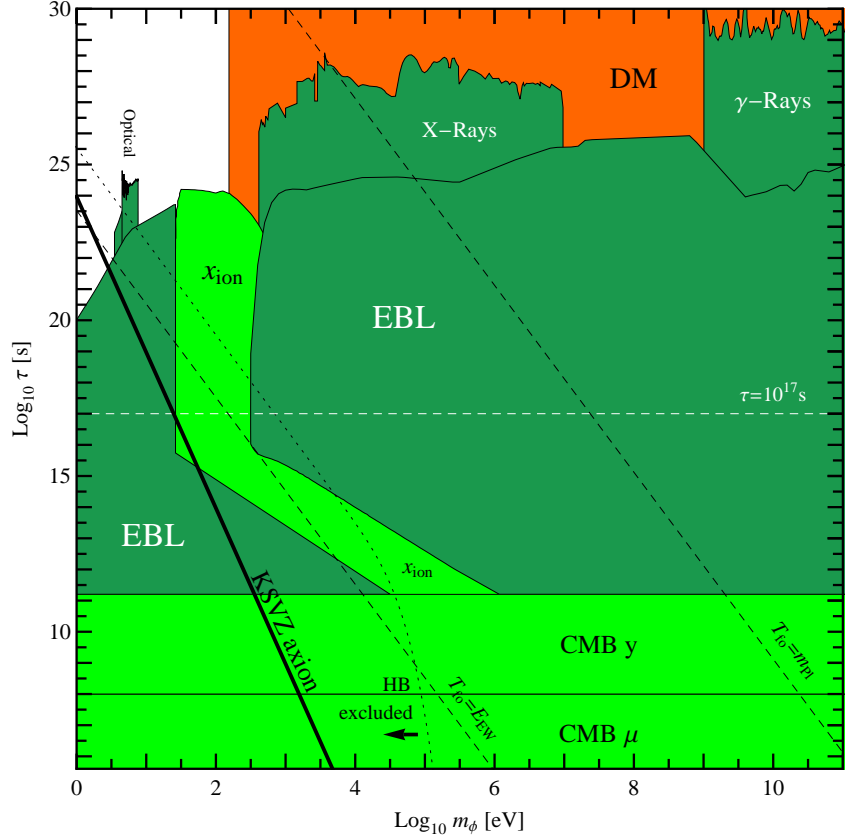}
   \caption{Summarising picture for this chapter in the $m_\phi$-$\tau$ parameter space. All the bounds are described in the text. In light green are plotted the bounds related to CMB distortion, respectively labelled {\bfseries CMB y}, {\bfseries CMB $\boldsymbol{\mu}$} and $\boldsymbol{x_{\rm ion}}$. In dull green are the limits from the lack of direct observation of decay photons. They are {\bfseries EBL} and { \bfseries Optical}, {\bfseries X-Ray} and {\bfseries $\boldsymbol{\gamma}$-Ray}. We considered the ALP abundance provided by the thermal production mechanism for all the bounds in all the parameter space but direct decay photon detection in the $m_\phi>154$ eV range. Here, ALPs would be overproduced with respect to the measured DM abundance, therefore $\Omega_{\rm DM}$ is taken for ALP abundance in the direct photon detection limits. The orange region labelled {\bfseries DM} covers the exclusion bound provided by ALP overproduction.}
   \label{fig:photons}
\end{figure}

We have seen in the previous section that ultraviolet light is absorbed by atoms.
If the photons from ALP decay do not lie in the UV region described above, the universe is transparent to them and they can in principle be detected.
The decay photons should be manifest in spectral observations as a peak at energy $\omega=m_\phi/2$, broadened by redshift.
In the parameter space we can constrain, the decay happens at rest in the comoving frame. 
The spectral flux of photons produced in the decay of a diffuse pseudoscalar population is~\cite{Masso:1997ru,Masso:1999wj}
\begin{subequations}\label{eq:flux}
\begin{align}
\frac{dF_E}{dEd\Omega}&=\frac{1}{2\pi}\frac{\Gamma}{H(z)}\frac{n_{\phi}(z)}{(1+z)^3}=\\
&\simeq \frac{\bar{n}_{\phi 0}}{2\pi\tau H_0}\(\frac{E_0}{m_\phi/2}\)^{3/2}\exp\[-\frac{t_0}{\tau}\(\frac{E_0}{m_\phi/2}\)^{3/2}\]\; ,\label{eq:fluxMD}
\end{align}
\end{subequations}
where the subscript 0 means quantities at present time, $\bar{n}_{\phi 0}$ is the putative number density if $\phi$ would be stable, and $E_0$ is the energy at which the photon would be seen today. 
The photon initial energy is $m_\phi/2$ and thus the redshift of the decay is given by $1+z=(m_\phi/2)/E_0$.
For simplicity we assumed matter domination neglecting $\mathcal{O}(1)$ corrections due to the cosmological constant. 
To take into account the opacity of the universe to the ultraviolet radiation, the flux has to be corrected multiplying it by the survival probability
\begin{align}
P(z)&=e^{-\kappa(z,E)} \label{eq:absfactor}\\
\kappa(z,E)&=\int_0^z \frac{dz^\prime}{H(z^\prime)(1+z^\prime)}n_H(z^\prime)\sigma_{\rm ph\mbox{-}ion}(E)\; ,
\end{align}
where $E=E_0(1+z)$, $n_H$ is the hydrogen number density and $\sigma_{\rm ph\mbox{-}ion}$ is the hydrogen photoelectric cross-section given by equation \eqref{eq:ph-ion}, with $E_{1s}=13.6\,{\rm eV}$ and $Z=1$.
The contribution from helium photoionization is not relevant.
Once we corrected the spectrum  by this factor, we compared it with the extragalactic background light (EBL) spectrum reviewed in~\cite{Overduin:2004sz}.
We corrected also the microwave range of the spectrum subtracting the CMB contribution, which is known extremely well after the measurement of COBE-FIRAS. 
The EBL spectrum does not present sharp features that eventually could be recognised to be the contribution of a decaying particle.
Therefore we require the decay photon spectrum to be lower than the EBL.
The region of parameter space excluded by this comparison is plotted in dull green in figure \ref{fig:photons}, where it is labelled {\bf EBL}. 
The same approach was already used in references \cite{Overduin:2004sz,Ressell:1991zv} to look for axions in the optical EBL. 
In the part of parameter space that can be constrained in this way, the ALP abundance would be the minimum considered $n_\phi/n_\gamma=(3.9/106.75)/2\simeq 0.019$.
Since the thermal ALPs would provide too much DM for $m_\phi>154$ eV, in our analysis we found more convenient \emph{to assume that ALPs provide the right amount of DM above this mass}, while below it the standard thermal abundance is considered. 
In this way, we avoid to duplicate the limits in the region which is already excluded, and provide a bound that only implies ALPs to constitute the DM. 
Of course this requires some non-standard dilution of the ALP number density by additional degrees-of-freedom above the electroweak scale.
The bound gets severely degraded in the $m_\phi=13.6$--$300$ eV range, where not only absorption is very strong but also the 
experimental data are extremely challenging and the EBL spectrum has only an upper bound estimate.
Of course these facts are closely related.

In order to gain sensitivity, a more effective strategy to search for decay photons is to examine the light emitted in galaxies and large scale structures. 
There, the dark matter density is above the average, providing therefore an enhanced signal for the decay products. 
However, also the backgrounds are above the EBL, but anyway the decay photons should stand above the background appearing like a peak. 
Searches of axions in the visible have been presented in~\cite{Ressell:1991zv,Bershady:1990sw} and more recently in~\cite{Grin:2006aw}.
The limits from these references are plotted in figure \ref{fig:alimits} in the blue band labelled \textbf{Telescope}.
In~\cite{Boyarsky:2006fg,Boyarsky:2009ix} sterile neutrino dark matter decays in a photon plus an active neutrino were searched in X-rays with the same methods. 

We followed these references, rescaling the results for the ALP decay mode when exploiting neutrino searches, and abundance when the thermal ALP yield gives $\Omega_\phi<\Omega_{\rm DM}$.
The exclusion bounds we obtained are plotted in figure \ref{fig:photons}, labelled respectively Optical and X-Rays.
We have also plotted in figure~\ref{fig:photons} the limit obtained by galactic line searches in the $\gamma$-ray range for decaying DM, labelling it $\gamma$-rays. 
We took the data from~\cite{Vertongen:2011mu} and again we rescaled them for the ALP decay mode.
Anyway, considering that the region constrained by $\gamma$-ray observations is well above the $T_{\rm fo}>m_{\rm Pl}$ line, it makes probably not too much sense in our thermal-ALP picture, but we showed it anyway for the sake of completeness.
In this $\gamma$-ray region, a very suggestive yet tentative claim of a detection at $\omega\simeq 130$ GeV appeared recently~\cite{Weniger:2012tx}. 
Unfortunately, we can not relate it to decaying pseudoscalars, since it can only be attributed to annihilating DM.

In these spectral analyses, the monochromatic emission line caused by the decay --- or annihilation --- is reconstructed, and its intensity tracks the density profile of the galaxy or the cluster taken into account, which can be measured through lensing observations.
The high resolution of modern instruments permits to extract different spectra from different zones of an object. 
The emission line are thus searched privileging the higher density regions.

The radiation produced by ALP decay, being proportional to the relic ALP density, would be linearly sensitive to the particle content above the EW scale through $g_*(T_{\rm fo})$. 
The constraint on $g_\phi$ relaxes as $\sqrt{1+{\rm new\, dof}/106.75}$ with the new thermal degrees of freedom with masses between the EW and $T_{\rm fo}$ in the $m_{\phi}<154$ eV range.
The new degrees of freedom do not affect the direct decay photon detection bound for $m_{\phi}>154$ eV, since in this mass range we assumed $\Omega_\phi=\Omega_{DM}$.


\chapter{Diluting neutrinos and nucleons: \\limits from dark radiation and big-bang nucleosynthesis}\label{chap:dilution}
\chaptermark{Diluting neutrinos and nucleons}
If the pseudoscalar decay happens long before matter-radiation decoupling, the primordial plasma has enough time to regain thermodynamic equilibrium.
This does not mean that is not possible to observe any consequence of this event.
The decay modifies the photon abundance, and leaves unaltered the baryon and the neutrino abundances --- if neutrinos have already decoupled from the bath.
The baryon-to-photon ratio $\eta$ and the effective number of neutrinos $N_{\rm eff}\propto\rho_{\nu}/\rho_{\gamma}$ are effectively \emph{diluted}, and they can be measured through the analysis of CMB anisotropies and the observation of primordial elements abundance.

A fast estimate of the entropy increment due to the decay is a useful guideline to understand the topics of this chapter.
Since neutrinos are not affected by the decay if already decoupled, it proves convenient to normalise abundances in terms of the neutrino temperature which redshifts as $T_\nu\propto R^{-1}$. 
So we define
\begin{equation}\label{eq:AB}
\mathcal{T}\equiv\(\frac{T_\gamma}{T_\nu}\)^3 \propto \frac{n_\gamma}{n_\nu}
\quad\hbox{and}\quad
\h\equiv\(\frac{T_\phi}{T_\gamma}\)^3=\frac{n_\phi}{n_\phi^{\rm eq}}\; .
\end{equation}
The initial condition we use is $T_{\gamma,0}=T_{\nu,0}=T_{\phi,0}=2$~MeV, when neutrinos have decoupled and $e^+e^-$ annihilation has not yet begun. 
A first stage is defined by $T_{\gamma,1}\simeq m_e/10$, when electron-positron annihilation is over. 
Entropy conservation during $e^+e^-$ annihilation implies
\begin{equation}\label{B0}
\mathcal{T}_0\(\frac{7}{2}+2+\h_0\)=\mathcal{T}_1(2+\h_1)\,,
\end{equation}
where $7/2$ and 2 are the $e^+e^-$ and $\gamma$ entropy degrees of freedom. 
Because of the initial condition we have chosen, $\mathcal{T}_0$ and $\h_0$ are equal to 1.
Considering the axion case, from figure~\ref{fig:axionTd} we see that we are describing an epoch far below the dashed line, which means that they are relativistic in the mass range we are interested.
We assume kinetic equilibrium, even if axions decouple during the $e^+e^-$ annihilation.

If axions or ALPs decouple before $e^+e^-$ annihilation, like neutrinos, they are not affected by the entropy released in the annihilation.
Photons are heated by the standard amount, thus $\mathcal{T}_1=11/4$, while the product $\h \mathcal{T}$ is conserved.
On the other hand, if they decouple during or after $e^+e^-$ annihilation they are heated, sharing some of the electron entropy, and thus we have
\begin{equation}
\label{B1}
\mathcal{T}_1 = \frac{13}{2(\h_1+2)}\; .
\end{equation}
If pseudoscalars are fully coupled during $e^+e^-$ annihilation, then $\h_1=1$ and $\mathcal{T}_1=13/6$. 
In general we have $11/4<\mathcal{T}_1<13/6$.
Figure~\ref{fig:axionabu} plots the value of $\h$ for the axion at $T_1$.

The eventual decay of a population of pseudoscalars makes the entropy they carry to be transferred to the photon bath.
In the case that $\phi$ never leaves thermal equilibrium, we easily find that the final photon abundance is $13/11$ times the standard value.
In the axion case, we have pointed out that this happens for $m_a\agt20$~keV.
In general the entropy transfer to photons depends on two parameters, the initial pseudoscalar abundance, parametrised by $\h$, and the effectiveness of inverse decay when axions and ALPs become non-relativistic. 

If the decay freezes-in when $T_\gamma>m_\phi/3$, the pseudoscalar temperature catches up with the photon one which has to decrease because of energy conservation.
In this process, radiation is simply shuffled from one form to another and comoving energy remains conserved.
Using again neutrinos as a ruler, in our units the energy density is $\rho_{\gamma+\phi}/\rho_\nu\propto (2+{\h_1}^{4/3})\mathcal{T}_1^{4/3}$ before the recoupling.
After recoupling at $T_*$, the pseudoscalar temperature catch up with the photon one, thus we have $\rho_{\gamma+\phi}/\rho_\nu\propto (2+1)\mathcal{T}_*^{4/3}$, and, because of energy conservation, the photon abundance is reduced to 
\be
\mathcal{T}_*=\mathcal{T}_1\,\(\frac{2+{\h_1}^{4/3}}{2+1}\)^{3/4}\; .
\ee 
Comoving entropy increases by the factor $[(2+1)/(2+\h_1)]\mathcal{T}_*/\mathcal{T}_1$.
At the later temperature $T_2$, when pseudoscalars become nonrelativistic and their abundance gets Boltzmann suppressed, the decay transfers the pseudoscalar entropy to the photon bath, heating it according to 
\be
2\mathcal{T}_2=(2+1)\mathcal{T}_*\; .
\ee
Putting all together, the final photon heating by pseudoscalars recoupling and adiabatic decay is 
\begin{equation}\label{eq:equisumma}
\frac{S_{2}}{S_{1}}=\frac{\mathcal{T}_2}{\mathcal{T}_1}=\frac{3}{2}\(\frac{2+\h_1^{4/3}}{3}\)^{3/4}\; ,
\end{equation}
which is also the ratio of the entropy density after and before the decay.
We refer to this case as \emph{equilibrium decay}, since it happens in a local thermal equilibrium condition.

If pseudoscalars recouple non-relativistically and they dominate the energy budget of the universe just before recoupling, their \emph{out-of-equilibrium decay} has more dramatic consequences.
The out-of-equilibrium decay can produce a vast increase in the entropy density, because the energy which is first stored in the mass of a non-relativistic particle is suddenly released as radiation. 
This case is illustrated by an analytic approximation to the entropy generation~\cite{Kolb:1990vq}
\begin{equation}\label{eq:entropy}
\frac{S_{2}}{S_{1}}=\frac{\mathcal{T}_2}{\mathcal{T}_1}=1.83\,\langle g_{*S}^{1/3}\rangle^{3/4}
 m_\phi Y_{\phi}(T_1)\sqrt{\frac{\tau}{m_{\rm Pl}}}\;,
\end{equation}
where $\langle g_{*S}^{1/3}\rangle$ denotes an average over the decay time and $Y_\phi\equiv n_\phi/s$. 
This formula is valid if the decay produces relativistic particles, and the energy densities of all the species other than $\phi$ are negligible before the decay.
The energy density scales as $\rho \propto R^{-3}$ for non-relativistic species, while $\rho \propto R^{-4}$ for radiation. 
A pseudoscalar dominated universe requires non-relativistic particles which are long-lived enough to survive until the pseudoscalar-radiation equality. 
This condition is more easily met in the intermediate mass range for a given coupling constant: very heavy pseudoscalars decay early because of the $m_\phi^{-3}$ dependence of the lifetime, while very light particles becomes not-relativistic late.

\begin{figure}[tbp]
   \centering
   \includegraphics[width=7.3cm]{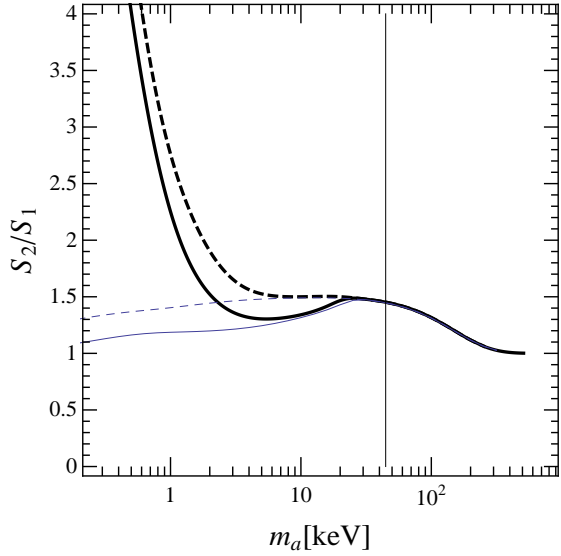}
   \caption{Photon density increase in our modified cosmology as
     expressed by $S_2/S_1$ for $C_\gamma=1.9$.
     Solid and dashed lines stand for hadronic and non-hadronic 
     axions respectively. In the latter case $C_e=1/6$. 
     The thin blue lines show the value if we assume that no
     entropy is generated in axion decay. 
     The vertical line signals the end of $e^+e^-$ annihilation epoch.}
   \label{fig:B1B2}
\end{figure}

In figure~\ref{fig:B1B2} we show using black lines the resulting $S_2/S_1$ from a numerical solution of the set of Boltzmann equations for the axion phase space distribution and the neutrino, photon and electron abundances as a function of $m_a$ in the case $C_\gamma=1.9$, with and without the electron coupling.
The blue thin lines represent just the increase in the entropy density due to reshuffling of entropy between axions and photons, which follows equation~\eqref{eq:equisumma}, ignoring the entropy generated in the out-of-equilibrium decay.
It is interesting to observe how dramatic is the effect of the axion domination on the comoving entropy density.
Starting from the very right, in the high mass range, we can observe how negligible is the effect of the decay, since in this case the axion population has already disappeared from the bath when $T=T_1$.
The value of $S_2/S_1$ increases moving to the left, as the equilibrium decay happens more and more after the electron-positron annihilation.
The maximum value allowed in this situation is provided by equation~\eqref{eq:equisumma} and is $3/2$.
Continuing to the left, especially if no direct coupling with the electron is present, axions experience a period of decoupling followed by recoupling due to inverse decay.
The value of $S_2/S_1$ decreases because $\h_1$ is increasingly less than one in equation~\eqref{eq:equisumma}.
Finally, the out-of-equilibrium range begins, and the black and blue lines diverge. 
Here, $\h_1$ tends to zero, and indeed the blue lines approach two. 
But the actual value of $S_2/S_1$ is affected by entropy generation, and it is highly boosted if the decay happens late enough for the axion to dominate the universe.   
Again, the additional coupling to electrons makes the transition from equilibrium to non-equilibrium decay shift towards lower mass.

\section{Effective number of neutrinos}

From the previous discussion, we have seen that the pseudoscalar decay can increase $\mathcal{T}$, i.e.~the abundance of photons relative to neutrinos.
However, the cosmological point of view is a bit different: since all cosmic parameters are defined relative to the observed CMB properties, this means that effectively the neutrino abundance is reduced. 
Precision observables measure the cosmic radiation density at decoupling, and this \emph{neutrino dilution} should make a difference.

To extract cosmological information on the radiation density at CMB decoupling, in~\cite{Cadamuro:2010cz} we analysed the usual 8-parameter standard $\Lambda$CDM model described in reference~\cite{Hannestad:2010yi}, extended in two ways. 
We allowed the effective number of neutrino degrees of freedom to vary, assuming a flat prior on the interval $0< N_{\rm eff}<3.0$. 
The radiation energy density is traditionally expressed as
\begin{equation}\label{eq:rhorad}
\rho_{\rm rad}=
\left[1+\frac{7}{8}\,\left(\frac{4}{11}\right)^{4/3}N_{\rm eff}\right]
\,\rho_\gamma\; ,
\end{equation}
where $N_{\rm eff}$ is the effective number of thermally excited neutrino degrees of freedom. 
The standard value is $N_{\rm eff}=3.046$ instead of 3 because of residual neutrino heating by $e^+e^-$ annihilation~\cite{Mangano:2005cc} but, given the experimental uncertainty, we neglected this tiny correction. 
In addition, we allowed for neutrino masses, assuming a common value $m_\nu$ for all flavours and a flat prior on $0<\Omega_\nu/\Omega_{\rm m}<1$. 
The other parameters and their priors are identical with those provided in reference~\cite{Hannestad:2010yi} and are listed in table~\ref{tab:hannestad}. 
They are the cold dark matter density $\Omega_{\rm cdm}$, the baryon density $\Omega_{\rm b}$, the present day normalised Hubble parameter $h$, the optical depth to reionization $\tau_{\rm opt}$, the amplitude of the primordial scalar power spectrum $A_{\rm s}$, and the scalar spectral index of the primordial fluctuation spectrum $n_s$.
Moreover, we use the same set of cosmological data, which are the WMAP 7-year CMB measurements, the 7th data release of the Sloan Digital Sky Survey, and the Hubble constant from Hubble Space Telescope observations.

\begin{table}[tb]
\centering
{\footnotesize
\begin{tabular}{lll}
\hline
Parameter&~~ &  Prior\\
\hline
$\Omega_{\rm cdm}h^2$  &~~ & $0.01$--$0.99$ \\
$\Omega_{\rm b}h^2$    &~~ & $0.005$--$0.1$ \\
$h$                    &~~ & $0.4$--$1.0$\\
$\tau_{\rm opt}$       &~~ & $0.01$--$0.8$ \\
$\ln(10^{10}A_{\rm s})$&~~ & $2.7$--$4.0$ \\
$n_{\rm s}$            &~~ & $0.5$--$1.5$ \\
\hline
\end{tabular}
}
\caption{Priors for the cosmological fit parameters~\cite{Hannestad:2010yi}. 
All priors are uniform in the given intervals.\label{tab:hannestad}} 
\end{table}

\begin{figure}[tb]
\hspace{25mm}
\includegraphics[width=10.cm,angle=0]{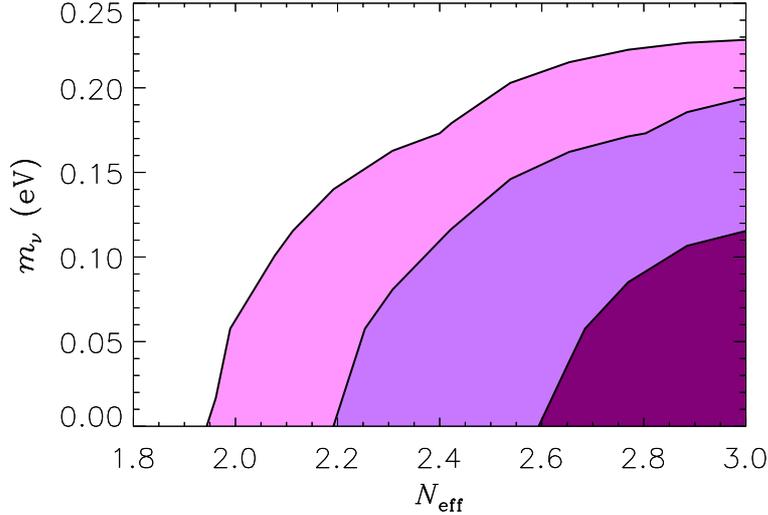}
\caption{2D marginal 68\%, 95\% and 99\% contours in the $m_\nu$--$N_{\rm eff}$
plane, where $m_\nu$ is the individual neutrino mass
(not the often-used sum over masses).
\label{fig:potato}}
\end{figure}

Marginalizing over all parameters but $m_\nu$ and $N_{\rm eff}$ we find the 2D credible regions shown in figure~\ref{fig:potato}.
Marginalizing in addition over $m_\nu$ we find the limits
\begin{equation}\label{eq:neffconstraint}
N_{\rm eff}>
\begin{cases}
{2.70}&\hbox{at 68\% C.L.}\\
{2.39}&\hbox{at 95\% C.L.}\\
{2.11}&\hbox{at 99\% C.L.}
\end{cases}
\end{equation}
These limits are very restrictive, because on present evidence cosmology actually prefers extra radiation beyond $N_{\rm eff}=3$~\cite{Hamann:2007pi,Hamann:2010pw,Komatsu:2010fb,GonzalezGarcia:2010un,Hamann:2010bk}.
The favourite value is $N_{\rm eff}=4.34^{+0.86}_{-0.88}$ considering only WMAP7 data, and $N_{\rm eff}=4.78^{+1.86}_{-1.79}$ if also SDSS observations are taken into account~\cite{Hamann:2010pw}.

The PLANCK satellite, currently taking CMB data, is expected to boost precision determinations of cosmological parameters. It will measure the cosmic radiation content at CMB decoupling with a precision of about $\Delta N_{\rm eff}=\pm0.26$ or better and thus will clearly decide if there is extra radiation in the universe~\cite{Hamann:2007sb}. 
If it finds convincing evidence for extra radiation, much in cosmology will have to be reconsidered besides our axion-ALP limits.

\subsection{ALP bounds}

At the beginning of this chapter we have described some limiting cases to understand the matters related to dilution and photon injection.
The key quantity was $\mathcal{T}$.
But considering the large ALP parameter space and all the intermediate cases it includes, it is unavoidable to numerically solve the problem.
The active species to consider are photons, electrons and neutrinos. 
Electrons acts as messengers between photons and neutrinos.
Photons thermalise very fast with electrons, which, through weak reactions, heat the neutrinos to keep track of the temperature changes in the electromagnetic bath. 
The most relevant of these energy redistribution processes is $e^+ e^-\to \bar \nu \nu$. 
We neglect scattering processes such as $e^\pm \nu \to e^\pm \nu$, since they cannot change the neutrino number and are less effective in transferring energy to the neutrino bath\footnote{
Assume that electrons have temperature larger than neutrinos like in our case. 
Then the speed of the energy transfer per unit volume in scattering processes is proportional to $ \langle \delta E \rangle T_e^4 T_\nu^4$, with $\langle \delta E \rangle\sim (T_e-T_\nu)$ a thermal-averaged energy transfer per scattering. 
For annihilations it is $\propto T_e(T_\nu^8-T_e^8)$, which is much more sensitive to nonequilibrium situations.}. 
The energy flow from ALPs to neutrinos can then be modelled by a set of Boltzmann equations for comoving energies, defined as
\be
X_i=\rho_i R^4  \; ,
\ee
and the ALP phase space distribution function $f(K_\phi)$ as a function of the comoving momentum, $K_\phi=R k_\phi$,
\begin{subequations}\label{eq:boltzmannSet}
\begin{align}
&\frac{d}{d\,t} f(K_\phi)=-(c_\gamma+c_{\rm P})(f-f^{\rm eq})\; ,\label{eq:boltzgALP}\\ \nonumber
&\frac{d}{d\,t}(X_\gamma+X_e)= 3 H \delta p_e R^4 + \int \!\!\frac{d^3 K_\phi}{(2\pi)^3} W_\phi (c_\gamma+c_{\rm P})(f-f^{\rm eq})\label{eq:boltzEM}\\
&\quad\quad\quad\quad\quad  +\frac{\Gamma_{e\nu}}{R^4} \[c_e\(X_{\nu_e}^2- X_{\nu_e}^{{\rm eq}\,2}\)+c_{\mu\tau}\(X_{\nu_{\mu\tau}}^2-X_{\nu_{\mu\tau}}^{{\rm eq}\,2}\)\]\; ,\\
&\frac{d}{d\,t}X_{\nu_e}=-\frac{\Gamma_{e\nu}}{R^4}  c_e\(X_{\nu_e}^2-X_{\nu_e}^{{\rm eq}\, 2}\) \; ,\label{eq:boltzgNUE}\\
&\frac{d}{d\,t}X_{\nu_{\mu\tau}}=-\frac{\Gamma_{e\nu}}{R^4} c_{\mu\tau}\(X_{\nu_{\mu\tau}}^2-X_{\nu_{\mu\tau}}^{{\rm eq}\,2}\)\; ,\label{eq:boltzgNUMU}\\
&\frac{d\, }{d\,t} R= \frac{1}{R}\sqrt{\frac{8\pi}{3 m_{\rm Pl}^2}\(X_{\gamma}+X_e+X_{\nu_e}+X_{\nu_{\mu\tau}}+\rho_\phi R^4\)}\; \label{eq:boltzR}, 
\end{align}
\end{subequations}
where $\omega_\phi=\sqrt{{k_\phi}^2+{m_\phi}^2}$ is the ALP energy and $W_\phi=\omega_\phi R $. 
The set of equations~\eqref{eq:boltzmannSet} describes the evolution of the comoving energy density stored in $\gamma$ together with $e^\pm$, the three species of $\nu$'s, the ALPs and the cosmic scale factor $R$. 
The solution of equation~\eqref{eq:boltzgALP} is the distribution function $f$ for the pseudoscalar particle.
The collision terms for the decay and Primakoff processes are~\cite{Cadamuro:2010cz}
\begin{align}
c_\gamma&=\frac{m_\phi^2-4m_\gamma^2}{m_\phi^2}\frac{m_\phi}{\omega_\phi}
\[1+   \frac{2T}{k_\phi}\log
\frac{1-e^{-(\omega_\phi+k_\phi)/2T}}{1-e^{-(\omega_\phi-k_\phi)/2T}}\]\,
\frac{1}{\tau}\; ,\\
c_{\rm P}&\simeq \frac{\alpha \, g_\phi^2}{16}n_e \log\[1+\frac{\[4\omega_\phi(m_e+3T)\]^2}{m_\gamma^2\[m_e^2+(m_e+3T)^2\]}\]\;  . 
\end{align}
Since the energy stored in electrons and positrons is transferred directly to the photon bath, we kept $X_\gamma$ and $X_e$ together in equation~\eqref{eq:boltzEM}.
The term $\delta p_e=p_e-\rho_e/3$ accounts of the comoving energy density gain as electrons become increasingly non-relativistic, experiencing the transition from $\rho_e\propto R^{-4}$ to $\rho_e\propto R^{-3}$.
Here, $p_e$ and $\rho_e$ are the pressure and energy density of $e^\pm$.
The second term in the right hand side reckons up the photon-ALP energy exchange, while in the last last line of equation~\eqref{eq:boltzEM} are the neutrino-electron interactions.
The equations~\eqref{eq:boltzgNUE} and~\eqref{eq:boltzgNUE} provide the evolution of neutrino comoving energies.
Neutrino flavour influence the electron-neutrino energy exchange rate.
We have separated the three neutrino species in two unknowns, according to the presence of not of charged current interactions with electrons.
The first is the case of $X_{\nu_e}$ for electron neutrinos, while in the second one we have $X_{\nu_{\mu\tau}}$ for muon and tau flavours together.
The energy exchange rate is given by the factor $\Gamma_{e\nu}\equiv G_{F}^2T_{\gamma}$, with $G_{F}$ the Fermi constant, multiplied by $c_e\simeq 0.68$ for the electron flavour and $c_{\mu\tau}\simeq 0.15$ for the muon and tau ones, which follow from the appropriate thermally averaged cross section. 
We assume that neutrinos always have a thermal distribution, determined only by an effective temperature, which should be a reasonable first approximation and accurate enough for our purposes. 
We neglect the energy reshuffling between different neutrino species, which does not influence the total neutrino density at leading order.
Finally, through equation~\eqref{eq:boltzR} we calculate the expansion rate of the universe through the evolution of the cosmic scale factor $R$.
The initial conditions are specified at $T\gg {\rm MeV}$ by having all species at a common temperature and the ALP number density given by equation~\eqref{ALPyield}. 

For values $g_\phi\lesssim10^{-7}~{\rm GeV}^{-1}$, the Primakoff process is decoupled in the temperature range of interest and can be neglected. 
If the inverse decay is also negligible, we can integrate the ALP phase space distribution explicitly and directly compute the evolution of the number density
\begin{equation}
\frac{d}{d\,t}\(n_\phi R^3\)=-\frac{n_\phi R^3}{\tau} \; 
\end{equation}
instead of using equation \eqref{eq:boltzgALP}.
In this way we recover the exponential decay law $N_\phi \propto  e^{- t/\tau}$. 
The integral in equation \eqref{eq:boltzEM} is then 
\begin{equation}\label{eq:boltzmann2}
\int\frac{d^3 K_\phi}{(2\pi)^3} W_\phi (c_\gamma+c_P)(f-f^{\rm eq})\simeq 
m_\phi  n_\phi R^4/\tau \; .
\end{equation}

\begin{figure}[tbp] 
   \centering
   \subfloat[]{\includegraphics[height=200pt]{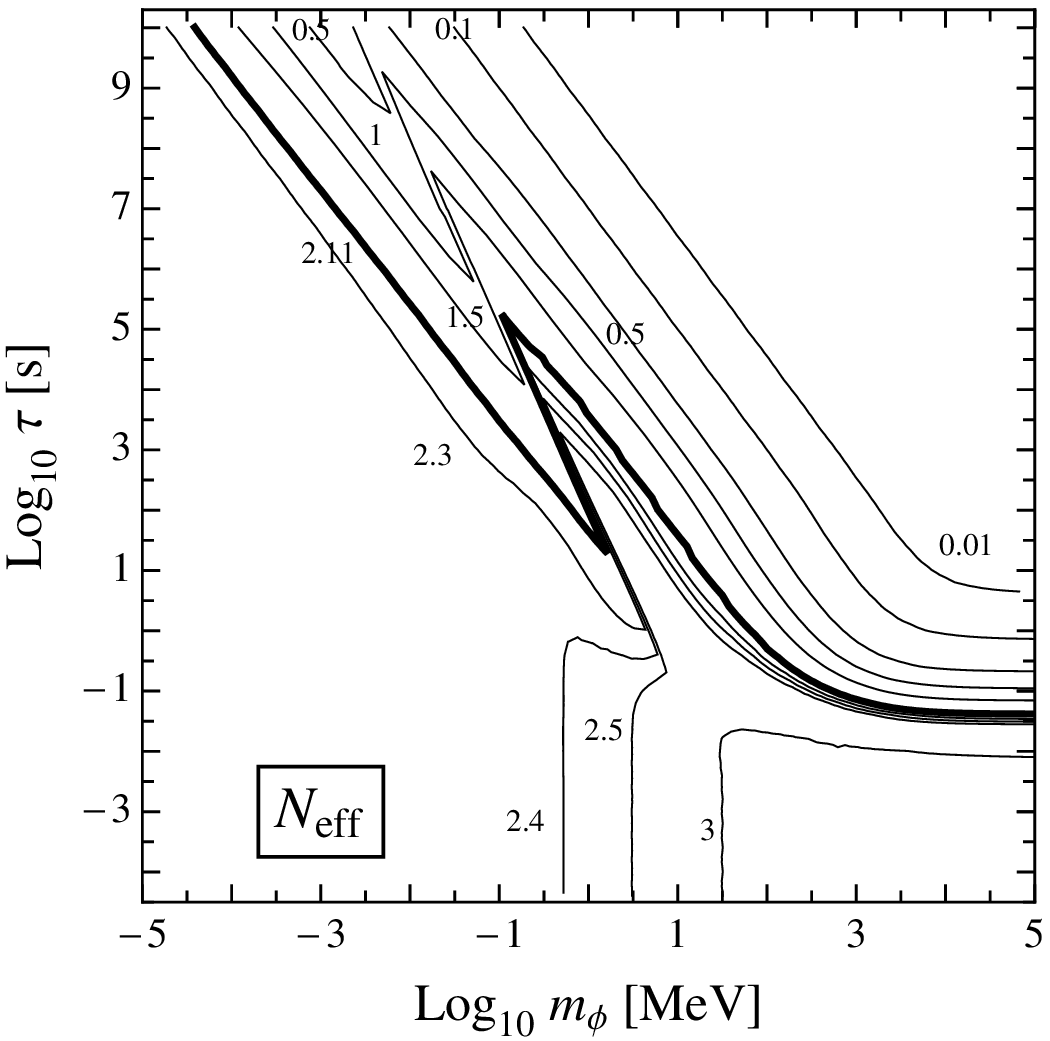}\label{fig:NeffA}}
   \subfloat[]{\includegraphics[height=202pt]{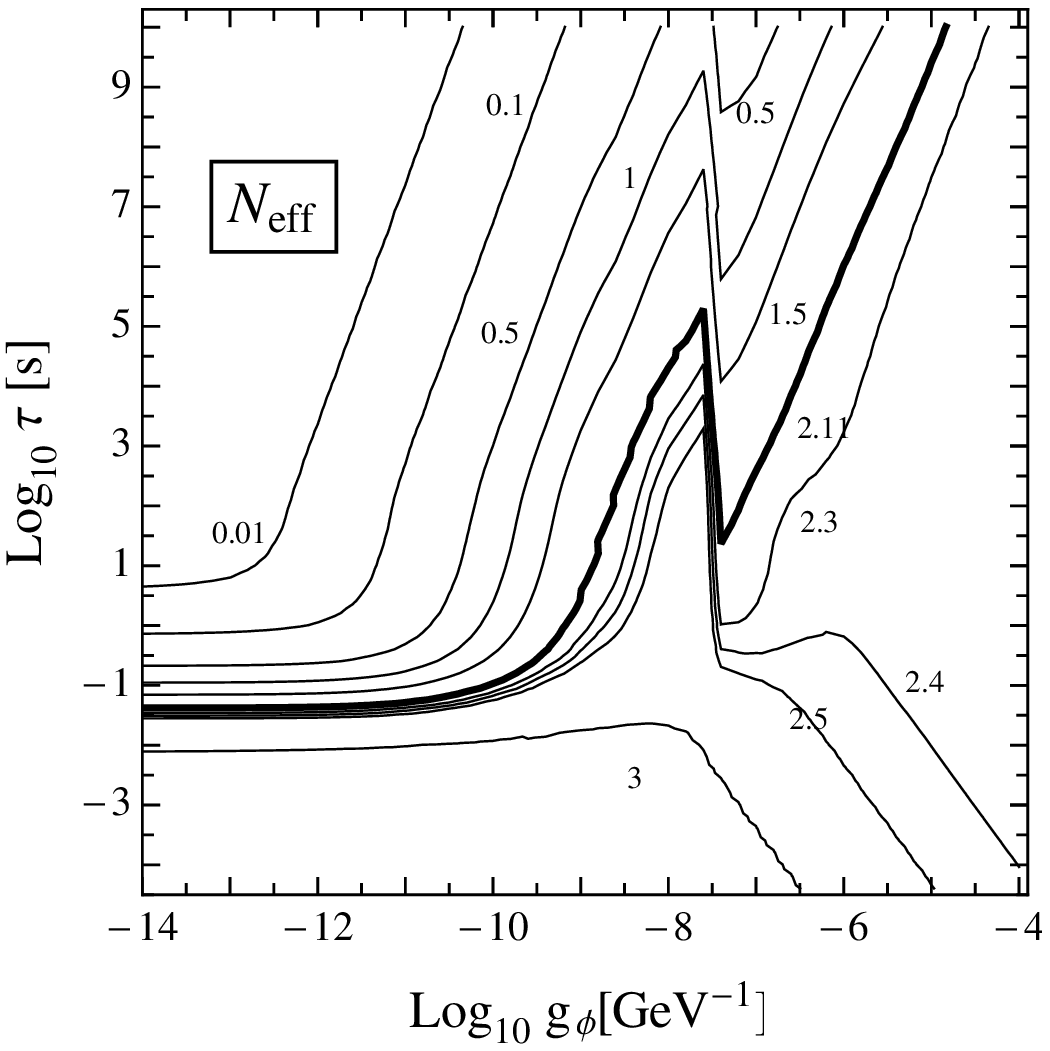}\label{fig:NeffB}}
   \caption{Contour plots of the effective number of neutrinos $N_{\rm eff}$ as a function of the ALP mass and lifetime (left) and of the coupling parameter $g_\phi$ and lifetime $\tau$ (right). 
   ALP cosmologies leading to $N_{\rm eff}<2.11$ can be safely excluded.}
   \label{fig:Neff}
\end{figure}

We have scanned the ALP parameter space in~\cite{Cadamuro:2011fd}, and we present our results for $N_{\rm eff}$ in the $g_\phi$--$\tau$ and $m_\phi$--$\tau$ planes in figure~\ref{fig:Neff}. 
We required a standard cosmology scenario at temperatures below the standard matter-radiation equality $T_{\rm mr}\sim 1$~eV.
Thus we have analysed only the part of parameter space where $T_{\rm d}>T_{\rm mr}$, in order to not perturb the standard picture with the decay.
From equation~\ref{decayT}, we obtain a mass lower bound for the validity of our calculations
\be
\label{MReq}
\frac{g_\phi}{{\rm GeV}^{-1}}\gtrsim 10^{-3}\(\frac{\rm eV}{m_\phi}\)^{3/2} \; , 
\ee 
which largely overlaps with the limits obtained from CMB spectral distortions obtained in the previous chapter. 

The sharp features of the isocontours in figure~\ref{fig:Neff} are due to the abrupt and sizeable decrease of $g_{*S}$ during the QCD phase transition.
The relic abundance of ALPs depends on $g_\phi$ through the freezing out temperature, and for ALPs with coupling around $g_\phi\sim 10^{-7.5}\ {\rm GeV}^{-1}$ we have $T_{\rm fo}\simeq \Lambda_{QCD}$ from equation~\eqref{eq:T_fo}.
In a small range of $g_\phi$ the ALP abundance considerably changes, according if the decoupling is late enough for ALPs to share the entropy transferred from the QCD degrees of freedom or not.
The larger is the ALP abundance, the larger is the subsequent entropy release at the time of the decay, hence the peculiar shape of the isocontours. 
In figure~\ref{fig:NeffA} we observe the same features, as the $g_\phi$ dependence is hidden in $\tau$. 

\begin{figure}[tbp]
\begin{center}
\subfloat[]{\includegraphics[height=6.9cm]{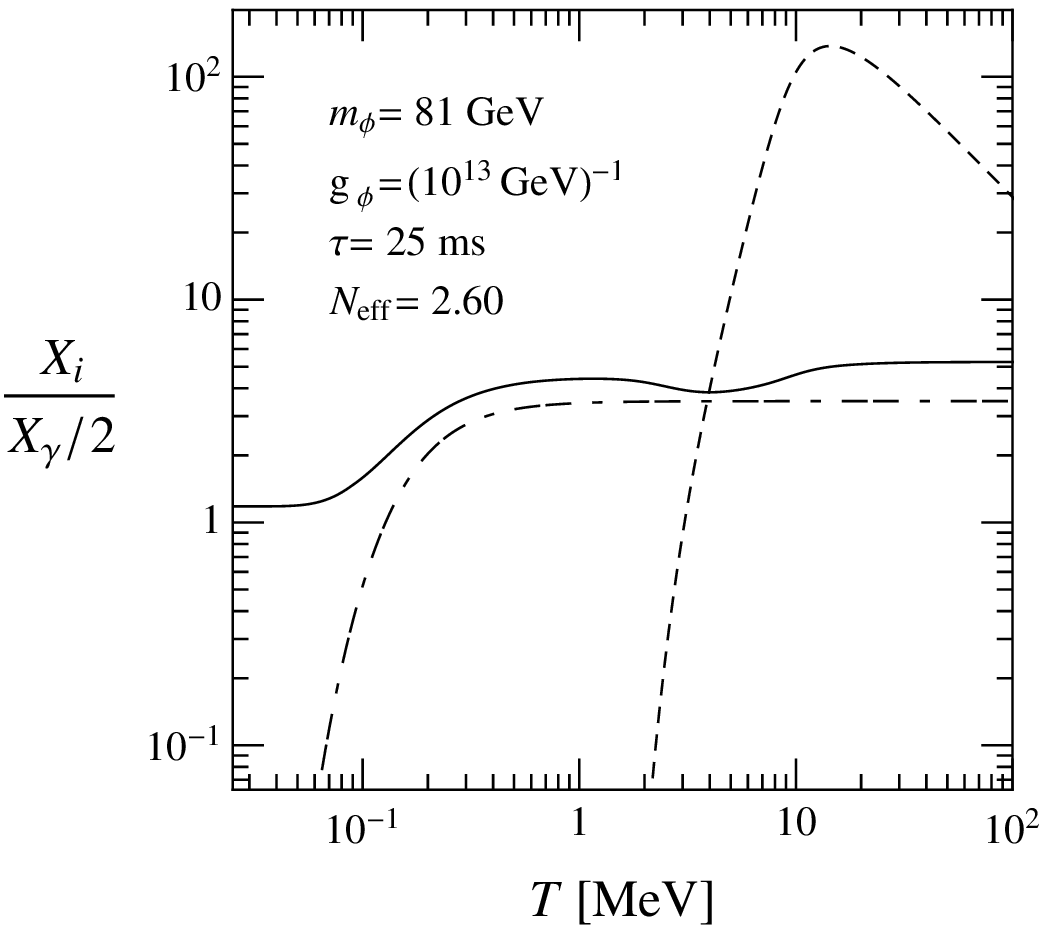} \label{fig2a}}
\subfloat[]{\includegraphics[height=6.9cm]{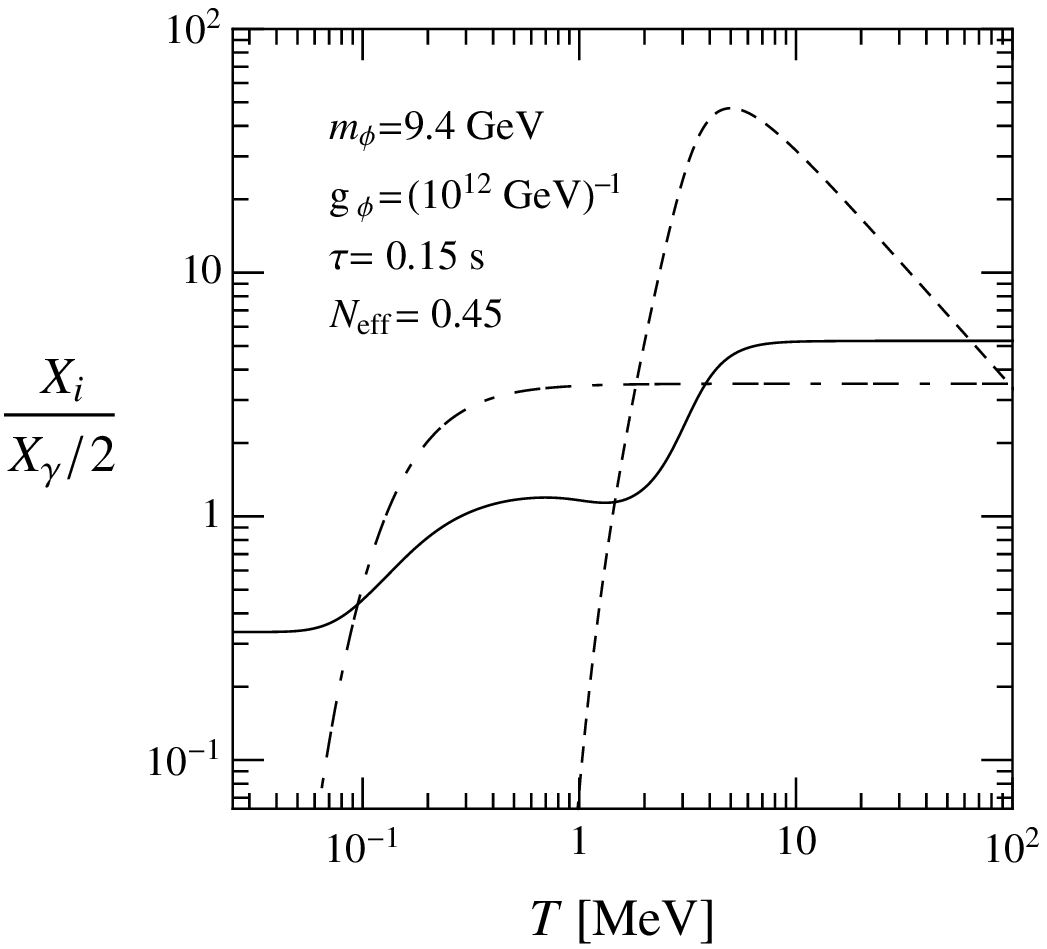} \label{fig2b}}
\hspace{.1cm}
\subfloat[]{\includegraphics[height=6.9cm]{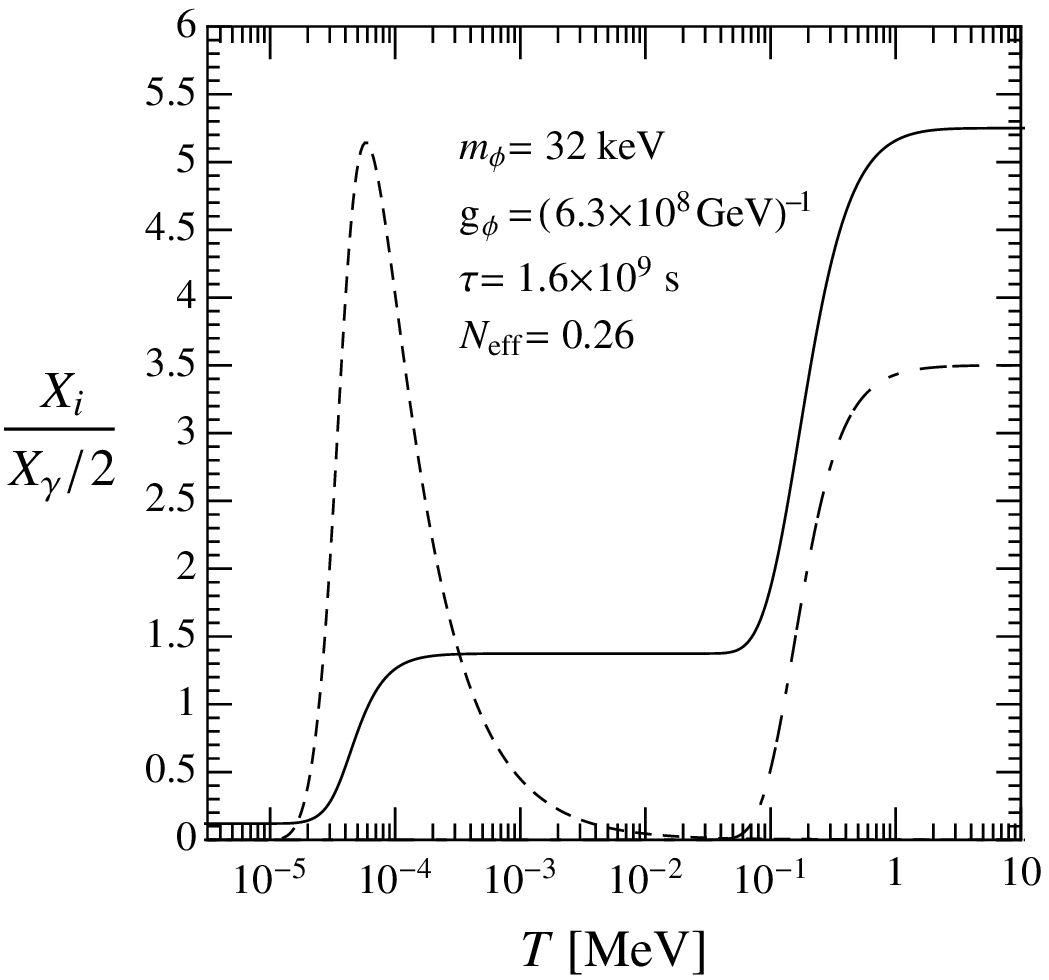} \label{fig3a}}
\subfloat[]{\includegraphics[height=6.9cm]{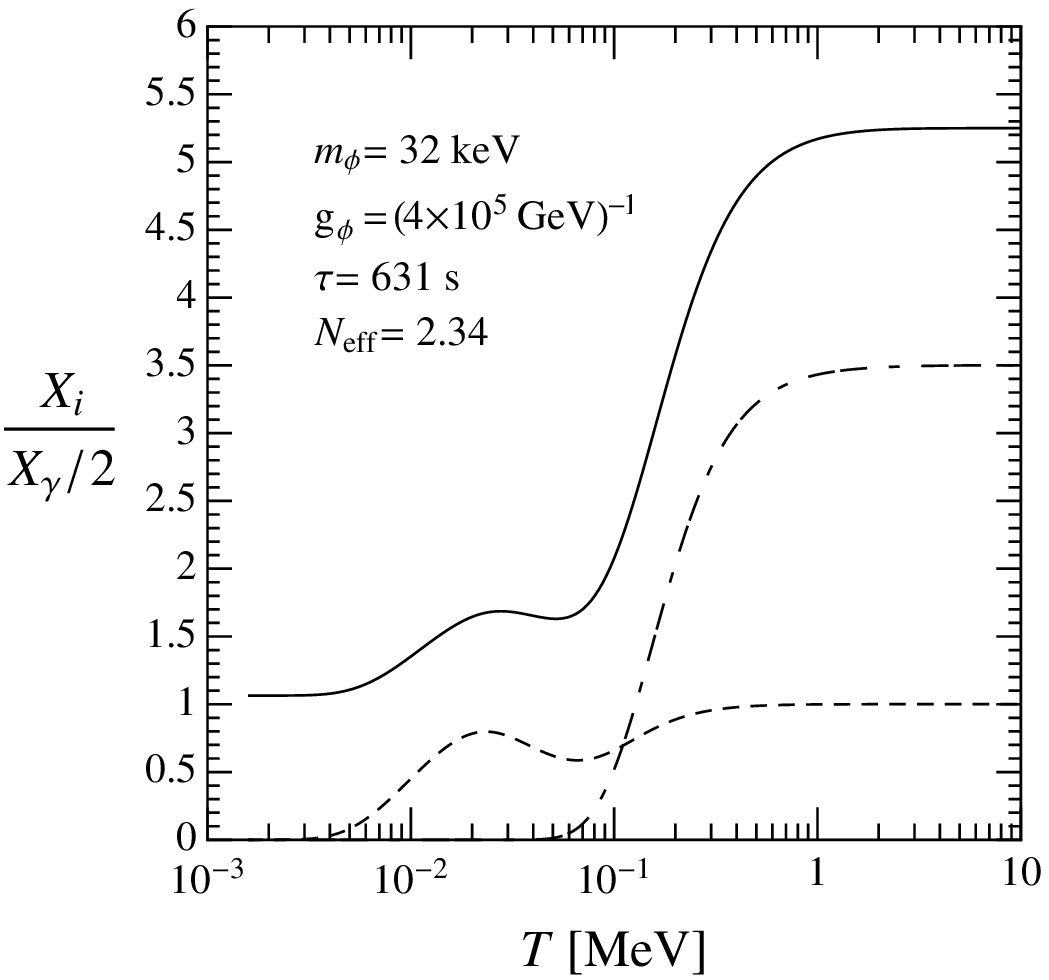} \label{fig3b}}
\end{center}
   \caption{Examples of the evolution of the comoving energy of all species of neutrinos (solid), ALPs (dashed) and electrons (double-dashed), as functions of the temperature of the universe. All energies are normalized to one thermal bosonic degree of freedom.}
   \label{fig:examples}
\end{figure}

In figure~\ref{fig:examples}, we present some illustrative examples depicting the evolution of the $X$s of electrons, neutrinos and ALPs as a function of the temperature. 
Note that in all of them, when ALPs become non-relativistic, the ratio $X_\phi/X_\gamma$ rises because it becomes proportional to $m_{\phi}/T$, until the age of the universe becomes comparable with $\tau$. 

Figure~\ref{fig2a} shows a typical case of very early decay of a massive ALP, when neutrinos are still partially coupled to electrons.
This example depicts the ALP behaviour in the lower right corner of figure~\ref{fig:NeffA}, and correspondingly in the lower left one of figure~\ref{fig:NeffB}. 
Here $N_{\rm eff}=2.6$, marginally different from the standard value of $3$.
Even if the ALP energy dominates the universe and the entropy injected du\-ring the decay is huge, the reheating temperature is large enough for neutrinos to almost fully recover their thermal abundance. 
In general, the final value of $N_{\rm eff}$ is related not to the total entropy released, but only to the part of it injected after the freeze out of the neutrino-electron interactions. 
In this particular region of the parameter space, the neutrino dilution is mainly sensitive to the ALP lifetime and not to $m_\phi$ or $g_\phi$ individually.
The outcome of a decay earlier than $t\sim10^{-2}$~s is undistinguishable from standard cosmology, since neutrinos regain completely their thermal abundance.
Around $T\sim m_e$, electrons and positrons become nonrelativistic and annihilate, heating the photon bath but not the 
neutrinos, which have decoupled. 
The ratio $X_\nu/X_\gamma$ therefore decreases.
In this period, the photon temperature increases with respect to the neutrino temperature by the standard factor $(4/11)^{1/3}$ due to entropy conservation.

The picture changes considering later decays, after neutrino decoupling: photons and electrons get all the ALP enthropy, as shown in figure~\ref{fig2b}.
This is what happens in figure~\ref{fig:NeffA} and in figure~\ref{fig:NeffB} above the thick isocontour. 
The neutrino dilution is computable in this case, as the ratio of the final and initial comoving entropies of the photon-electron bath is given by equation~\eqref{eq:entropy}.
The temperature of the electromagnetic bath increases with respect to the neutrino one by a factor 
$(S_{f}/S_{i})^{1/3}$, making $N_{\rm eff}=3(S_{f}/S_{i})^{-4/3}$ because also the electron entropy ends up in photons. 
The neutrino energy density is therefore strongly diluted by the energy gain of photons plus electrons. 
Note that this is mainly a function of $m_\phi\sqrt{\tau}\propto \({g_\phi\sqrt{m_\phi}}\)^{-1}$ --- as we can evince from equation~\eqref{eq:entropy} --- which produces the characteristic slope of the isocontours at long ALP lifetimes --- $\tau\propto {m_\phi}^{-2}$ in figure~\ref{fig:NeffA} and $\tau\propto g_\phi^{4}$ in figure~\ref{fig:NeffB}. 
For $g_\phi\lesssim 10^{-9}\ {\rm GeV}^{-1}$, this is the only dependence on $g_\phi$, since $Y_\phi$ is constant.
Another example of late ALP decay, but with a smaller mass, is shown in figure~\ref{fig3a}. 
In this case we observe first the $e^{\pm}$ annihilation, which heats photons with respect to neutrinos. 
A sizeable neutrino dilution is observable after the ALP decay. 

Finally, in figure~\ref{fig3b} we show an example for which inverse decays are relevant, which is on the left side of figure~\ref{fig:NeffA} and on the right in figure~\ref{fig:NeffB}.
As the temperature drops, we observe a first decrease of the ALP energy density due to the electrons heating the photon bath. 
The inverse decay channel opens around $T\sim 70$~keV and helps ALPs to regain equilibrium before disappearing at $T\sim m_\phi$. 
During rethermalization, the photon energy \emph{decreases}, which can be seen as a slight rise in $X_\nu/X_\gamma$.
Due to this mechanism, in the small-mass and short-lifetime region of the parameter space we have seen that entropy conservation gives $N_{\rm eff}=3(11/13)^{4/3}\simeq 2.4$. 
If $m_\phi$ is larger than a few MeV, the decay-in-equilibrium happens when neutrinos are still coupled, so $N_{\rm eff}$ approaches~3. 
The disappearance from the thermal bath is governed only by the mass, and the isocontours of $N_{\rm eff}$ exactly follow the isocontours of $m_\phi$.

The excluded region of the parameter space are determined comparing these numbers with the limits \eqref{eq:neffconstraint}, taking the 99\% C.L. value of $N_{\rm eff}>2.11$.
We are compelled to this conservative choice because the 95\% C.L.~value of $N_{\rm eff}>2.39$ is just below the $N_{\rm eff}\simeq 2.4$ obtained from the case of decay-in-equilibrium.
Given all the uncertainties of this calculation, we can consider this large part of parameter space disfavoured but not excluded.
In figure \ref{fig:Neff}, the excluded region is above the thick line, which is coloured in yellow in the summarising plot for this chapter, figure~\ref{summa}.

\subsection{Axion bounds}

\begin{figure}[tbp]
   \centering
   \includegraphics[width=7.3cm]{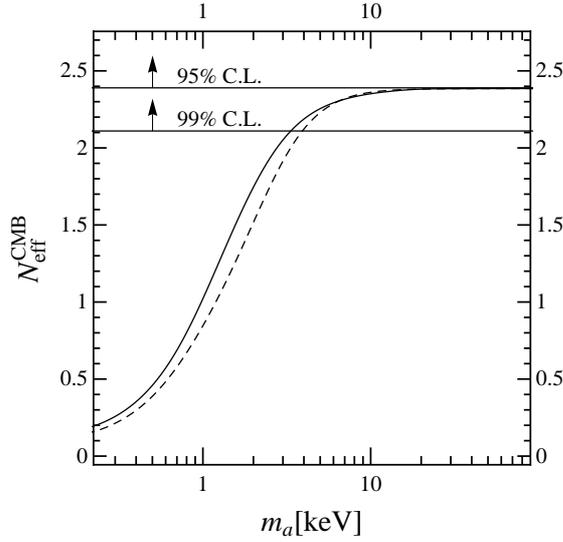}
   \caption{Radiation density during the CMB epoch ($C_\gamma=1.9$).
   Solid and dashed lines are for hadronic and non-hadronic axions ($C_e=1/6$).
   The 95\% and 99\% C.L.\ lower limits from Eq.~(\ref{eq:neffconstraint})
   are shown as horizontal lines. }
   \label{fig6}
\end{figure}

Pertaining to neutrino dilution, axion phenomenology is simpler than that of ALP.
The cosmic energy density in neutrinos is modified by the factor $(T_\nu^{\rm ax}/T_\nu^{\rm std})^4$ between the axion and standard cosmology. 
The results of the introduction imply that this ratio can be expressed in terms of the quantity $\mathcal{T}_2$, the modified $(T_\gamma/T_\nu)^3$ value after axions have disappeared, and
\begin{equation}\label{eq:Neff}
N_{\rm eff}=3\(\frac{11}{4\mathcal{T}_2}\)^{4/3}\; .
\end{equation}
The variation of $N_{\rm eff}$ with $m_a$ for $C_\gamma=1.9$ is shown in figure~\ref{fig6}. 
At sufficiently high $m_a$, the inverse decay process keeps thermal equilibrium during the decay. 
In this case, we have an analytical expression
\begin{equation}
N_{\rm eff} = 3\(\frac{11}{4}\frac{2(\h_1+2)}{13}
\frac{2}{3}\)^{4/3}\(\frac{3}{2+\h_1^{4/3}}\)\; .
\end{equation}
For $m_a>20$~keV we have $\h_1\simeq 1$ and $N_{\rm eff}$ reaches asymptotically the minimum neutrino dilution $3\,(11/13)^{4/3}=2.401$.
At much larger masses, $m_a\sim$ MeV, axions would disappear in local thermal equilibrium before neutrino decoupling, leaving no trace in cosmology. 
Therefore, at $m_a\sim$ MeV the value of $N_{\rm eff}$ shown in fi\-gure~\ref{fig6} reaches a plateau at the the standard value $N_{\rm eff}=3$.
Comparing with the cosmological limits of equation~\eqref{eq:neffconstraint} we see that even the minimum neutrino dilution is only barely allowed at 95\% C.L.\ and, again, disfavoured, but not credibly excluded.

For $m_a\alt 20$~keV, axions decay increasingly out of equilibrium, creating entropy and reducing the final $N_{\rm eff}$ further.
Comparing the calculated $N_{\rm eff}$ in figure~\ref{fig6} with the observational limits implies
\begin{equation}\label{eq:dilutionconstraint}
m_a>3~\rm keV~~\hbox{at~~99\% C.L.}\; .
\end{equation}

\section{Influence on big-bang nucleosynthesis}

\begin{figure}[tbp]
   \centering
   \includegraphics[width=13cm]{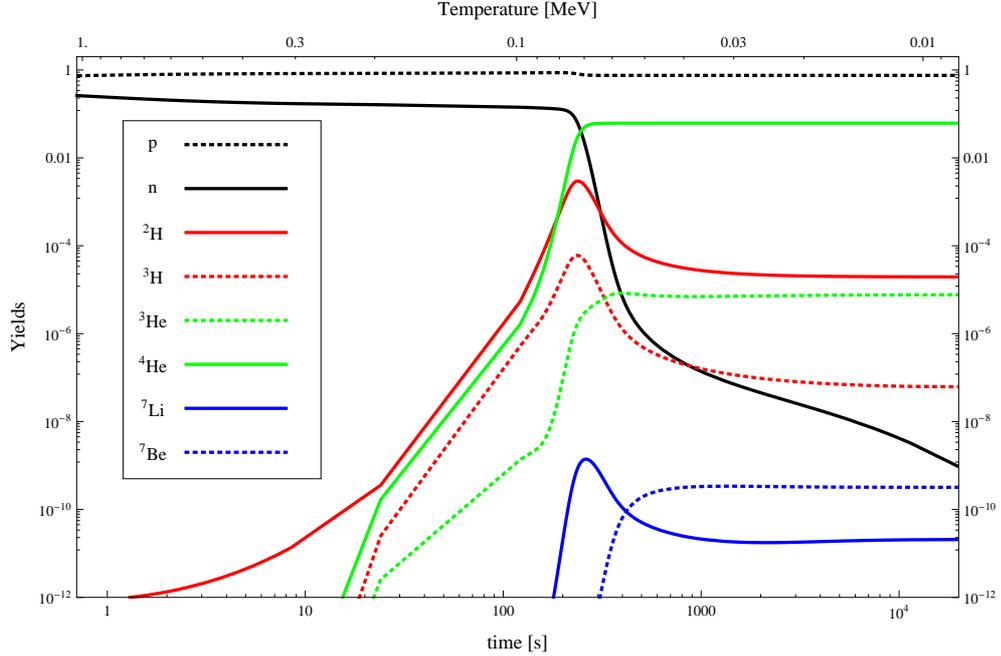}
   \caption{The evolution of primordial nucleosynthesis, after a \parte~si\-mulation~\cite{Pisanti:2007hk} with $\eta_{10}=6.23$ and $N_{\rm eff}=3$.
All the plotted yields but the proton and the $^4$He ones are normalised to hydrogen, i.e.~$n_i/n_p$. 
The ratio $n_p/n_B$ is plotted for protons, while in the helium case we plot the mass fraction $Y_P\equiv4n_{\rm He}/n_B$, where $n_B$ is the baryonic number density. }\label{fig:stdbbn}
\end{figure}

During the primordial nucleosynthesis neutrons and protons combine to form nuclei.  
Because of the absence of stable nuclear combinations with mass number 5 or 8, only nuclei lighter than beryllium can form.
Heavier elements will appear only after the formation of the first stars, where the density is high enough to make triple $\alpha$ collisions possible. 
We can follow the development of BBN and the evolution of the primordial nuclear abundances in figure~\ref{fig:stdbbn}.

The ratio of nucleon abundances set the initial condition for BBN. 
For $T\gg 1$~MeV and $t\ll 1$~s, the weak reactions
\begin{align}\label{eq:np_reactions}
p^+ + e^- &\leftrightarrow n + \nu_e \\
p^+ + \bar{\nu}_e &\leftrightarrow n + e^+ 
\end{align}
keep the neutron-to-proton ratio equal to its equilibrium value
\be\label{eq:eq_np_ratio}
\left. \frac{n_n}{n_p} \right|_{\rm eq}=\exp\(-\frac{Q}{T}\)\; ,
\ee
where $Q\equiv m_n - m_p=1.293$~MeV is the difference between the nucleon masses, $n_n$ is the neutron number density and $n_p$ the proton one. 
Relativistic electrons and neutrinos are more abundant than nucleons of a factor close to the baryon-to-photon ratio $\eta$, and eventual deviations of $n_n/n_p$ from the equilibrium value~\eqref{eq:eq_np_ratio} are rapidly erased at this stage.
Around $T\simeq 1$~MeV and $t\simeq 1$~s, the weak reactions~\eqref{eq:np_reactions} freeze out and neutrinos decouple.
The neutron-to-proton ratio departs from equilibrium too, with a value $n_n/n_p=\exp\(Q/1\ {\rm MeV}\)\sim1/6$.
After this time, $n_n/n_p$ decreases because of the $\beta$-decay of the neutron
\be
n\rightarrow p^+ + e^- + \bar{\nu}_e\; .
\ee 
At this point, the combination of nucleons experiences a break.
All the main BBN nuclear reactions involve a deuterium nucleus to be combined with other nuclei. 
The reaction rate depends on the cross section and on the target abundance, so a certain critical deuterium reservoir has to pile up to proceed on. 
Deuterium is produced through the reaction
\be\label{eq:D_prod}
p^+ + n \rightarrow {\rm D} + \gamma\; .
\ee
However, because of $\eta$, the number of photons with energy greater than the binding energy of deuterium $B_D=2.22$~MeV is easily greater than the number of deuterium nuclei, and the inverse reaction is favoured. 
When the photon temperature has decreased below $T\sim0.3$~MeV, the abundance of deuterium can finally start to rise --- as we can see in figure~\ref{fig:stdbbn} --- and below $T\sim0.1$~MeV the deuterium critical density is reached: the \emph{deuterium bottleneck} opens and heavier nuclei like $^{3}$H and helium isotopes can be synthesised.
The reaction network is 
\begin{align}
{\rm D}+{\rm D}&\leftrightarrow\ ^3{\rm H }+p^+\\
{\rm D}+{\rm D}&\leftrightarrow\ ^3{\rm He }+n\\
^3{\rm H }+p^+ &\leftrightarrow\ ^3{\rm He }+n\\
^3{\rm H }+{\rm D} &\rightarrow\ ^4{\rm He }+n\\
^3{\rm He }+{\rm D} &\rightarrow\ ^4{\rm He }+p^+
\end{align}
while the production of $^7$Be and $^7$Li involves the combination of $^4$He with $^3$He or tritium. 
In the lapse of time till the deuterium bottleneck opening the neutron-to-proton ratio decreases to $n_n/n_p\sim1/7$.
Because helium is the most tightly bond among the light nuclei, almost all the available neutrons are combined into $^{4}$He.
After the nuclear reactions freeze out around $T\sim30$~keV, only very few neutrons are left out of $^{4}$He nuclei, mainly in form of deuterium.

The outcome of standard BBN depends basically on the baryon-to-photon ratio and on the effective number of neutrino species.
The former measures the nucleon density, while the latter essentially affects the expansion rate, so these two numbers determine the effectiveness of the nuclear reactions rates.
Again, through a set of Boltzmann equations it is possible to predict the yield of the different nuclear species after BBN. 
Comparing the results of BBN simulations with the primordial elemental abundances which are measured thanks to some astro\-physical observations, it is possible to have an estimate of $\eta$ and $N_{\rm eff}$.
The overall agreement of the predicted light-element abundances with observations implies $5.1<\eta_{10}< 6.5$ at 95\% C.L.~\cite{Nakamura:2010zzi}, where $\eta_{10}=10^{10} \eta$. 
The independent determination derived from the CMB temperature fluctuations, $\eta_{10} = 6.23\pm 0.17$, agrees rather well with the light-element abundance estimate.
A recent estimate at 95\% C.L.~of the number of effective neutrino species from elemental observations gives $N_{\rm eff}=3.68^{+0.80}_{-0.70}$ and $N_{\rm eff}=3.80^{+0.80}_{-0.70}$ respectively for neutron lifetime $\tau_n=885.4$~s and $\tau_n=878.5$~s~\cite{Izotov:2010ca}.
Again, these estimates agree with CMB ones.
In the following, we will use $\eta^{\rm BBN}$ and $\eta^{\rm CMB}$ to refer to the baryon-to-photon ratio estimates coming respectively from elemental abundance measurements and from CMB analysis. 

The expansion of the universe influences the weak interactions freeze out and the deuterium bottleneck opening, and thus the value of the neutron-to-proton ratio.
Since almost all the neutrons belong to helium nuclei, $^4$He was in the past the favourite indicator of the presence of extra degrees of freedom in the very early stages of BBN.  
It is current practice to express the abundance of primordial $^4$He through its mass fraction 
\be
Y_p\equiv\frac{ 4 n_{\rm He}}{n_p+n_n}\sim \frac{ 2 \(n_n/n_p\)}{1+\(n_n/n_p\)}\; .
\ee
The value $Y_p$ can be estimated from an extrapolation to zero metallicity of the measured $^4$He content of metal-poor extragalactic HII regions. 
The systematics of the measurements have caused the estimate of $Y_p$ to vary significantly over the years.
We are therefore called for extreme caution when quoting this bound. 
We shall be conservative and adhere to the proposal made in~\cite{Mangano:2011ar}. 
Here the authors set a robust upper bound on $Y_p$ based on the assumption that the helium content is an increasing function of the metallicity of the cloud. 
They find
\be
\label{Yp}
Y_p < 0.2631 \quad {\rm at}\ 95\%\ {\rm C.L.}\ \;  . 
\ee

\begin{figure}[tbp]
   \centering
   \includegraphics[width=7.4cm]{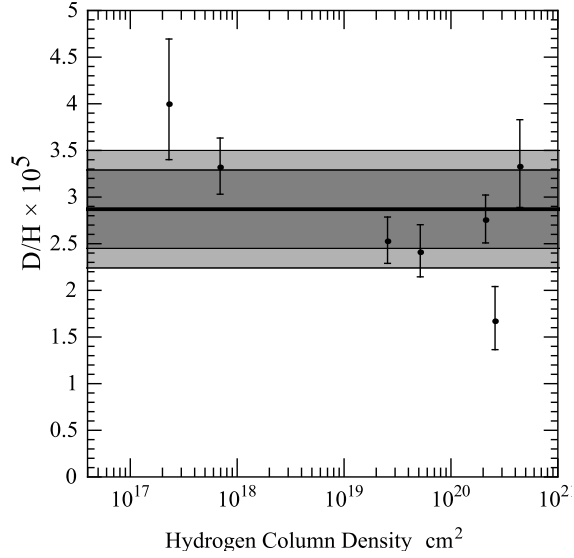}
   \caption{Measurements of primordial deuterium-over-proton ratio D/H. The grey bands represent the 95\% and
     99\% range for the observed D abundance. It is derived from the 
     7 high redshift Ly-$\alpha$ clouds shown as a function of the hydrogen column
     density~\cite{Pettini:2008mq}.}
   \label{fig:YD}
\end{figure}

Compared with the standard BBN picture, a higher value of the baryon-to-photon ratio implies an earlier opening of the deuterium bottleneck, and intermediate nuclei --- like D itself --- are processed for a longer time.
From the abundances of D, $^3$H and $^3$He we can obtain the value of $\eta$.
However, tritium is unstable and decays into $^3$He, plus an electron and an antineutrino.
The estimate of the primordial $^3$He from astrophysical observation is very difficult, since the only data come from the solar system and high metallicity HII regions in our galaxy~\cite{Nakamura:2010zzi}. 
Moreover, the matter is complicated by the $^3$He processing inside stars. 
Thus the abundance of deuterium is the remaining observable. 
Unfortunately, also the data about D are scarce. 
At the time of the analysis carried on  in~\cite{Cadamuro:2010cz}, we only had a reliable estimations from 7 high redshift low metallicity clouds absorbing the light of background quasars~\cite{Pettini:2008mq}, which are plotted in figure~\ref{fig:YD} as a function of the H column density. 
Other two observations recently joined these seven measurements~\cite{Fumagalli:2011iw}. 
The results of these estimations agree well at first glance but there is a scatter of the measurements beyond the expectations from the quoted systematics. 
The PDG quotes D/H$ =(2.82\pm0.21)\times 10^{-5}$ for the primordial ratio of deuterium to hydrogen, where the error has been enlarged to account for the still unexplained scatter. 
Such a large scatter in the measurements of the D abundance may be a signal of some not yet understood processing of D inside these high-redshift clouds. 
Whether this is the case or not, there are no known astration processes that increase the D concentration, so the primordial deuterium yield should be \emph{larger} than this estimation. 
Any measurement of deuterium constitutes therefore a lower limit to the primordial value.  
In order to be conservative, in this study we use 
\be
{\rm D/H}|_p >2.1\times 10^{-5}\;  .  
\ee   
The dark (light) grey band in figure~\ref{fig:YD} is the experimental 95\% (99\%) C.L.~region. 

Also the $^7$Li abundance could be in principle used to determine the baryon-to-photon ratio, but it requires a value of $\eta$ rather different from the deuterium estimate.
However, WMAP measurements of $\eta$ agree very well with the deuterium value, while the $^7$Li one lies 4--5$\sigma$ out~\cite{Cyburt:2008kw,Fields:2011zzb}.
Because of this \emph{primordial lithium problem}, we are not going to use the $^7$Li abundance to constrain axion and ALP decays since at the moment observations do not agree with standard BBN predictions. 

The agreement between independent estimates of $\eta$ from BBN calculations and WMAP measurements provides one of the most beautiful tests of standard cosmology. 
A word of caution is however in order since the WMAP value depends on cosmological priors such as the spectral index of primordial fluctuations. 
The quoted value stems on a scale-free power-law which we carry as a further assumption.

\subsection{Axions and BBN}

\begin{figure}[tbp]
   \centering
   \includegraphics[width=7.4cm]{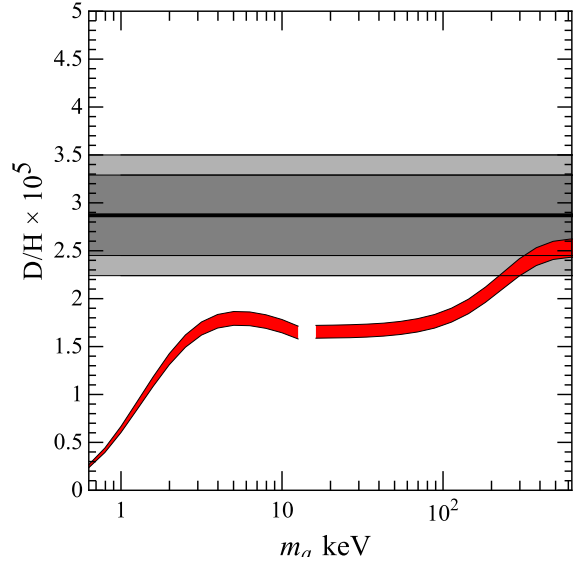}
   \caption{D/H as a function of $m_a$ for $C_\gamma=1.9$ (Hadronic axions).
     The width of the red band represents the $1\sigma$ uncertainty
     of the CMB determination of $\eta$. To the right of the break,
     axions are treated as being in LTE, to the left they are assumed
     to be decoupled during BBN. The grey bands represent the 95\% and
     99\% range for the observed D abundance as in figure~\ref{fig:YD}.}
   \label{fig:axionD}
\end{figure}

In our decaying pseudoscalar scenario all the BBN parameters can be heavily perturbed.
In the previous section we have seen the effect on $N_{\rm eff}$, and here we will show how also the concordance between the BBN and CMB values of $\eta$ is disturbed.
The baryon abundance is diluted by the axion decay, as the photons injected in the bath increase $n_\gamma$ suppressing $\eta$.
If the decay takes place after BBN but before recombination, the concordance between $\eta^{\rm BBN}$ and $\eta^{\rm CMB}$ is compromised.  
In our scenario is
\be
\eta^{\rm BBN}=(S_2/S_1)\eta^{\rm CMB}\; .
\ee 
Taking for granted the WMAP 7 measurement of $\eta$, it means that during the BBN time the baryon to photon ratio was enhanced by the factor given by the black curves depicted in figure~\ref{fig:B1B2}. 
This is the main effect influencing BBN.
The primordial deuterium abundance is the key observable. 
The increased expansion rate plays only a sub-leading role in the axion case.
Moreover, here the decay-in-equilibrium case provides a dilution factor $S_2/S_1=3/2$ which is large enough to obtain an elemental abundance completely different from the observed one, and the equilibrium decay can be excluded too.
The bound vanishes for early enough decays, which do not perturb too much the standard BBN picture. 

In order to quantify our arguments we have modified the publicly available BBN code \parte~\cite{Pisanti:2007hk} to include the effects of axions, taking into account their impact on the Friedmann equation and the modified densities of different radiation species.
Our results are the main outcome of~\cite{Cadamuro:2010cz} and are shown in figure~\ref{fig:axionD}, where we plot the D/H as a function of the axion mass. 
For $m_a>20$~keV we treat axions as being in thermal equilibrium throughout BBN. 
For $m_a\lesssim 10$~keV we use their abundance from our numerical freeze-out calculation, assuming that they are decoupled during BBN. 
We do not treat the intermediate case, leaving a gap in the predicted deuterium yield as a function of $m_a$ that is shown as a red band in figure~\ref{fig:axionD}. 
To calculate this curve we have adjusted the baryon abundance such that in the very end it matches the CMB-implied value. 
Its $1\sigma$ range is represented by the width of the red band.

We compare the predicted D yield with the measured value, and the results are plotted in figure~\ref{fig:axionD}. 
As we see in figure~\ref{fig:axionD}, the presence of axions reduces D/H, so our bounds appear to be conservative regarding the possibility of some unknown stellar production process.

The deuterium abundance is reduced below its 2$\sigma$ observation if $m_a$ is below 300~keV. 
Therefore, BBN constrains axions to have masses
\begin{equation}\label{bound:axionBBN}
m_a > 300\ {\rm keV}\;.
\end{equation}
Axions with mass above this limit have almost completely disappeared from the thermal bath before they can affect BBN.
In this case the predictions approach standard BBN.
This mass upper bound closes the red \textbf{Hot DM} band in figure \ref{fig:alimits}.
Since this bound corresponds to axions that are always in thermal equilibrium, even only via the Primakoff and inverse decay processes, it also applies to non-hadronic axions which would interact more strongly. 

\subsection{ALPs and BBN}

Again, the phenomenology of ALP decay leads to a broader sample of situations and requires a scan of the parameter space.
In order to numerically evaluate the impact of decaying ALPs in the BBN predictions, we have used a BBN code that includes the modified cosmology driven by ALP decays computed with the tools of the previous sections. 
We have written a simple BBN code in Mathematica to compute the primordial abundances of D, $^3$He, $^4$He, $^7$Li and $^7$Be. 
We have used the minimal reaction network relevant for $\eta\sim \eta^{\rm CMB}$ and $N_{\rm eff}\sim 3$ as detailed for instance in~\cite{Mukhanov:2003xs,Esmailzadeh:1990hf,Cuoco:2003cu}. 
When facing standard ALP-less cosmology, our results are in very good agreement with standard BBN calculations obtained with the KAWANO~\cite{Olive:1999ij} or \parte~\cite{Cuoco:2003cu} codes, given the theoretical and experimental uncertainties, which gives us confidence in our results. 
This allows us to easily compute the outcome of BBN when ALPs have a non trivial role during BBN itself, and gives the right trend when ALPs decay much later than BBN, enhancing enormously the value of $\eta^{\rm BBN}$ with respect to $\eta^{\rm CMB}$.  
In the latter case, the minimal reaction network we used might be not sufficient to have a precise BBN outcome, but anyway it gives a first order approximation that leads clearly to exclude a too large $\eta^{\rm BBN}$. 
The effect of relic particle decays on the abundances of light elements predicted by BBN is largely discussed in the literature~\cite{Sarkar:1995dd,Iocco:2008va,Pospelov:2010hj}, usually considering very large mass particles, with $m\ll{\rm GeV}$. 
In our analysis instead we focused also on lighter masses, in the keV and MeV range.

The impact of ALPs on BBN, which was discussed in~\cite{Cadamuro:2011fd}, depends strongly on the $m_\phi$, in particular whether ALPs are heavy enough to induce electromagnetic or hadronic cascades. 
In the following these cases are discussed separately.

\subsubsection*{Small masses}

If the ALP mass is smaller than a few MeV, the decay products cannot induce nuclear reactions and their effect on BBN is only indirect. 
The injected photons --- and perhaps a small amount of electron-positron pairs --- dilute both the neutrino and baryon densities. 
The impact then depends on whether the decays happen before or after BBN.  

For decays happening after BBN, the injected photons heat the bath, decreasing the baryon to photon ratio $\eta$.  
The outcome of this high-$\eta^{\rm BBN}$ scenario is the same of the axion case: intermediate nuclei like D or $^3$He are more easily consumed, and the final abundance of heavier nuclei like Li increases. 

However, the effect of ALP decay can have a strong effect on the $^4$He yield too.
In the late ALP decay scenario, the bottleneck opens earlier because of the enhanced $\eta$, so neutrons have less time to decay, enhancing the final  $^4$He yield. 
But this is not the only effect. 
The presence of ALPs makes the universe expand faster, which has two additional implications.
It induces an earlier freeze out of the $p\leftrightarrow n$ conversion reactions, and therefore a larger $n$ abundance.
Moreover, the time between this freeze-out and BBN is even more shortened, and therefore the amount of neutrons that decay is lower. 
Thus three mechanisms are responsible for the enhancement of the $^4$He yield.  

We have $\eta^{\rm BBN}=\eta^{\rm CMB}$ if ALPs decay \emph{before} BBN, i.e.~before the opening of the deuterium bottleneck. 
The main trends mentioned before disappear. 
However, the ALP decay can still modify BBN indirectly, if it happens between the freeze-out of weak interactions and BBN. 
There are three effects that we should take into account. 
First, ALPs are present during the freeze-out of the $p\leftrightarrow n$ reactions, so the $n$ abundance is in principle larger because of the faster expansion. 
Second, when ALPs decay they reduce $N_{\rm eff}$, as shown in the previous section. 
The cosmic expansion is slow and neutrons have more time to decay.
The neutron concentration is thus affected in two \emph{opposite} ways which tends to compensate each other. 
The time that neutrons have to decay depends on how close to BBN the decays happen. 
In fact we find that $^4$He grows as the ALP decay happens closer to BBN. 
Only in a small $^4$He region at $m_\phi\sim$ MeV and $\tau \sim 30$ s, neutron decay plays a role inducing low $^4$He. 
Because of the two opposing effects, the $^4$He abundance is therefore not a sensitive indicator for ALPs in this region.
Anyway, this scenario also implies low D and $^3$He and high Li, even if for a completely different reason.
The slower expansion gives more time for the consumption of intermediate nuclei, and the final D yield is again lower than the standard BBN one. 

\begin{figure}[tbp] 
   \centering
   \subfloat[]{\includegraphics[width=2.8in]{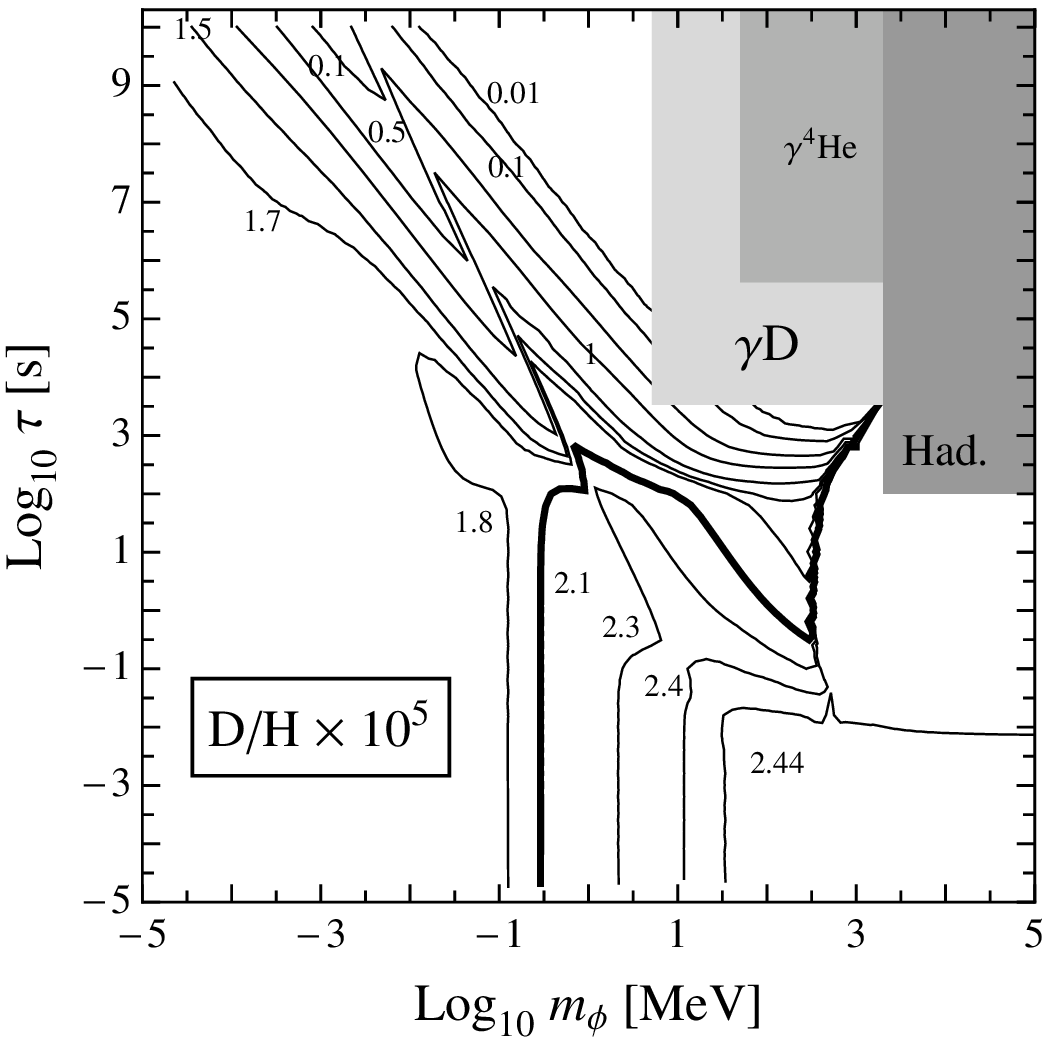}    \label{fig:D}}
   \subfloat[]{\includegraphics[width=2.8in]{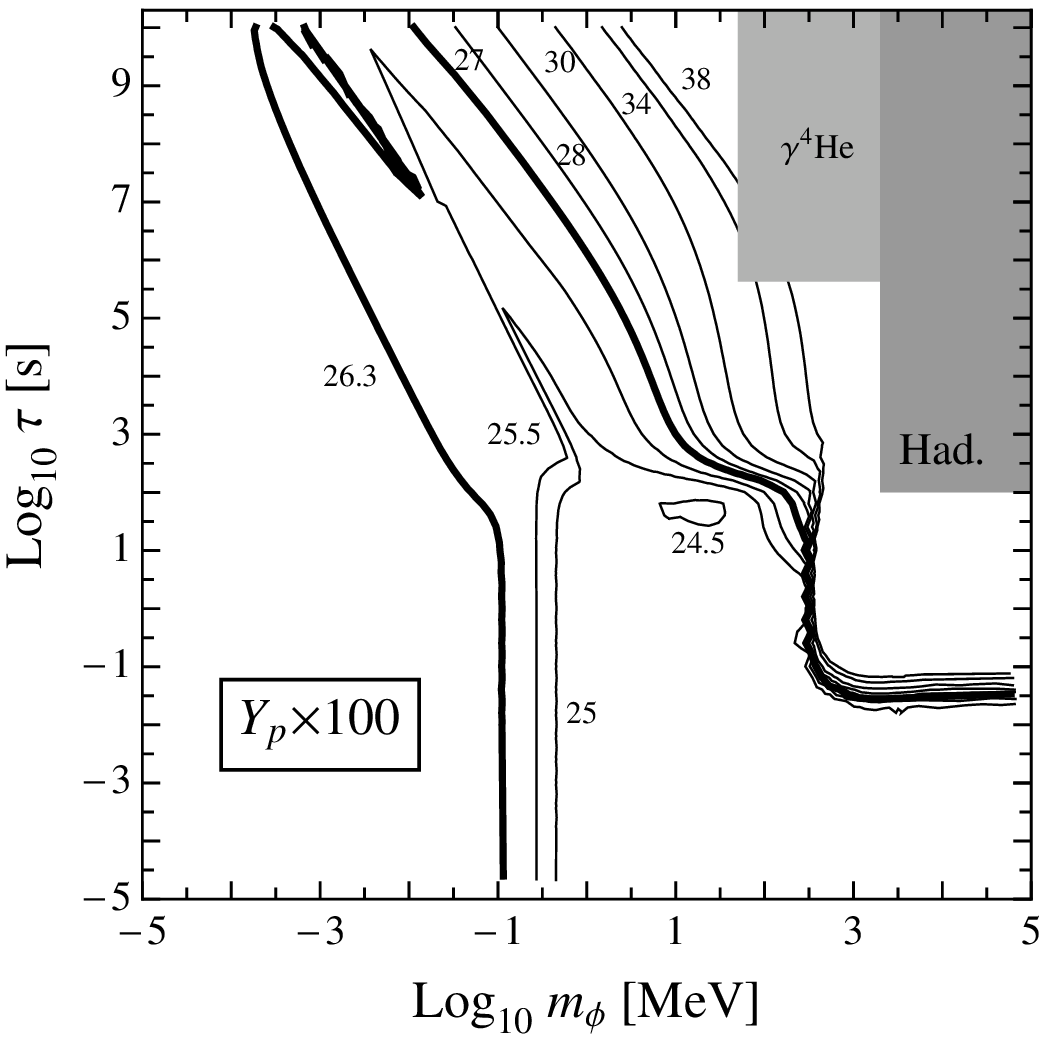}    \label{fig:He}}
   \caption{Isocontours of the primordial abundance of deuterium normalized to protons D/H (left) and helium $Y_p$ (right) in the decaying ALP cosmology, as a function of the ALP mass and lifetime.}
   \label{fig:DHe}
\end{figure}

The predictions for the primordial mass fraction of $^4$He, $Y_p$, and the deuterium-over-proton ratio D/H in the ALP-decay scenario are shown in figure~\ref{fig:DHe}. 
The excluded regions are above the thick isocontour.
The isocontours very much resemble those of $N_{\rm eff}$ because the outcome of BBN is mostly sensitive to the value of $\eta^{\rm BBN}$ and therefore to the baryon dilution, which qualitatively follows the same logic as the neutrino dilution.  
ALPs with small mass and fast decay disappear from the bath in local thermal equilibrium with photons at temperatures around $m_{\phi}$. 
Therefore here BBN only depends on the ALP mass, not on the lifetime, and the outcome is the same of equilibrium-decay of axions that provide the bound \eqref{bound:axionBBN}.  
When the lifetime is longer, the isocontours are parallel to the lines of constant entropy production as in the $N_{\rm eff}$ case.

The exclusion bound from $^4$He is depicted as a purple region in the chapter summarising picture \ref{summa}, while the deuterium bound is shown in red.
In the light masses range is deuterium that provides the main exclusion bound.
The bound corresponds to ALPs that inject a a fraction of order 10\% of the total entropy in the electromagnetic bath after BBN. 
Under this circumstances the baryon and neutrino dilution is $\cal O$(1), and, since deuterium is the most sensitive observable to $\eta^{\rm BBN}$, it is also the most constraining argument. 
Note that the $^4$He abundance depends only logarithmically on $\eta^{\rm BBN}$ while D/H $\propto 1/(\eta^{\rm BBN})^{\sim 1.6}$~\cite{Cyburt:2008kw}. 

\subsubsection*{Large masses}

The decay of large mass ALPs affects BBN through $N_{\rm eff}$ and $\eta$, which continues along the trend of the analysis for small masses.
The additional ingredient in this case is the direct dissociation of nuclei due to energetic enough decay pro\-ducts, which can strongly modify the predictions of BBN.
We have depicted the areas where these effects are relevant as grey regions in figures~\ref{fig:DHe}.
Moreover, the creation of pions in hadronic cascades has also a peculiar way of affecting the primordial nucleosynthesis.

We first comment on photo-dissociation, since the main ALP decay channel we are analysing is the two-photon one. 
The photo-dissociation of deuterium --- which requires a threshold energy $E_{\rm th}=B_D=2.22$ MeV --- of course reduces its abundance further, making our bounds even stronger. 
On the other hand, the photo-dissociation of $^4$He can create deuterium, which reverses the trend of our constraints. 
This requires higher-energy photons since the threshold energy is $E_{\rm th}=19.81$ MeV.
We have followed the methods of~\cite{Cyburt:2002uv} to check in which region of parameter space this phenomena can change our predictions. 
At high densities and temperatures, high-energy photons interact very fast with the thermal bath creating electromagnetic cascades.
The energy injected is rapidly redistributed and the resulting cascade spectrum is~\cite{Kawasaki:1994sc}
\be
\frac{d\,n_\gamma}{d\,E}(E)=
\begin{cases}
K_0\(\frac{E_X}{E}\)^{3/2}\; 	&\mbox{if\ } E<E_X\\
K_0\(\frac{E_X}{E}\)^{2}\; 	&\mbox{if\ } E_X<E<E_C\\
0					&\mbox{otherwhise}
\end{cases}\; ,
\ee
where the normalisation constant $K_0$ is determined equating the total and the injected energies.
The spectrum features a knee and a very sharp cut-off at high energies which depend on the plasma features and temperatures, $E_X\sim m_e^2/(80 T)$ and $E_C\sim m_e^2/(22 T)$. 
At high $T$, this cut off lies below the photo-dissociation threshold of nuclei and these effects are negligible.
Therefore, requiring $E_C(T_{\rm d})>E_{\rm th}$ and converting the decay temperature in seconds, we see that ALP decay cannot dissociate deuterium for $\tau\lesssim 3000$~s and Helium for $\tau\lesssim3\times 10^5$~s.
A second key quantity is the injected energy per photon in the bath
\be
\zeta_\phi\equiv m_\phi \frac{n_\phi}{n_\gamma}=\frac{m_\phi}{2} \frac{g_{*S}\(T_{\rm d}\)}{g_{*S}(T_{\rm fo})} \; .
\ee   
The fractional change for a nuclear abundance is~\cite{Cyburt:2002uv}
\be
\frac{\delta X_i}{X_i}\sim \frac{\zeta_\phi}{2 \eta \langle E \rangle}\(\frac{X_T}{X_i}f_{T\rightarrow i}-f_{i\rightarrow P}\)\;,
\ee
where $\langle E \rangle$ is the average energy in the electromagnetic cascade and $f_{T\rightarrow P}$ is the relative strength of the cross section for the photo-destruction of the target $T$ into the product $P$ compared with the Thomson cross section.
Taking the deuterium case as an example, a fractional change of 1/2 is provided by~\cite{Cyburt:2002uv}
\be
\zeta_\phi^{\rm prod}\sim3.2\times10^{-11}\ {\rm GeV}\(\frac{5000}{X_T/X_D}\)\(\frac{5.0\times10^{-4}}{f_{T\rightarrow D}}\)\(\frac{\eta_{10}}{6}\)
\(\frac{E_{\rm th}}{20\ {\rm MeV}}\)^{1/2}\!\(\frac{\tau}{10^8\ {\rm s}}\)^{1/4}
\ee
if production dominates, while if destruction does it is~\cite{Cyburt:2002uv}
\be
\zeta_\phi^{\rm dest}\sim6.3\times10^{-8}\ {\rm GeV}\(\frac{5.0\times10^{-4}}{f_{D\rightarrow A}}\)\(\frac{\eta_{10}}{6}\)
\(\frac{E_{\rm th}}{2.22\ {\rm MeV}}\)^{1/2}\(\frac{\tau}{10^8\ {\rm s}}\)^{1/4}\; .
\ee
In our analysis, the ALP thermal origin makes the fractional change for D and $^4$He always too strong, and the regions where photo-dissociation is allowed can be safely excluded unless some extra dilution of the ALP population is taken into account.
 
For ALP masses above a few GeV,  ALP decays produce radiatively quark-antiquark pairs that will hadronise. 
Hadronic cascades can dissociate nuclei in a similar fashion of the electromagnetic ones, if they are happening after a typical time of $\tau\sim10^2$~s~\cite{Cyburt:2009pg,Kawasaki:2004qu,Kawasaki:2004yh}. 
For earlier decays, the electromagnetic interactions in the plasma make the hadrons lose rapidly their energy and the nuclear destruction is suppressed.
The effects of hadronic cascades are again necessarily dramatic because of the large ALP relic abundance.  
We consider extremely unlikely that the combined effect of nonstandard BBN with the post-BBN processing gives similar results to standard BBN. So we exclude all the regions where electromagnetic and hadronic cascades play a role, see figure~\ref{fig:DHe}.  
In the literature, cascade constraints are usually presented in terms of $\zeta_{\phi}$, plotted in function of lifetime $\tau$. 
In the case in exam, it is possible to refer to this representation almost directly from figure~\ref{fig:DHe}, considering that the ratio $n_\phi/n_\gamma$, given by equation~\eqref{nphionngamma}, is constant in all the plotted parameter space in which cascades play a role and does not change much outside.

A crucial difference between electromagnetic and hadronic cascades lies in the peculiar effect of pions produced in the latter.
If $m_\phi$ is larger than twice the charged pion mass $m_{\pi^+}=139.57$ MeV, the decay channel $\phi\to \gamma \pi^+\pi^-$ opens up.  
Even if its branching ratio is very small, the abundance of relic ALPs is such that a huge amount of pions --- compared with that of the present nuclei --- can be produced. 
If the decay happens before or during BBN --- typically $\tau\lesssim 100$~s --- the universe is dense enough to make pions induce neutron-proton interconversions, $\pi^+ + n \to \pi^0+p^+$ and $\pi^- + p^+ \to \pi^0+n$, before decaying~\cite{Reno:1987qw}. 
The second reaction is favoured because of the Sommerfeld enhancement and the typical overabundance of protons over neutrons at $T<Q$. 
These reactions will therefore tend to increase the neutron to proton ratio. 
They can do it much even more drastically than the mere presence of the ALP during the $p\leftrightarrow n$ freeze out commented in the previous subsection.  
The higher neutron abundance would aid the heavy element production but it also increases D/H, which is the most important effect.  
Since almost all neutrons end up in $^4$He nuclei taking protons with them, a higher initial neutron abundance yields a smaller final proton abundance and thus a larger D/H ratio. 
If $n_n/n_p\simeq 1$ at the onset of BBN, all protons end up in $^4$He and D/H would be arbitrarily large!\footnote{Of course including neutron decay during BBN would still give a finite, albeit very large, result.}

We have included the effects of pions in our BBN code following reference~\cite{Kawasaki:2004qu}.  
In figure~\ref{fig:DHe} we can see that the low D/H trend of low mass ALPs is drastically changed when the mass gets above $2m_{\pi^+}$ and the effect on $^4$He gets strongly boosted when crossing this boundary. 
The effects of pions are hampered if the ALPs decay very early, for $\tau\lesssim 10^{-2}$ s. 
In this case the electroweak reactions $p^++e^-\leftrightarrow n +\nu_e$ can still re-establish the $n_n/n_p$ equilibrium.
Also very late decay, $\tau\agt 10^{2}$ s, does not affect BBN through pions injection, because $\pi$s fail to interact before decaying.
In this last tiny region our results cannot be taken quantitatively on trust, since we have not taken into account the possibly ineffective slowing down of pions after $e^+e^-$ annihilation.

\begin{figure}[tbp]
\begin{center}
\includegraphics[width=9cm]{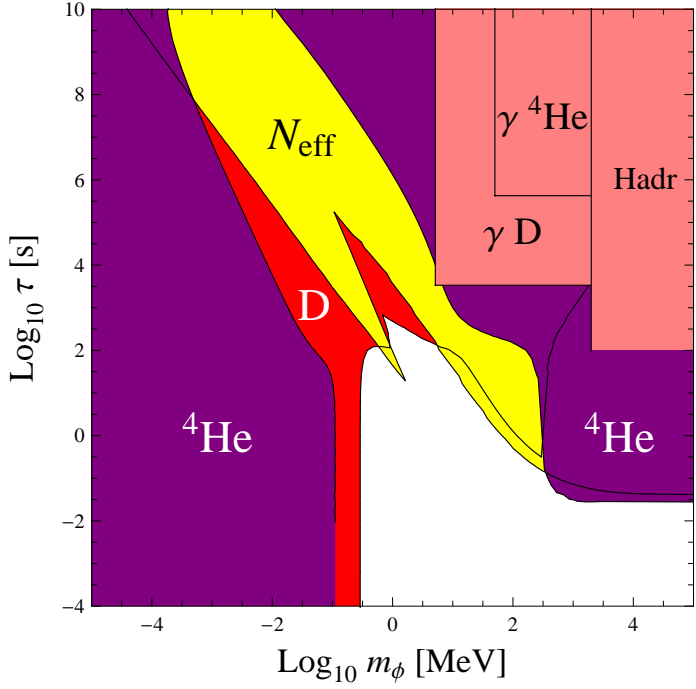}
\end{center}
   \caption{Bounds on early ALP decays from deuterium underproduction (D, red), helium overproduction (He, purple), 
helium photodissociation ($\gamma$He, pink), hadronic cascades (Hadr, pink) and neutrino dilution ($N_{\rm eff}$, yellow).}
   \label{summa}
\end{figure}

The region of the parameter space where cascades are protagonists are depicted in pink in figure~\ref{summa}.
Again, the purple region is excluded by $^4$He overproduction.
Together with the neutrino dilution limit this is the best bound for the $m_\phi>2m_\pi$ region of ALP parameter space.
The parametric dependence of the bound is the same of the $N_{\rm eff}$ limit.
Like in the axion case, short-lived ALPs that decay before BBN can not be constrained.
In this large mass region, it corresponds to $\tau\lesssim10^{-1.5}$~s.

\section{Effect of other couplings}

A few words are in order to consider the effects of other couplings, besides the two-photon one, in the ALP case.
Additional couplings imply in general a higher interaction rate and therefore a later decoupling and a larger number density.
Of course, the more abundant is a population the more important are the consequences of its decay.
However, in our bounds these considerations are only partially true. 
Since we are not directly taking into account possible degrees of freedom beyond the standard model, a population of ALPs with additional couplings would be more abundant only if the new decoupling temperature is lower than $E_{\rm EW}$, otherwise the dilution factor does not change, because $g_*(T>E_{\rm EW})=106.5$.

For a generic PNGB a CP-odd coupling to $G\tilde{G}$ and derivative couplings to fermions are allowed, 
\be\label{eq:ALPfermions}
\mathcal{L}_{\phi f f^\prime}=\frac{C_{ff^\prime}}{f_\phi}\partial_\mu\, \phi \bar{\psi}_f \gamma^\mu \gamma_5 \psi_{f^\prime}\; ,
\ee
the latter even flavour nondiagonal.
The coupling to fermions~\eqref{eq:ALPfermions} allows the decay $\phi\to \bar f f' $ at a rate
\begin{equation}
\Gamma_{\phi\to \bar f f'} =
\( \frac{C_{ff'}}{f_\phi}\)^2  
\frac{\(m_f+m_{f'}\)^2m_\phi}{16\pi}
\sqrt{1-\(\frac{m_f+m_{f'}}{m_\phi}\)^2}
\[1-\(\frac{m_f-m_{f'}}{m_\phi}\)^2\]^{3/2} . 
\end{equation}
which is suppressed with respect to the two-photon decay for small fermion masses.
Writing $g_\phi\equiv C_{\gamma}\alpha/(2\pi f_\phi)$ the $\phi\to \bar f f'$ can dominate only in an interval near the kinematic threshold $1>(m_f+m_f')/m_\phi\gtrsim \alpha C_{\gamma}/4\pi C_{ff'}$.  
For ALP masses above few GeV, the coupling to gluons allows the ALP decay into two gluons at a rate $\Gamma_{\phi\to gg}= 8 (C_{gg}/C_{\gamma})^2 \Gamma_{\phi\to \gamma\gamma}$.

If the ALPs are cosmologically stable, the bounds from direct detection of ALP decay photons and the DM overproduction --- which are described in the previous chapters --- still hold. 
These limits depends mainly on the ALP lifetime and abundance.
The pre-decay abundance can only be increased by additional couplings, but the area constrained by these limits lies all in the $T_{\rm fo}>E_{\rm EW}$ region, where $g_{*S}$ is constant according to our assumption on the primordial particle content.
For a given photon coupling, the lifetime is shortened if more decay channels open.
Thus the limits on the long lifetime range are only mildly affected by the abundance in the $m_\phi$--$\tau$ parameter space, while in the $m_\phi$--$g_\phi$ plane they have to be additionally adjusted according to the different relation between lifetime and photon coupling.  

The short lifetime region corresponds to the constrains described in this chapter.
At low masses, the deuterium and helium bounds come from ALPs in thermal equilibrium with the bath. 
Clearly, adding more couplings between the ALP and SM particles we cannot avoid these bounds. 
In the intermediate mass region 300 keV $\lesssim m_\phi\lesssim $ 2 $m_\mu$, where $m_\mu=105.7$ MeV is the muon mass, the D/H and $N_{\rm eff}$ bounds follow from the dilution of baryons and neutrinos with respect to photons. 
These bounds apply to ALPs decaying into photons or electrons, and indeed we have not made a difference between the two in our equations. 
Again, in the $m_\phi$--$\tau$ plane these bounds suffer only a mild change due to the different abundance.
When we translate the bounds in the $m_\phi$--$g_\phi$ plane they will show \emph{worst} if the decay into two electrons dominates than if we only consider the two photon coupling, because what is important for the bound is the total decay rate.
The lower bound on $g_\phi$ reduces by a factor $\sim \(2\pi C_{e e} m_e\)/\(\alpha C_{\gamma} m_\phi\)$. 

Since the direct decay into neutrinos is suppressed by $\sim (m_\nu/m_\phi)^2$, an ama\-zingly tiny number, the bounds are perfectly valid provided one interprets $\tau$ as the total lifetime, not only due to the two photon decay channel. 
Of course, this is valid unless one considers sterile neutrinos with $m_\nu\sim m_\phi$. 
Then in this case neutrinos have a strong tendency to constitute too much DM. 
A way to avoid this is to make them decay into a SM neutrino plus a photon, but this produces entropy and we expect a similar, slightly smaller, bound from D/H in this case. 
In these models the low $N_{\rm eff}$ tendency is reversed since the sterile neutrinos produce neutrinos in its decay.

If $2 m_{\pi^+}>m_\phi>2 m_\mu$ we have a somewhat different scenario where the ALP tends to favour the $\phi\to \mu^+ \mu^-$ decay. 
The upper limit on $\tau$ in this region comes from having too low $N_{\rm eff}$ already before BBN. 
But if the decay into muons dominates we will rather have a high $N_{\rm eff}$. 
In this case the amounts of energy released in electrons and in neutrinos by muon decay $\mu\to e \bar \nu\nu$ are similar.
Since data favours values larger than the standard $N_{\rm eff}=3$, the $N_{\rm eff}$ bound will relax somehow. 
Anyway, we do not expect them to disappear, because ALPs can still produce too many neutrinos. 
Also in this case the bound on deuterium should come from a too high D/H, which is less conservative a constraint. 
In any case the bound from He will stay since it mainly comes from a high $\eta^{\rm BBN}$ and the ALP contribution to the expansion at the freeze out of $p\leftrightarrow n$ weak reactions. 

Finally, for $m_\phi>2 m_{\pi^+}$ the most stringent bound comes from $^4$He overproduction due to the presence of charged pions before BBN, enhancing the neutron to proton ratio. 
As we commented, this bound does depend very little on the details and branching ratios of the ALP since only a minimal number of pions are sufficient for a drastic change. 
Therefore we expect it not to change very much. 
However, when quoting this constraint in the $m_\phi$--$g_\phi$ plane this bound would display in a lower position than in the case where only the two-photon coupling is considered.
Only in this region the coupling to two gluons can affect the ALP decay and will certainly increase the pion multiplicity of the decay making the bound on $\tau$ slightly better. 
The decay into muons can dominate if $m_\phi$ is not too far from $2m_\mu$ and all said in the above paragraph holds. 
It appears that the helium bound will still be the most relevant in this case.



\chapter{Summary and conclusion}\label{chap:conclusion}

\label{sec:introcosmology}

The axion is a side product of the elegant solution of the strong CP-problem proposed by Roberto Peccei and Helen Quinn in 1977.
The different realisations of the Peccei-Quinn idea produced a variety of axion models, which are tested in particle physics laboratories, and challenged in astrophysical and cosmological observations.
Moreover, it seems that particles with similar characteristics to the axion could arise in several extensions of the standard model of particle physics.
In particular, string theory seems a fertile environment that can provide plenty of these axion-like particles.
One day, the so craved experimental test of string theory could finally come from ALP-related observations.

In this dissertation we have depicted the constraints that cosmology puts on the existence of such particles.
In chapter~\ref{chap:intro}, we started with a presentation of the axion theory, together with a brief motivation for ALPs.
Then, we introduced the general limits on the parameter space of axions and ALPs.
We listed the direct experimental tests and the astrophysical observations that exclude the existence of these pseudoscalar for determined choices of mass and couplings.
But to give new hope, we also reviewed some astrophysical problems that could be solved by pseudoscalar particles in two different ranges of the parameter space.

Motivated by the possible solution of particle and astroparticle problems, we continued our analysis in chapter~\ref{chap:DM}, where we treated the possibility for a primeval population of pseudoscalars to arise in the early universe. 
This enquiry sets the basis for the subsequent discussion about cosmological limits, and most importantly provides the motivation to consider axions and ALPs as a constituent of the dark matter of the universe.
So the existence of these particles can provide the solution of a further problem of modern physics.
In this chapter we illustrated a very rich phenomenology.
Axions and ALPs can be created by thermal interactions with the particles of the primordial plasma.
But the non-thermal creation --- via the so called realignment mechanism --- is even more interesting, since it involves the physics of spontaneous symmetry breaking and phase transitions applied to the primeval universe, and may remarkably provide the explanation to the dark matter. 
A first limit on the parameter space can be put at this stage, if too much dark matter is produced.

Also the cosmological stability of pseudoscalars is discussed in chapter~\ref{chap:DM}.
Their peculiar two-photon coupling provides them with a decay channel. 
Therefore, a pseudoscalar population can decay in photons or be reabsorbed by the primordial thermal bath, if this electromagnetic interaction is active.
We briefly treated also the role primordial magnetic fields could have in this process, to further develop the discussion in the appendix~\ref{app:axphotmix}.

The decay of a population of particles during the early epochs of the universe can have dramatic consequences and leave an indelible imprint on cosmological observables.
This is the topic of the second part of this study.
We have divided the limits into two sets, according to the epoch of the decay.
Both of them are dealing with the influence of photons produced by the pseudoscalar decay on the successive evolution of the cosmos.

In chapter \ref{chap:photons} we collected the limits related to late decay and to photon detection.
These could either be the CMB photons, which could have imprinted in their spectrum the distortions caused by the decay products, or the decay photons themselves, which could have freely travelled towards our telescopes.
Observations of photon spectra leave very little room to pseudoscalar decay --- especially in the CMB case --- and the decay of a whole population can be safely excluded, unless it happened when electron-photon interactions were active and able to thermalise rapidly the injected photons.
In this chapter we constrained rather long lifetimes, since the distortion of the CMB requires the decay to happen just before the recombination era, which occurred when the universe was roughly 400,000 years old. 
The direct observation of the decay products excludes cosmologically stable ALPs and axions up to $\tau\sim10^{24}$--$10^{28}$~s.
A section in this chapter is dedicated to the ultraviolet radiation that can be eventually emitted in late decays.
The universe is very opaque to ultraviolet light.
This kind of photons can not appear in direct observations, but they can nevertheless be constrained because of their effect on the ionisation history of the universe. 

Early decays that do not affect the CMB spectrum can be constrained by the arguments of chapter~\ref{chap:dilution}.
The entropy transfer and increase due to the decay of a relic population have the leading role in the discussion. 
We first discussed the case of a population that decays in local thermal equilibrium.
Entropy is conserved and just transferred to the species in thermal contact with the disappearing population.
The effects of an out-of-equilibrium decay can be more dramatic, especially if the pseudoscalar population is dominating the energy density of the universe before the decay.
A large amount of entropy is created and transferred to the photon bath.
In both cases, the subsequent evolution towards thermal equilibrium makes the temperature of the species in thermal contact with photons to increase relatively to the decoupled ones.
In this sense we defined this event as a dilution of the decoupled species, the best example being neutrinos.
Solving numerically the set of Boltzmann equations that describes the evolution of pseudoscalar, photon, electron and neutrino populations, we calculated the effect of the decay on neutrinos.
We then compared our neutrino dilution scenario with cosmological neutrino observations.
Through CMB multipole analysis and LSS survey it is possible to measure the number of effective neutrino species $N_{\rm eff}$.
Present data prefer $N_{\rm eff}>3$, which works against the cosmological dilution of neutrinos.
After this observation, we were able to put some limits on pseudoscalar parameter space.

The outcome of primordial nucleosynthesis is also influenced by the pseudoscalar decay. 
Both directly, if pseudoscalars are massive enough to inject decay products energetic enough to break nuclear bonds, and indirectly, through the influence on the number densities of baryons and neutrinos relative to the photon one.
This topic is also part of chapter~\ref{chap:dilution}. 
The primordial yield of deuterium is very sensitive to the baryon-to-photon ratio, which is heavily perturbed in our scenario of early pseudoscalar decay.
It indeed provides the most restrictive cosmological lower bound on the axion mass, $m_a>0.3$ MeV. 
Also ALPs are severely constrained by BBN, since they affect the baryon-to-photon ratio, $N_{\rm eff}$ and can generate destructive electromagnetic and hadronic cascades.
The production of pion cascades in the decay of $m_\phi>2m_\pi$ ALPs has a peculiar effect on the primordial $^4$He outcome, which we constrained too.
Chapter~\ref{chap:dilution} concludes with some considerations on the changes in the cosmological bounds that further ALP couplings could provide. 
The conclusion here is that cosmological limits, although slightly modified, are very solid.

\begin{figure}[tbp]
\begin{center}
\includegraphics[width=13cm]{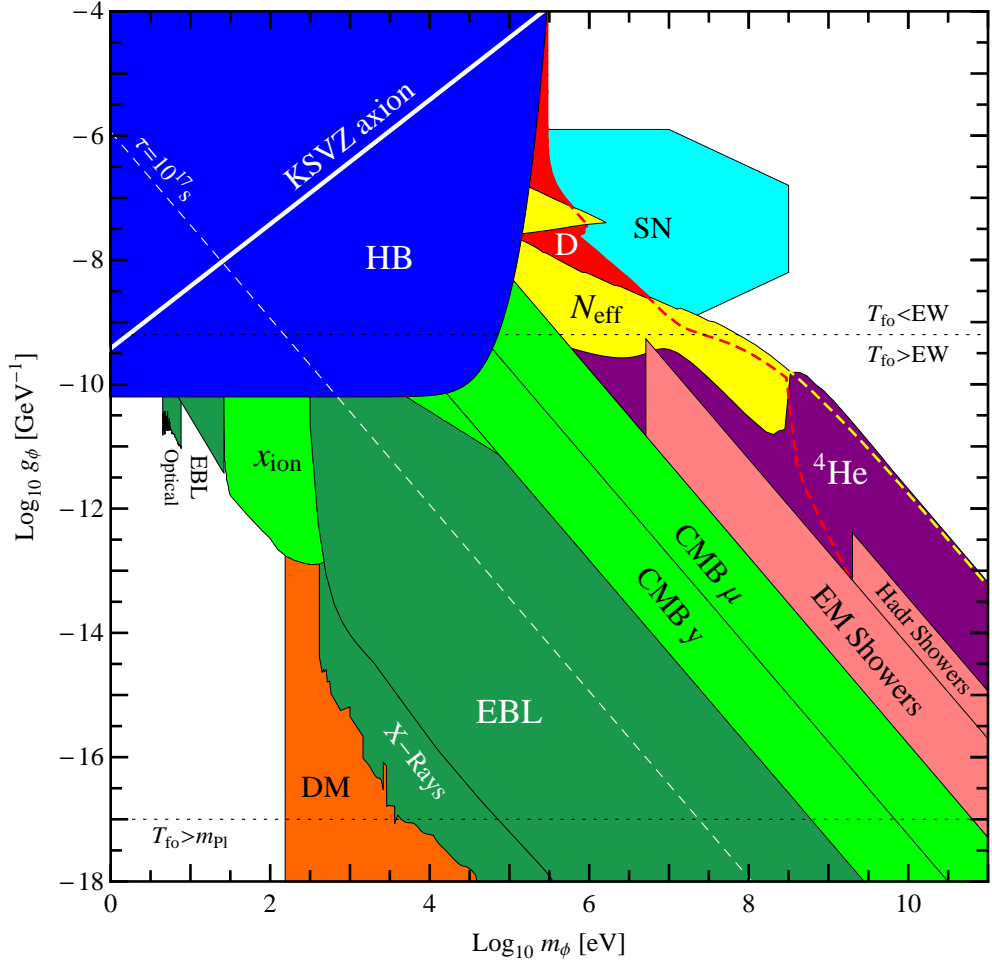}
\end{center}
\vspace{-.6cm}
   \caption{Cosmological ALP bounds in the $m_\phi$-$g$ parameter space. 
   The labelling is described in the text.}
   \label{fig:bounds}
\end{figure}

\begin{figure}[tbp]
\begin{center}
\includegraphics[width=13cm]{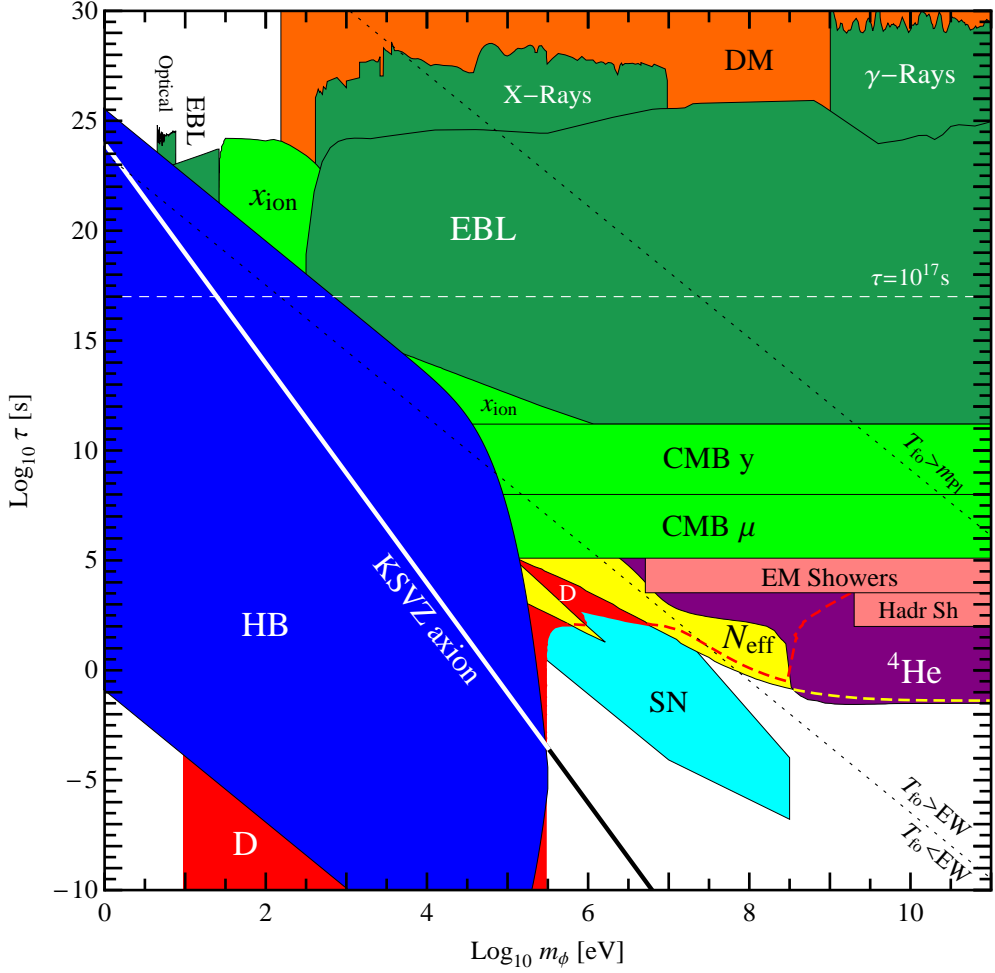}
\end{center}
\vspace{-.6cm}
   \caption{Same as figure~\ref{fig:bounds}, but in the $m_\phi$-$\tau$ parameter space.}
   \label{fig:boundsLife}
\end{figure}

In our summary plots~\ref{fig:bounds} and~\ref{fig:boundsLife} we collected all the bounds arising from co\-smology, considering a thermal origin for the primordial pseudoscalar population.
While figure~\ref{fig:bounds} is meant to be the reference for the particle physicist, since the ALP parameter space is plotted in function of the ALP mass $m_\phi$ and coupling $g_\phi$, figure~\ref{fig:boundsLife}, where the parameters are $m_\phi$ and the lifetime $\tau$, provides a better physical understanding. 
In this picture we can better follow the relation of the limits with the sequence of events in the history of the universe.
The cosmolo\-gical bounds critically depend on the ALP lifetime while they have only a milder dependence on other parameters. 
In figure~\ref{fig:bounds} they have indeed the characteristic slope $g_\phi\propto {m_\phi}^{-3/2}$.
As a reference, we plotted the line along which ALPs have the lifetime equal to $10^{17}$~s, i.e.~the age of the universe.
To warn the reader about our assumptions, we also plotted the lines for freezing out temperature $T_{\rm fo}$ equal to the electroweakscale $E_{\rm EW}$ and the Planck scale $m_{\rm Pl}$.
Most likely, $T_{\rm fo}>m_{\rm Pl}$ makes no physical sense, and for $T_{\rm fo}>E_{\rm EW}$ the actual abundance could be lower than that used in our calculations, since we have not taken into account other particles besides the ALP and the SM ones.  

Ordered by decreasing lifetimes, the excluded regions are: 
\begin{itemize}
\item {\bfseries DM} --- if ALPs are cosmologically stable and behave as dark matter they should not exceed the DM fraction measured by WMAP.
\item {\bfseries Optical, X-Rays, $\boldsymbol{\gamma}$-Rays} --- photons produced in pseudoscalar decays inside galaxies would show up as a peak in galactic spectra that must not exceed the known backgrounds.
\item {\bfseries EBL} --- photons produced in pseudoscalar decays when the universe is tran\-sparent must not exceed the extragalactic background light. 
\item {\bfseries $\mathbf{x_{\rm ion}}$} --- the ionization of primordial hydrogen caused by the decay photons must not contribute significantly to the optical depth after recombination.
\item {\bfseries CMB y, $\boldsymbol\mu$} --- if the decay happens when the universe is opaque, the decay photons must not cause spectral distortions in the CMB spectrum that cannot be fully rethermalised.
\item {\bfseries EM, Hadr showers} --- the decay of high mass ALPs produces electromagnetic and hadronic showers that must not spoil the agreement of big-bang nucleosynthesis with observations of primordial nuclei. 
\item {\bfseries $\mathbf{^4}$He, D} --- the ALP and axion decays produce photons that dilute the baryon and neutrino densities, whose values affect the outcome of BBN, in particular the deuterium and $^4$He yields. Again, this dilution should not compromise BBN.
\item {\bfseries $\mathbf{N_{\rm eff}}$} --- the neutrino density must not disagree with the value measured by WMAP and other large-scale-structure probes. 
Currently, data points to an effective number of neutrinos $N_{\rm eff}$ greater than 3, 
which is disfavoured in the cosmology of decaying pseudoscalars.
\end{itemize}

We have seen how cosmological observations can exclude a large part of the pseudoscalar parameter space. 
These limits are solid, and the prospects are good, since the amount and quality of cosmological observations is steeply rising.
This is very important, since the constrained region lies in a part of the parameter space presently inaccessible to direct experimental tests.

The investigation on the effects of axions and ALPs on astrophysical and co\-smological observables must proceed further, because the detection of one of several striking signatures could lead to their discovery.
The smoking gun could be hidden in the white-dwarf evolution or in the structure of galaxies, maybe influenced by the dark matter in a Bose-Einstein condensate, or in some different phenomena that we still have to analyse. 
But after the eventual and desired discovery, the direct detection and laboratory experiments will have the task of measuring precisely the new particle characteristics.
In the near future, direct detection experiments like the haloscopes and the helioscopes will finally reach the sensitivity to test the axion hypothesis and to explore two phenomenologically important regions of the parameter space, and maybe some long standing questions will find an answer.

Understanding the ultimate theory behind the laws of Nature is the final purpose of physics.
The large hadron collider is pushing the knowledge on particle physics towards unexplored energy scales. 
The recent discovery of the Higgs boson gives a bit more confidence about the axion theory, which requires the existence of scalar fields and the spontaneous breaking of symmetries at high energy scales.
But increasing the energy tested by colliders is not the only way to probe the physics beyond the SM. 
If axions and ALPs will be finally discovered, we will have the tempting possibility of exploring the physics related to very high energy scales through its low energy regime.
Still largely unexplored, the low energy frontier could hide important novelties and bring to fundamental advances in the understanding of Nature.

\appendix

\chapter{Axion-photon mixing}\label{app:axphotmix}
In the following we follow the treatment of \cite{Redondo:2008ec} and \cite{Redondo:2008aa}, where the hidden photon resonance case was studied.
The Lagrangian of the pseudoscalar-photon system in presence of a strong magnetic field $\vec{B}_{\rm ext}$, whose modulus is $B_{\rm ext}$, is
\be
\mathcal{L}=-\frac{1}{4}F^{\mu\nu}F_{\mu\nu}+A^\mu j_\mu+\frac{1}{2}\partial^\mu \phi \partial_\mu \phi - \frac{1}{2}{m_\phi}^2 \phi^2
-g_\phi \phi \vec{B}_{\rm ext}\cdot \partial_0 \vec{A}\; ,
\ee
where we expressed $F\tilde{F}/4$ as the scalar product of the external magnetic field and of the electric field component $-\partial_0 \vec{A}$ of the field strength. 
From the Lagrangian we obtain the equations of motions for the pseudoscalar field $\phi$ and the component of the vector potential parallel to the magnetic field $A_\sslash$
\begin{subequations}\label{eq:EOM1}
\begin{align}
&\Box A_{\sslash}-\sigma \partial_0 A_{\sslash} + g_\phi B_{\rm ext} \partial_0 \phi=0\; ,\label{eq:EOM1a}\\
&\Box \phi+m_\phi^2 \phi + g_\phi B_{\rm ext} \partial_0 A_{\sslash}=0\; ,\label{eq:EOM1b}
\end{align}
\end{subequations}
while the perpendicular component $A_{\bot}$ is not affected by the interaction with $\phi$.
In equation \eqref{eq:EOM1a} we have used Ohm's law for the current density, $j^{\mu}=-\sigma \partial_0 A^\mu$, assuming a linear response of the medium whose conductivity is $\sigma$ \cite{Ahonen:1995ky}.
In Fourier space the equations \eqref{eq:EOM1} become
\begin{subequations}\label{eq:EOM2}
\begin{align}
&\(-\omega^2+k^2-i \omega \sigma\) A_{\sslash} + i g_\phi B_{\rm ext} \omega \phi=0\; ,\label{eq:EOM2a}\\
&\(-\omega^2+k^2+{m_\phi}^2\) \phi + i g_\phi B_{\rm ext} \omega A_{\sslash}=0\; .\label{eq:EOM2b}
\end{align}
\end{subequations}
Assuming the quantity $\chi=\sqrt{g_\phi B_{\rm ext}\omega}$ to be negligible with respect to the other energy scales of the problem, this system of equations can be diagonalised by means of the redefinitions
\begin{subequations}\label{eq:EOMshift}
\begin{align}
A_{\sslash}=\hat{A}_{\sslash}-\frac{ i g_\phi B_{\rm ext} \omega}{i \omega \sigma-{m_\phi}^2}\hat{\phi}\; ,\\
\phi=\hat{\phi}-\frac{ i g_\phi B_{\rm ext} \omega}{i \omega \sigma-{m_\phi}^2}\hat{A}_{\sslash}\; .
\end{align}
\end{subequations}
The new states $\hat{A}_{\sslash}$ and $\hat{\phi}$ are decoupled, and if the mixing parameter
\be\label{eq:EOMmixing}
\chi_{\rm eff}^2=\left|\frac{ i g_\phi B_{\rm ext} \omega}{i \omega \sigma-{m_\phi}^2}\right|^2\; 
\ee
is very small, they are almost photon and pseudoscalar states.
They can oscillate one into another, if $m_{\phi}\simeq \omega$.
Their dispersion relations are respectively
\begin{subequations}\label{eq:EOMdispersion}
\begin{align}
&\omega^2-k_{A_{\sslash}}^2=-i \omega \sigma+\mathcal{O}(\chi^2)\; ,\\
&\omega^2-k_{\phi}^2=m_\phi^2+\mathcal{O}(\chi^2)\; ,
\end{align}
\end{subequations}
where the $\mathcal{O}(\chi^2)$ factors are complex.
In the case of zero mode DM scalars, $k_\phi=0$.
The original states $A_\sslash$ and $\phi$ can be obtained inserting the $\hat{A}_\sslash$ and $\hat{\phi}$ solutions into the \eqref{eq:EOMshift}. 

For the conductivity of the medium we use the relation~\cite{Arias:2012mb}
\be\label{eq:conductivity}
i\sigma\omega=\omega_{\rm P}^2\frac{\(\omega \tau_{\rm coll}\)^2+i \omega\tau_{\rm coll}}{1+\(\omega \tau_{\rm coll}\)^2}+i \omega \Gamma_{\rm Th}\equiv m_\gamma^2(T)+i\omega D(\omega,T)\; ,
\ee
that provides an interpolation of the classical approximation and the quantum regime.
In this formula, $\omega_{\rm P}$ is the plasma frequency and $\Gamma_{\rm Th}=\sigma_{\rm Th}n_e$ is the photon absorption width due to Thomson scattering off electrons, while $\tau_{\rm coll}$ is the average time between electron collisions, which sets the time scale.
If we call $p$ and $E$ the momentum and the energy of electrons, the plasma frequency is~\cite{Raffelt:1996wa} 
\be
\omega_{\rm P}^2=
\frac{4\alpha}{\pi}\int_0^\infty dp\frac{p}{\exp\(E/T\)+1}\[\frac{p}{E}-\frac{1}{3}\(\frac{p}{E}\)^3\]\; .
\ee 
We next define the squared plasmon mass
\be
{m_\gamma}^2=\mbox{Re} \(i\sigma\omega\)=
\begin{cases}
4\pi \alpha n_e/m_e\ &\mbox{if}\ T\ll m_e\, \\
\frac{2}{3}\alpha\pi T^2 \ &\mbox{if}\ T\gg m_e
\end{cases}\;.
\ee
The damping factor
\be
D(\omega,T)=\mbox{Im} \(i\sigma\)\; 
\ee 
expresses the net-rate at which photons are created or absorbed.
The full expressions for these quantities can be found in~\cite{Redondo:2008ec}.
Using these definitions, the oscillation probability \eqref{eq:EOMmixing} becomes
\be\label{eq:EOMmixing2}
\chi_{\rm eff}^2 \equiv \left|\frac{ i g_\phi B_{\rm ext} \omega}{i \omega \sigma-{m_\phi}^2}\right|^2=\frac{ \(g_\phi B_{\rm ext} \omega\)^2}{ \(m_\gamma^2-m_\phi^2\)^2+\(\omega D\)^2}\; .
\ee
Because of this magnetic field-mediated mixing with the photon, the pseudoscalar field has the same interactions of the photon, but with a rate suppressed by $\chi_{\rm eff}^2$.
In particular, it can be absorbed or emitted during Thomson-like interactions $\phi+e \leftrightarrow \gamma +e$ and annihilations-like events $\phi+\gamma\leftrightarrow e^+ +e^-$.
If thermal contact is regained by the pseudoscalar condensate, its distribution tends towards equilibrium.
The amount of pseudoscalars that survives the magnetic field induced evaporation is provided by 
\be\label{eq:survivingB}
n_\phi(t_0)=n_\phi(t_i)\(\frac{R_i}{R_0}\)^3\exp\(-\int_{t_{i}}^{t_{0}}\chi_{\rm eff}^2 D \,dt\)\; ,
\ee
where the subscripts $i$ means quantities measured just after the realignment me\-chanism ended and $0$ are today quantities.
We used in the integral the damping factor $D$, that provides the net-rate of absorbed photons.
Because we deal with temperature-depending quantities, for future convenience we can write the integral in the exponential of equation \eqref{eq:survivingB} as
\be\label{eq:tempB}
-\int_{t_{i}}^{t_{0}}\chi_{\rm eff}^2 D \,dt\simeq\int_{T_{B}}^{T_{0}}\frac{ \(g_\phi B_{\rm ext} \omega\)^2}{ \({m_\gamma}^2-{m_\phi}^2\)^2+\(\omega D\)^2}\frac{D}{H} \frac{dT}{T}\; .
\ee

The primordial magnetic field evolves with time, since it is stretched by the expansion of the universe.
The conservation of magnetic flux implies $B_{\rm ext}=B_{{\rm ext},*}(R_*/R)^2\sim B_{{\rm ext},*}(T_*/T_B)^2$, where the subscript $*$ refers to quantities measured at the reference time $t_*$, after the creation of the primordial magnetic field at $T_B$.

The temperature dependence of the conductivity~\eqref{eq:conductivity} is mainly affected by $\tau_{\rm coll}\sim\(\sigma_{\rm e\gamma} n_e\)^{-1}\propto T^{-1}$.
For $T\gg\omega$ the conductivity is dominated by the damping factor, thus $\chi_{\rm eff}^2\sim\(g_\phi B_{\rm ext} \)^2/D^2$ if also $m_\phi/\omega\ll D$. 
The damping factor is $D= T/(3\pi\alpha)$ in a relativistic plasma of electrons~\cite{Ahonen:1995ky}, and equation~\eqref{eq:tempB} becomes
\begin{align}
\int_{T_{B}}^{T_{0}}\frac{ \(g_\phi B_{\rm ext}\)^2}{ D^2}\frac{D}{H} \frac{dT}{T}\sim
\frac{g_\phi^2 B_0^2}{T_0^4}\int_{T_{B}}^{T_{0}}\frac{T^4}{T/(3\pi\alpha)} \frac{m_{\rm Pl}}{1.66 \sqrt{g_*(T)}T^2}\frac{dT}{T}\approx \nonumber\\
\approx\(\frac{g_\phi}{10^{-10}\ {\rm GeV}^{-1}}\)^2\(\frac{B_0}{ {\rm nG}}\)^2\(\frac{106}{g_*(T_B)}\)^{1/2}\frac{T_B}{10^{9}\ {\rm GeV}}\; .
\end{align}
For axion cold DM this is the approximation to use, because the axion mass is much lower than the damping factor~\cite{Ahonen:1995ky}.
Axion cold DM is not affected by primordial magnetic field: its two photon coupling strength is much lower than $10^{-10}\ \mbox{GeV}^{-1}$, while the present value of the magnetic field $B_0$ can not be stronger than few nG~\cite{Kahniashvili:2010wm,Paoletti:2010rx}.
Even if $T_B$ is very large, it cannot affect the condensate at temperatures larger than $f_a$, because in this case there is no condensate.
Moreover, in the case of the hadronic axion, there is no two-photon coupling above $\Lambda_{QCD}$.

When the damping factor does not suppress $\chi_{\rm eff}$ and it is much smaller in magnitude than the plasmon mass, we can distinguish three different regimes of this quantity, according to the relative magnitude between $m_\gamma(T)$ and $m_\phi$. The resonant regime for $m_\gamma(T)\sim m_\phi$ is the most important, since the mixing gets enhanced by the cancellation of the mass terms in the denominator.

\begin{figure}[tb] 
   \includegraphics[width=100mm]{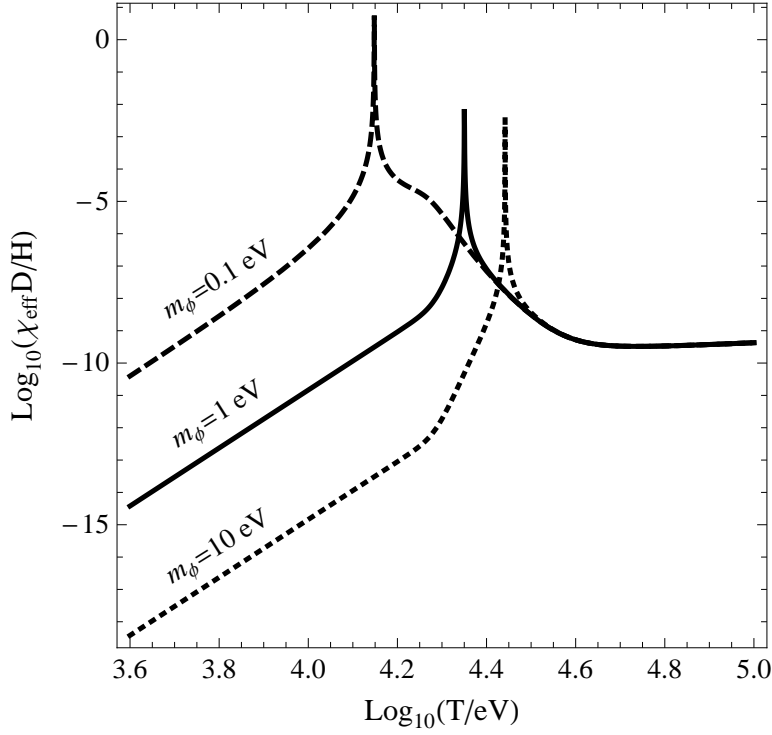}   
   \centering
   \caption{The integrand of equation~\eqref{eq:tempB} for $g_\phi=10^{-10}\ {\rm GeV}^{-1}$, $B\sim T^2$ and $m_\phi=0.1$~eV (dashed), 1~eV (solid) and 10 eV (dotted).}
   \label{fig:resonance}
\end{figure} 

At temperatures higher than the resonant one, equation~\eqref{eq:tempB} is
\be
\frac{ \(g_\phi B_{\rm ext} \omega\)^2}{ \({m_\gamma}^2-{m_\phi}^2\)^2+\(\omega D\)^2}\frac{D}{H T} \sim
\begin{cases}
\frac{ \(g_\phi B_{{\rm ext},*} \omega\)^2 \(T/T_*\)^4}{ \alpha^2{T}^4}\frac{T/\alpha}{T^3/m_{\rm Pl}} & T\gg m_e\\
\frac{ \(g_\phi B_{{\rm ext},*} \omega\)^2 \(T/T_*\)^4}{ \alpha^2{\(n_e/m_e\)}^2}\frac{T^{3/2}m_e^{-1/2}\alpha^{-1}}{T^3/m_{\rm Pl}} & T\ll m_e
\end{cases}\; ,
\ee
which in both cases is dominated by lower temperatures, so towards the resonance.
For the low temperature we used the conductivity provided by a hydrogen plasma~\cite{Ahonen:1995ky}.
The low temperature regime gives instead
\be
\frac{ \(g_\phi B_{\rm ext} \omega\)^2}{ \({m_\gamma}^2-{m_\phi}^2\)^2+\(\omega D\)^2}\frac{D}{H T} \sim
\begin{cases}
\frac{ \(g_\phi B_{{\rm ext},*} \omega\)^2 \(T/T_*\)^4}{ m_\phi^4}\frac{T/\alpha}{T^3/m_{\rm Pl}} & T\gg m_e\\
\frac{ \(g_\phi B_{{\rm ext},*} \omega\)^2 \(T/T_*\)^4}{ m_\phi^4}\frac{T^{3/2}m_e^{-1/2}\alpha^{-1}}{T^3/m_{\rm Pl}} & T\ll m_e
\end{cases}\; ,
\ee
which is dominated by the higher temperatures again towards the resonance.
The behaviour of the integrand can even better understood from figure~\ref{fig:resonance}, where this quantity is plotted for $m_\phi=0.1$, 1 and 10~eV.
Since this function is so peaked around the resonance, we can approximate the integral evaluating all the temperature dependent quantities around $T_{\rm res}$, expanding $m_\gamma^2\simeq m_\phi^2 + dm_\gamma^2/dT\(T-T_{\rm res}\)$,
and neglecting the rest
\begin{align}\label{eq:resonant}
\int_{T_{i}}^{T_{0}}&\frac{ \(g_\phi B_{\rm ext} \omega\)^2}{ \(m_\gamma^2-m_\phi^2\)^2+\(\omega D\)^2}\frac{D}{H} \frac{dT}{T}\sim \nonumber\\
&\sim\int_{T_{i}}^{T_{0}}\frac{\(g_\phi B_{\rm ext,res} \omega\)^2 D}{\(dm_\gamma^2/dT\)^2\(T-T_{\rm res}\)^2+\(\omega D\)^2}\frac{dT}{H_{\rm res}T_{\rm res}}\approx \nonumber\\
&\approx \frac{\pi}{2}\frac{\(g_\phi B_{\rm ext,res}\)^2}{H_{\rm res}m_\phi}\frac{ 1}{ \( d\ln m_\gamma^2/d\ln T\)_{\rm res}}\; .
\end{align}
To obtain the last line we used the approximation 
\be
\int_0^\infty \frac{dx} {y^2(x-1)^2+z^2}\simeq\frac{\pi}{2}\frac{1+\arctan(y/z)}{yz}\; ,
\ee
and $\omega\sim m_\phi$.
The pseudoscalar population is affected by the resonant magnetic field evaporation if the value of the integral~\eqref{eq:resonant} is larger than one.
This is plotted in figure~\ref{fig:Bevaporation}.

    \newpage

  \backmatter
\addcontentsline{toc}{chapter}{{Bibliography}}
\renewcommand*{\bibname}{B\MakeLowercase{ibliography}} 
\markboth{Bibliography}{Bibliography}
  \bibliographystyle{JHEPMS}
  \bibliography{main}
  \markboth{}{}

  \chapter*{Acknowledgement}

First of all, I would like to thank Dr.~Javier Redondo and Dr.~Georg Raffelt for their supervision of the Ph.D.~project and the careful reading of this thesis. 
A big ``thank you'' is also for my collaborators, it has been a pleasure to work with them. 
In particular I want to thank Dr.~J\"org J\"ackel for reading the manuscript and giving me feedback.

I want to thank my office mates, Peter Graf and Dr.~Sr{\dj}an Sarikas, and my colleagues, for the nice conversations about physics and the world around us.
Moreover, the German abstract would not have been in this shape if not for Peter.
Special thanks go also to the ``smokers \& friends club'': Dr.~Lorenzo Calibbi, Dr.~Javier Redondo, Dr.~Toshihiko Ota, Dr.~Daniel Greenwald  and Mr.~Davide Pagani. They alleviated the fatigues of the work with nice and refreshing afternoon breaks.

Finally, I am very indebted to my wife Sara for her support, her patience and most of all her choice to follow me here to Munich. Thanks.

\end{document}